\pdfoutput=1
\newif\ifarxivversion
\arxivversiontrue 
\ifarxivversion\else\PassOptionsToClass{linenumbers}{revtex4-2}\fi
\documentclass[aps,prd,10pt,superscriptaddress,twocolumn,amsmath,amssymb]{revtex4-2}

\usepackage{mathtools}
\usepackage{slashed}
\usepackage{tensor}
\usepackage{bbold}
\usepackage{amscd}
\usepackage{dsfont}
\usepackage{mathrsfs}
\usepackage{graphicx}
\usepackage[percent]{overpic}
\usepackage{multirow}
\usepackage{calrsfs}
\DeclareMathAlphabet{\pazocal}{OMS}{zplm}{m}{n}

\usepackage[dvipsnames]{xcolor}
\usepackage{tikz}
\usetikzlibrary{arrows.meta,calc,angles,quotes}

\definecolor{brickred}{rgb}{0.8, 0.25, 0.33}
\newcommand\myshade{85}
\colorlet{mylinkcolor}{BrickRed}
\colorlet{mycitecolor}{NavyBlue}
\colorlet{myurlcolor}{Aquamarine}

\usepackage{needspace}
\usepackage{hyperref}
\hypersetup{
  linkcolor  = mylinkcolor!\myshade!black,
  citecolor  = mycitecolor!\myshade!black,
  urlcolor   = myurlcolor!\myshade!black,
  colorlinks = true,
}
\usepackage{orcidlink}

\allowdisplaybreaks

\newcommand{\bv}[1]{\boldsymbol{#1}}

\begin{document}
  
\title{Curved geometry of collective neural activity across the mouse brain}

\author{Pablo A. Morales \orcidlink{0000-0002-0965-5199}}
\email{pablo\_morales@araya.org}
\affiliation{Research Division, Araya Inc., Tokyo, Japan}
\affiliation{Centre for Complexity Science, Imperial College London, London, UK
}

\author{Th\'eo Marchal \orcidlink{0009-0000-7311-8614}}
\affiliation{LMPS, ENS Paris-Saclay, Gif-sur-Yvette, France}

\author{Hideaki Shimazaki \orcidlink{0000-0001-7794-3064}}
\email{h.shimazaki@i.kyoto-u.ac.jp}
\affiliation{Graduate School of Informatics, Kyoto University, Kyoto, Japan}

\begin{abstract}
Neural populations exhibit statistical structure beyond firing rates and pairwise correlations, yet how its higher-order dependencies vary across brain regions remains unknown. We model this structure using curved neural networks, in which a single Rényi deformation parameter generates a constrained hierarchy spanning all interaction orders. In recordings of spontaneous neural activity from 39 cortical and subcortical regions in mice, third spike-count cumulants show widespread negative deviations from flat pairwise maximum-entropy predictions. Fits of the curved model identify predominantly negative curvature, most pronounced in visual populations and closer to the flat limit in midbrain and frontal populations. We also derive a deformed mean-field equation whose closed-form inverse recovers network parameters and thermodynamics under weak correlations and small deformation. The regional comparison probes limits of our analytical approximation, which remains applicable to most frontal populations but breaks down under stronger deformation and correlations in visual populations and heterogeneous pairwise correlations in the midbrain. Rényi curvature thus provides a compact measure of brain-wide higher-order structure, revealing nonlinear dependencies in neural population activity.

\end{abstract}

\maketitle

\makeatletter
\let\theo@acl\addcontentsline
\renewcommand\addcontentsline[3]{\def\@tempa{toc}\def\@tempb{#1}\ifx\@tempa\@tempb\else\theo@acl{#1}{#2}{#3}\fi}
\makeatother

\section*{Introduction}
Pairwise maximum-entropy models provide compact baseline descriptions of collective systems by matching one-body and pairwise observables. They are widely used in statistical physics~\cite{jaynes1957information}, biology~\cite{nguyen_inverse_2017}, and the social sciences~\cite{castellano2009statistical}, including analyses of neural populations~\cite{schneidman2006weak,Shlens_etal_2006,cocco2009neuronal,lynn2025exact}. Yet many such systems exhibit higher-order dependencies that manifest in triplet and larger-group statistics not fixed by the matched observables~\cite{battiston2020networks,battiston2021physics}. Models that are identical at the pairwise level can therefore make widely different predictions for rare collective events, coarse-grained behavior, and responses to perturbations.

In neural systems, departures from pairwise descriptions have been reported in cortical, hippocampal, and retinal recordings~\cite{ganmor2011sparse,Shimazaki_etal_2015,shomali2026inferring}.
Such higher-order structure is commonly associated with sparse activity~\cite{Ohiorhenuan_etal_2010,Shimazaki_etal_2015,rodriguez2025higher} and with the shape of population spike-count distributions, including their heavy tails~\cite{Amari_etal_2003,Tkacik_etal_2014,tkavcik2015thermodynamics,rodriguez2025higher}.
It can also affect population coding~\cite{Montani_etal_2009,cayco2015triplet,ZylberbergSheaBrown_2015}. However, these findings come from heterogeneous experimental conditions, both in vivo and in vitro, and generally concern individual brain areas.  How higher-order structure is organized across the brain therefore remains unknown.

Directly parameterizing higher-order interactions term by term is intractable as the number of couplings grows combinatorially and rapidly exceeds what recordings can reliably constrain~\cite{ganmor2011sparse,montani2013statistical,cayco2018moment}. A nonlinear energy function  offers a compact alternative, generating interactions at every order without assigning a separate parameter to each interaction~\cite{barra2018new,gardella2016tractable,humplik2017probabilistic,dicarlo2026extended}. Such nonlinear energy functions underpin dense associative memories~\cite{krotov2016dense,demircigil2017model}, a framework mathematically connected to transformer attention~\cite{ramsauer2021hopfield} and to the energy landscapes of generative diffusion models~\cite{ambrogioni2024search}.

Curved neural networks~\cite{aguilera2025explosive} derive such a nonlinearity through a R\'enyi maximum-entropy principle~\cite{morales2021generalization,wong2022tsallis}, which deforms the pairwise model obtained from Shannon entropy.
In the resulting curved model, fields, couplings, and a single curvature parameter determine interactions at every order, adding only one parameter beyond the pairwise baseline. In its recurrent-network formulation, this deformation changes how neurons integrate inputs: activation depends on combinations of inputs rather than their linear weighted sum alone. Curvature thus links higher-order population structure to nonlinear input integration. Previous work has shown that this deformation can produce explosive phase transitions, accelerate memory retrieval, and control storage capacity and robustness~\cite{aguilera2025explosive}. However, fitting the model to neural data remains an open problem.

Inverse Ising methods infer the fields and couplings from firing rates and correlations~\cite{schneidman2006weak,Shlens_etal_2006}, often using mean-field approximations~\cite{plefka_convergence_1982,kappen_efficient_1998,roudi2009ising,donner_approximate_2017}. In curved neural networks, however, curvature alters the distribution in ways that can be partly reproduced by adjusting these pairwise parameters, confounding their joint inference. We therefore develop a method that removes this shared component and retains the higher-order signal that identifies curvature.

Applying this method, here we examine higher-order structure across \(39\) cortical and subcortical regions using recordings of spontaneous neural activity in mice~\cite{steinmetz2019distributed}. We first compare the third cumulant of population spike counts with predictions from pairwise maximum-entropy models matched to firing rates and pairwise correlations. Widespread negative deviations reveal net-suppressive higher-order structure. We then characterize these departures through Rényi curvature, finding predominantly negative values, with stronger deformation in visual populations and values closer to the flat limit in midbrain and frontal populations.

Next, we ask which populations lie within reach of a simple analytical description. To this end, we derive a deformed Thouless--Anderson--Palmer mean-field equation~\cite{thouless_solution_1977} that provides a closed-form inverse for the fields and couplings of the curved model. Across regions, the approximation applies to most frontal populations but fails more often in hippocampal, sensorimotor, and visual populations. Post hoc analyses associate failures in visual populations with stronger deformation and mean correlation and those in midbrain populations with pairwise-correlation heterogeneity.

Together, these results provide a compact geometric description of brain-wide suppressive structure and a model-based link to the nonlinearities underlying neural dynamics.

\section*{Results}

\subsection*{Curved neural networks}

\begin{figure}[!t]
    \centering
    \includegraphics[width=\linewidth]{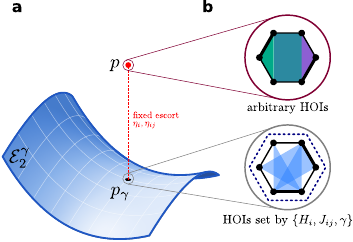}
    \caption{\textbf{Curved projection geometry and higher-order interactions.}
    \textbf{(a)} The distribution $p$ is projected onto the curved second-order family $\pazocal{E}_2^\gamma$ at $p_\gamma$ at fixed escort moments $\eta_i$, $\eta_{ij}$. 
    \textbf{(b)} Magnified view of $p$ (top) and $p_\gamma$ (bottom) shows their log-linear interactions~\eqref{eq:effective_hamiltonian} over the same $N$ neurons. 
    Lines mark pairwise terms and filled polygons mark higher-order terms among the enclosed neurons. In $p$, different colors are used to illustrate independently specified higher-order couplings, whereas in $p_\gamma$ higher-order couplings are determined by the fields $H_i$, pairwise couplings $J_{ij}$ and curvature $\gamma$. Selected triplets are shown explicitly in the lower inset, with the dotted enclosure indicating the full hierarchy through this $N=6$ instance. Edge thickness indicates pairwise coupling strength, and shaded groups denote higher-order interactions.
    }
    \label{fig:gamma_projection}
\end{figure}

\begin{figure*}[!tbp]
    \centering
    \includegraphics[width=\linewidth]{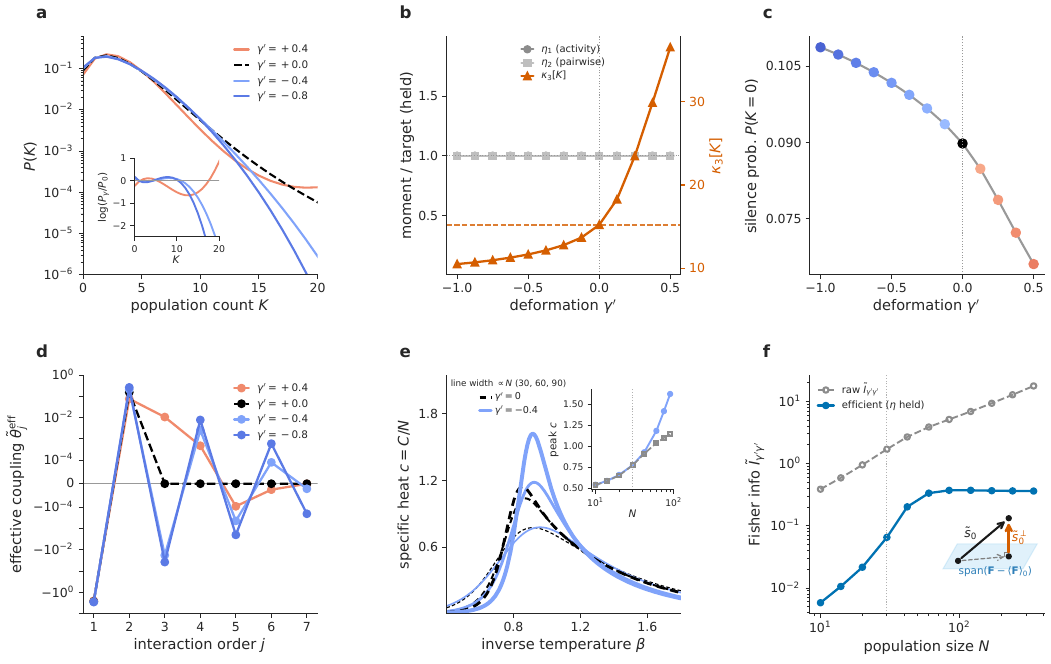}
    \caption{\textbf{R\'enyi curvature is a hidden thermodynamic control parameter.} Homogeneous curved network at $N=30$, mean activity $\eta_1=0.1$, and reference pairwise correlation $0.03$. At each rescaled deformation $\gamma'\equiv N\beta\gamma$, the field and coupling are recalibrated to keep the first two escort moments fixed at $\beta=1$. The zero-centered color map runs from blue at negative $\gamma'$, darker with larger $|\gamma'|$, black at zero, to warm red at positive $\gamma'$. All panels show escort quantities, with $\gamma'$ retaining the original deformation coordinate.
    \textbf{(a)}~Escort spike-count distribution $P(K)$ at log-scale and inset, $\log[P_\gamma(K)/P_0(K)]$, versus $K$.
    \textbf{(b)}~Held escort moments, mean activity $\eta_1$ and pairwise moment $\eta_2$, relative to their targets on the left axis, and escort third spike-count cumulant $\kappa_3[K]$ (vermilion) versus $\gamma'$, the dashed line for its $\gamma'=0$ value.
    \textbf{(c)}~Escort silence probability $\tilde p_\gamma(\bv{0})=P(K{=}0)$ versus $\gamma'$, colored marker denotes $\gamma'\neq 0$.
    \textbf{(d)}~Signed effective $j$-body coupling $\tilde\theta_j^{\mathrm{eff}}$ on a symmetric-log scale, obtained by the finite-difference M\"obius transform of the escort log-probability $\ln\tilde p_\gamma$, from the first-order field $j=1$ through higher orders at $\gamma'\in\{+0.4,0,-0.4,-0.8\}$; black dashed, undeformed reference.
    \textbf{(e)}~Specific heat $c(\beta)=C/N$ versus inverse temperature $\beta$ for the power-tempered escort family calibrated at $\gamma'=-0.4$ and $\beta=1$ (solid blue), and its undeformed counterpart (dashed black), for $N\in\{30,60,90\}$ with line width $\propto N$ and inset, peak $c$ versus $N$.
    \textbf{(f)}~Fisher information $\tilde I_{\gamma'\gamma'}$ at $\gamma'=0$ versus $N$. Raw (gray, open markers) and efficient $\tilde I^\perp_{\gamma'\gamma'}$ after projection off the pairwise sufficient statistics $\boldsymbol{F}=(x_i,x_ix_j)$ (blue). The inset illustrates the decomposition of score $\tilde s_0$ into its projection onto $\mathrm{span}(\boldsymbol{F}-\langle\boldsymbol{F}{\rangle_0})$ and common orthogonal residual $\tilde s_0^{\perp}=s_0^{\perp}$.}
    \label{fig:curvature_orderparameter}
\end{figure*}

Curved neural networks~\cite{aguilera2025explosive} model neural population activity with a R\'enyi maximum-entropy distribution, a deformed exponential family constrained by first- and second-order escort moments and parameterized by curvature $\gamma$. For $N$ binary variables $\bv{x}=(x_1,\ldots,x_N)$, where $x_i\in\{0,1\}$ denotes an inactive or active neuron,
\begin{align}
    p_{\gamma}(\bv{x})
    &= \left(1 - \gamma \beta E(\bv{x}) \right)^{1/\gamma} e^{-\psi_{\gamma}},
    \label{eq:def_exp_energy}
\end{align}
where $\beta>0$ is the inverse temperature and $\psi_\gamma$ the log partition function. The energy inside this nonlinear function remains pairwise,
\begin{align}
    E(\bv{x}) = - \sum_{i=1}^{N} H_i x_i - \sum_{i<j} J_{ij} x_i x_j,
    \label{eq:energy}
\end{align}
where $H_i$ and $J_{ij}$ are the fields and pairwise couplings. The ordinary pairwise (Ising) maximum-entropy model is recovered as $\gamma\to0$. To keep the effective energy extensive, we hold $\gamma'\equiv N\beta\gamma$ fixed as the population grows~\cite{aguilera2025explosive}.

The model is obtained by maximizing R\'enyi entropy while constraining escort moments rather than ordinary moments (Supplementary Note~\ref{supp_sec:max_ent_principle}). These are expectations of $F_i\in\{x_i, x_ix_j\}$ under the escort distribution
\begin{equation}
    \label{eq:escort_dist}
    \tilde{p}_\gamma (\bv{x}) = \frac{p_\gamma(\bv{x})^{1-\gamma}}{\sum_{\bv{y}} p_\gamma(\bv{y})^{1-\gamma}},
\end{equation}
namely
\begin{equation}
    \eta_i = \sum_{\bv{x}} \tilde{p}_\gamma(\bv{x})F_i(\bv{x}),
    \label{eq:eta_escort}
\end{equation}
giving activities $\eta_i$ for $F_i=x_i$ and pairwise moments $\eta_{ij}$ for $F_i=x_ix_j$. 
An interior R\'enyi projection onto a restricted family preserves the escort moments fixed by the projection (Fig.~\ref{fig:gamma_projection}; Supplementary Notes~\ref{supp_sec:escort_dual_coordinates} and \ref{supp_sec:projection}).

When fitting data, we equate observed firing rates and pairwise moments with these escort expectations. We therefore use the fitted escort distribution to describe the data. Its properties can be analyzed within the original model family through the exact reparameterization $\tilde{\gamma}=\gamma/(1-\gamma)$, $\tilde{H}_i=(1-\gamma)H_i$, and $\tilde{J}_{ij}=(1-\gamma)J_{ij}$ (\hyperref[sec:escort_fitting]{Methods}).

Deformation reorganizes the pairwise model into an effective higher-order network. The $\gamma$ expansion shows how the deformation induces many-body interactions from the field and pairwise couplings (\hyperref[sec:effective_interactions]{Methods}).
The three-body coefficient of $\ln p_\gamma$ is
\begin{equation}\label{eq:three_body_coupling}
\begin{split}
    \theta_{ijk}^{\mathrm{eff}} = -\gamma\beta^2\bigl[&H_iJ_{jk}+H_jJ_{ik}+H_kJ_{ij} \\
    &+ J_{ij}J_{ik}+J_{ij}J_{jk}+J_{ik}J_{jk}\bigr] + \pazocal{O}(\gamma^2).
\end{split}
\end{equation}
For the escort distribution, the same expressions apply with $\gamma$, $H_i$, and $J_{ij}$ replaced by $\tilde{\gamma}$, $\tilde{H}_i$, and $\tilde{J}_{ij}$, respectively.
In the homogeneous model $H_i=H$, $J_{ij}=J/N$ (Supplementary Note~\ref{supp_sec:meanfield_phase_calibration}), this coefficient reduces to $\theta_{ijk}^{\mathrm{eff}}
\simeq-3\gamma'\beta HJ/N^2$. Under the silence-dominated regime at $H<0$ and positive couplings $J>0$, the negative $\gamma'$ induces a negative triplet interaction. This sign pattern is consistent with previously reported triplet interactions in visual, premotor, and hippocampal neurons in primates, cats and rodents~\cite{Ohiorhenuan_etal_2010,Ohiorhenuan10b,Yu_etal_2011,Shimazaki_etal_2015,shomali2026inferring}.

In the homogeneous network, deformation can broaden or narrow the count distribution when only the rate is fixed, depending on the correlation regime (Supplementary Fig.~\ref{fig:supp_broaden_atlas}). To isolate higher-order reshaping, we therefore recalibrate the field and coupling at each $\gamma'$ to keep the escort activity $\eta_i$ and pairwise moment $\eta_{ij}$~\eqref{eq:eta_escort} fixed, preserving the mean and variance of the escort count distribution (Fig.~\ref{fig:curvature_orderparameter}a).

While the fitted moments remain fixed, the third spike-count cumulant varies with curvature (Fig.~\ref{fig:curvature_orderparameter}b), and population silence rises above baseline on the suppressive branch (Fig.~\ref{fig:curvature_orderparameter}c). For the negative deformations shown, the induced couplings form an alternating, shrinking hierarchy (Fig.~\ref{fig:curvature_orderparameter}d; \hyperref[sec:effective_interactions]{Methods}), consistent with sign-alternating higher-order interactions reported in an in vitro study and reproduced in a population model of threshold units~\cite{Shimazaki_etal_2015}.

A fit to first- and second-order moments cannot identify this structure because matching those moments
reabsorbs deformation into the fields and couplings (\hyperref[sec:escort_fitting]{Methods}).
Deformation nevertheless persists in higher-order statistics and thermodynamic fluctuations.

\subsection*{Curvature as a hidden control parameter}

\begin{figure*}[!t]
    \centering
    \includegraphics[width=0.95\linewidth]{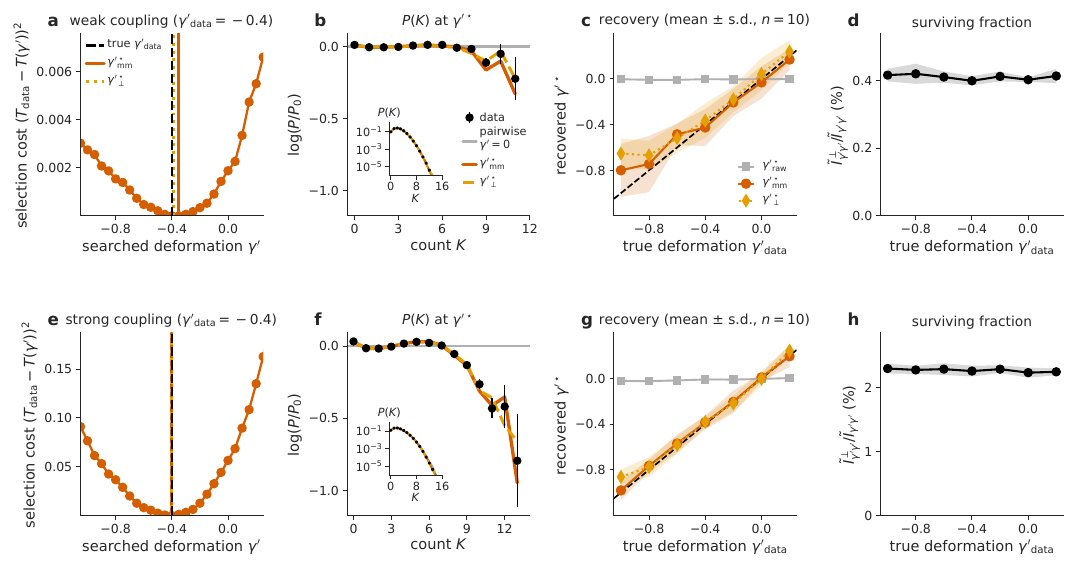}
    \caption{\textbf{Efficient-score estimation recovers deformation across coupling regimes.} Synthetic heterogeneous populations ($N=30$, $\beta=1$) span true deformations $\gamma'_{\mathrm{data}}$ under weak (LGd-matched, couplings drawn with $N\,\mathrm{sd}(J)=6.7$ at $\gamma'_{\mathrm{data}}=0$; (a)--(d)) and strong (VISam-matched, drawn with $N\,\mathrm{sd}(J)=14.8$ at $\gamma'_{\mathrm{data}}=0$; (e)--(h)) coupling. At each $\gamma'_{\mathrm{data}}$, the fields and couplings are recalibrated to keep the first two escort moments fixed. Synthetic data, fitted count distributions, and score estimates use the escort family, with $\gamma'=N\beta\gamma$ retaining the original deformation coordinate. Each realization uses $5\times10^{4}$ synthetic-data draws and $5\times10^{4}$ draws to evaluate the pairwise model and each candidate-model statistic.
    \textbf{(a)}~Squared profile cost versus searched $\gamma'$ at $\gamma'_{\mathrm{data}}=-0.4$, vertical lines mark truth (black dashed), the moment-matching estimate $\gamma'^\star_{\mathrm{mm}}$ (vermillion), and the one-shot estimate $\gamma'^\star_\perp$ (orange dotted).
    \textbf{(b)}~Spike-count distribution as $\log[P(K)/P_0(K)]$ for data (points, $\pm1$ counting s.e.), the pairwise ($\gamma'=0$) full-Boltzmann model $P_0$ (gray zero line), and deformed fits at $\gamma'^\star_{\mathrm{mm}}$ (vermillion) and $\gamma'^\star_\perp$ (orange dashed); inset, $P(K)$ on a log scale. Histograms use $3\times10^{6}$ draws each.
    \textbf{(c)}~Recovered versus true $\gamma'$ for the raw escort score without projection, $\gamma'^\star_{\mathrm{raw}}=N\beta\langle\tilde s_0\rangle_{\mathrm{data}}/\tilde I_{\gamma\gamma}$ (gray), moment-matching $\gamma'^\star_{\mathrm{mm}}$ (vermillion), and one-shot $\gamma'^\star_\perp$ (orange), dashed, identity.
    \textbf{(d)}~Surviving deformation-information fraction $\tilde I^{\perp}_{\gamma'\gamma'}/\tilde I_{\gamma'\gamma'}$ versus true $\gamma'$.
    \textbf{(e)}~Selection cost, \textbf{(f)}~$P(K)$ fit, \textbf{(g)}~recovery, and \textbf{(h)}~surviving information under strong coupling, with the conventions in (a)--(d). Values in (c), (d), (g), and (h) are mean $\pm$ s.d. across $n=10$ independent realizations. Efficient score and selection are defined in \hyperref[sec:efficient_score]{Methods}.}
    \label{fig:efficient_score_recovery}
\end{figure*}

Energy fluctuations can reveal curvature hidden from fitted low-order moments. We examine the specific heat, or heat capacity per neuron~\cite{tkavcik2015thermodynamics}, $c(\beta)=\beta^2\mathrm{Var}(\tilde E_\gamma)/N$, where $\tilde E_\gamma$ is the effective energy of the calibrated escort distribution and temperature is varied by raising its probabilities to the power $\beta$ (\hyperref[sec:thermo_observables]{Methods}). The specific-heat curves of the deformed and undeformed models separate increasingly with population size (Fig.~\ref{fig:curvature_orderparameter}e).

To estimate curvature, we next introduce the deformation score, the sensitivity of log probability to $\gamma$. At $\gamma=0$, $\tilde E_\gamma$ reduces to $E$, and the original-model score $s_0(\bv{x}) \equiv \left.\partial_\gamma\ln p_\gamma(\bv{x})\right|_{\gamma=0}$ at fixed fields, couplings, and $\beta$ is proportional to the centered squared energy under the fitted pairwise model, $E^2-\langle E^2\rangle_0$~\eqref{eq:deformation_score_null}. With the energy function and temperature fixed and ordinary mean energy matched, changes in $\langle E^2\rangle$ directly reflect changes in the specific heat $c=C/N=\beta^2\mathrm{Var}(E)/N$. The escort score differs from $s_0$ by a centered pairwise term. Projecting out the score component that can be absorbed by refitting the fields and couplings gives the same \emph{efficient score} for both families at $\gamma=0$ (Fig.~\ref{fig:curvature_orderparameter}f; Supplementary Note~\ref{supp_sec:score_geometry}).
The two signatures scale differently with population size: the specific heat increasingly separates the deformed and undeformed families
(Fig.~\ref{fig:curvature_orderparameter}e), whereas the efficient information about the intensive deformation rises up to $N\approx80$ and then levels off at a nonzero value (Fig.~\ref{fig:curvature_orderparameter}f). The plateau reflects the scaling $\gamma=\gamma'/(N\beta)$, under which the
information about the unscaled deformation, $(N\beta)^2 I^{\perp}_{\gamma'\gamma'}$, keeps growing as $N^2$.

In synthetic heterogeneous $N=30$ populations with escort moments held fixed across deformations, we estimated deformation from the efficient escort score evaluated in the pairwise ($\gamma'=0$) full-Boltzmann model, without fitting a finite deformation (the one-shot estimate $\gamma'^\star_\perp$). Profiling the same statistic with non-perturbative Monte-Carlo fits (the moment-matching estimate $\gamma'^\star_{\mathrm{mm}}$) recovered deformation across the searched window, with weak-coupling selection costs and count-distribution fits in Fig.~\ref{fig:efficient_score_recovery}a,b. The one-shot estimate $\gamma'^\star_\perp$ (\hyperref[sec:efficient_score]{Methods}) was accurate near $\gamma'=0$ but underestimated the magnitude of strongly negative deformations (Fig.~\ref{fig:efficient_score_recovery}c). Little raw escort information survived projection (Fig.~\ref{fig:efficient_score_recovery}d), showing how strongly pairwise refitting conceals deformation. Strong-coupling results appear in Fig.~\ref{fig:efficient_score_recovery}e--h, with about $0.4$--$2.3\%$ surviving across both regimes.

\subsection*{Suppressive curvature across the mouse brain}

\begin{figure*}[p]
    \centering
    \begin{minipage}[t]{0.25\linewidth}
        \centering
        \makebox[\linewidth][l]{\hspace*{8pt}\begin{overpic}[width=\linewidth]{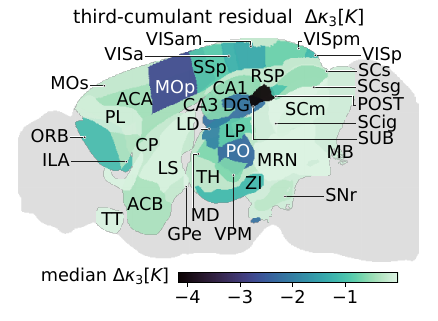}
                \put(1.0,90.0){\hspace*{-8pt}\raisebox{-3.28pt}{\makebox(0,0)[lt]{{\fontsize{10}{12}\selectfont\bfseries\sffamily a}}}}
            \end{overpic}}
    \end{minipage}\hfill
    \begin{minipage}[t]{0.72\linewidth}
        \centering
        \begin{overpic}[width=\linewidth]{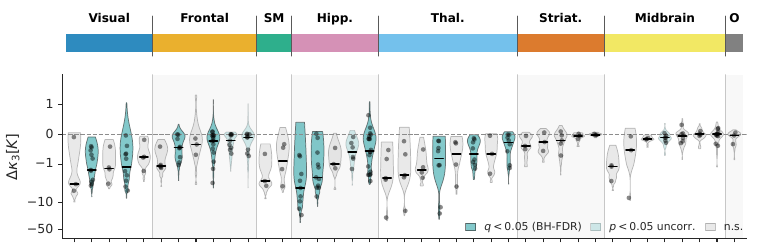}
            \put(0.5,31.0){\makebox(0,0)[lt]{{\fontsize{10}{12}\selectfont\bfseries\sffamily b}}}
        \end{overpic}
    \end{minipage}
    \par\vspace{-3pt}
    \begin{minipage}[t]{0.25\linewidth}
        \vspace{0pt}
        \centering
        \raisebox{-19pt}[\height][\depth]{\parbox{\linewidth}{\makebox[\linewidth][l]{\hspace*{8pt}\begin{overpic}[width=\linewidth]{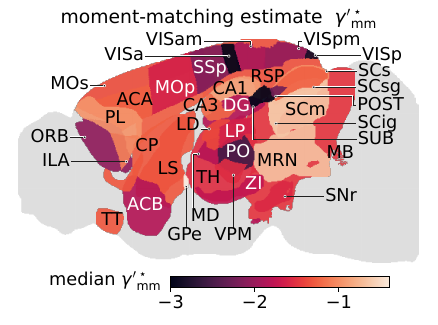}
                \put(1.0,73.0){\raisebox{14.53pt}{\hspace*{-8pt}\makebox(0,0)[lt]{{\fontsize{10}{12}\selectfont\bfseries\sffamily c}}}}
            \end{overpic}}
        }}
    \end{minipage}\hfill
    \begin{minipage}[t]{0.72\linewidth}
        \vspace{0pt}
        \centering
        \begin{overpic}[width=\linewidth]{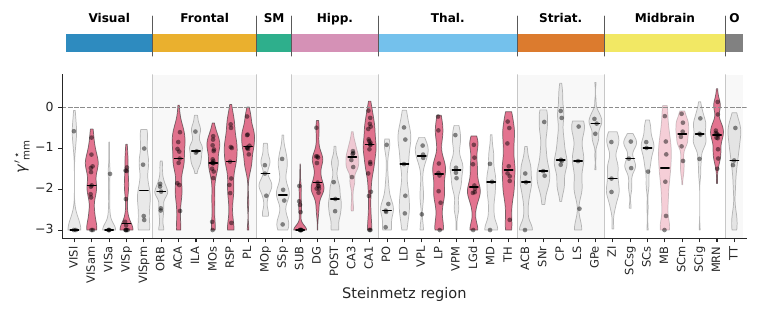}
            \put(0.5,39.0){\hspace*{-0.074pt}\raisebox{1.389pt}{\makebox(0,0)[lt]{{\fontsize{10}{12}\selectfont\bfseries\sffamily d}}}}
        \end{overpic}
    \end{minipage}
    \par\vspace{-3pt}
    \begin{minipage}[t]{0.44\linewidth}
        \centering
        \includegraphics[width=\linewidth]{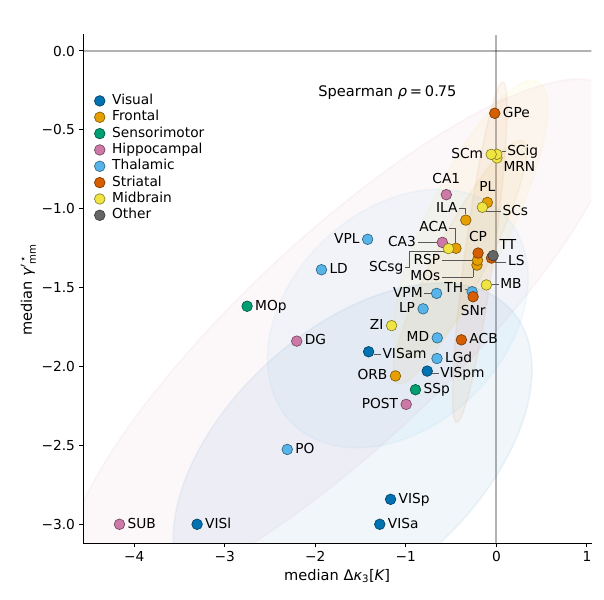}\llap{\makebox[\linewidth][l]{\hspace*{0.021261\linewidth}\hspace*{-0.31pt}\raisebox{0.958309\linewidth}[0pt][0pt]{{\fontsize{10}{12}\selectfont\bfseries\sffamily e}}}}
    \end{minipage}\hfill
    \begin{minipage}[t]{0.44\linewidth}
        \centering
        \includegraphics[width=\linewidth]{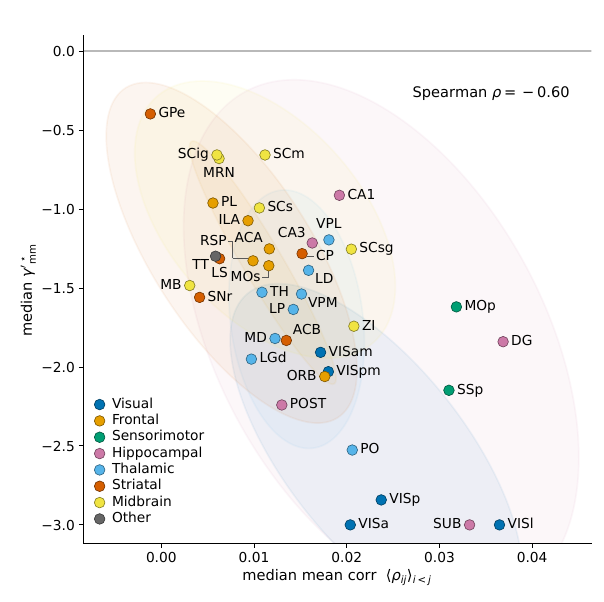}\llap{\makebox[\linewidth][l]{\hspace*{0.020415\linewidth}\hspace*{-0.30pt}\raisebox{0.961136\linewidth}[0pt][0pt]{{\fontsize{10}{12}\selectfont\bfseries\sffamily f}}}}
    \end{minipage}
    \caption{\textbf{Inferred curvature and third-cumulant residuals agree across the mouse brain.} Random $N=30$ subsamples from $39$ regions were analyzed in $20$~ms bins of spontaneous activity.
    \textbf{(a)} Allen CCFv3 sagittal section through a $\pm0.5$ mm slab centered at mediolateral position $0.9$~mm, showing $36$ regions labeled by Allen acronyms. Shading gives the median of per-session medians of $\Delta\kappa_3[K]$ (bars in b), with darker colors indicating more negative residuals. \textbf{(b)} $\Delta\kappa_3[K]$ against the variance-matched pairwise model on a symmetric-log scale, arranged by anatomical group, identified by the top strip (SM, sensorimotor; O, other). Violin width shows the density of subsample values. Points mark session medians and horizontal bars their median with sessions being the analysis units. Colors show two-sided cluster sign-flip significance with equal session weights and Benjamini--Hochberg false-discovery-rate correction across $39$ regions. Darker colors mark $q<0.05$, lighter colors nominal $p<0.05$ only, and gray $p\ge0.05$. \textbf{(c)}~Section in (a), shaded by the median of per-session medians of the moment-matching estimate $\gamma'^\star_{\mathrm{mm}}$ (bars in d)~(\hyperref[sec:efficient_score]{Methods}). \textbf{(d)} Moment-matching estimate $\gamma'^\star_{\mathrm{mm}}$, plotted as in (b) on a linear scale. Colors show results of sign-flip tests of the displayed estimates, assessed as in (b). Search range $\gamma'\in[-3,1]$, with $|\gamma'|=3$ the edge of the search range, Supplementary Figs.~\ref{fig:supp_kappa3_bins_violin} and~\ref{fig:supp_gammamap_bins}.
    \textbf{(e)} Regional median $\gamma'^\star_{\mathrm{mm}}$ against median $\Delta\kappa_3[K]$, both as in b and d. Colors denote anatomical groups and ellipses their $2\sigma$ Gaussian covariance contours. Spearman $\rho=0.75$, $p<0.001$. \textbf{(f)} Regional median $\gamma'^\star_{\mathrm{mm}}$ against median mean pairwise correlation, $\langle\rho_{ij}\rangle_{i<j}$, plotted as in (e). Spearman $\rho=-0.60$, $p<0.001$.
    }
    \label{fig:effscore_kappa3}
\end{figure*}

\begin{figure*}[!tbp]
    \centering
    \includegraphics[width=0.94\linewidth]{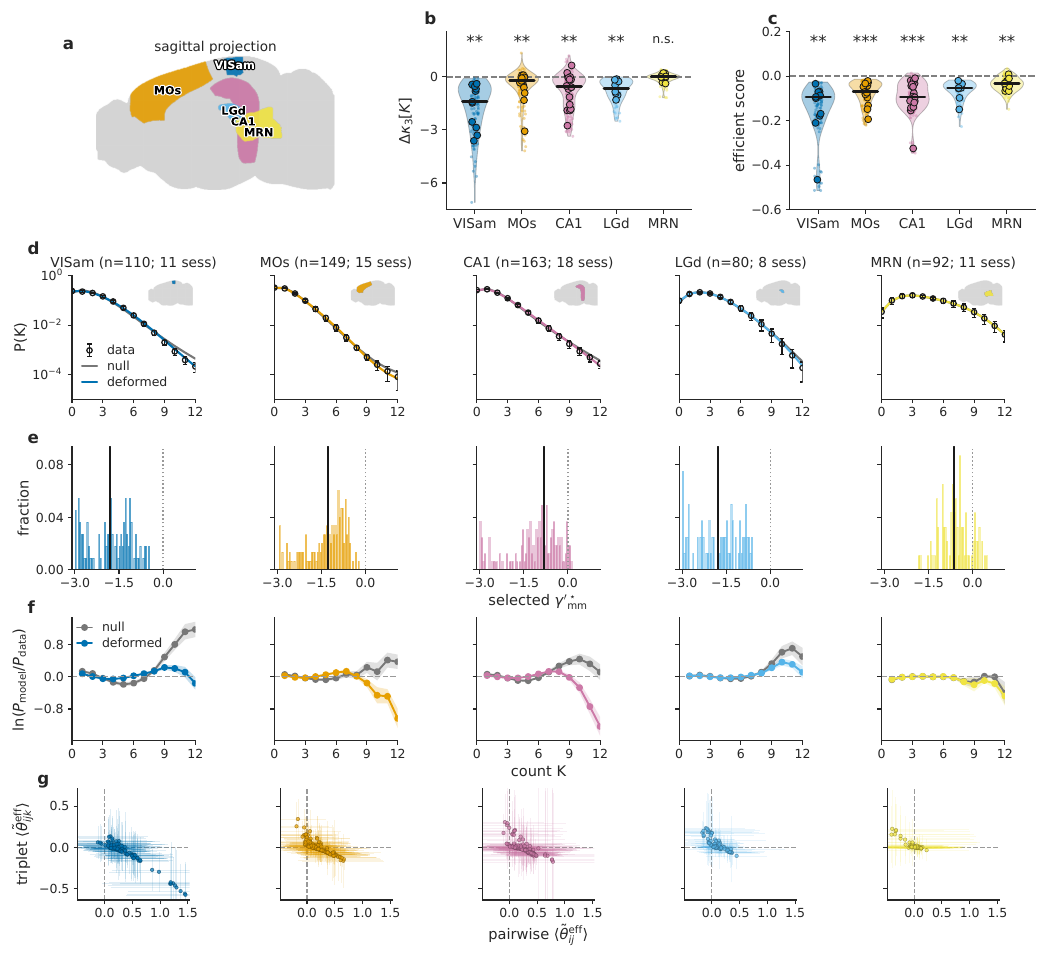}
    \caption{\textbf{Efficient scores identify suppressive count structure across five Steinmetz regions.} The panels summarize random $N=30$ subsamples from five regions selected by coverage (VISam, MOs, CA1, LGd, and MRN; $20$~ms spontaneous-activity bins~\cite{steinmetz2019distributed}). Pairwise models are variance matched at $\gamma'=0$, moment-matching estimates $\gamma'^\star_{\mathrm{mm}}$ use efficient scores and sample-based inverses (\hyperref[sec:efficient_score]{Methods}; \hyperref[sec:regional_selection]{regional selection}). Sessions are the analysis units; panel (d) gives regional subsample and session counts.
    \textbf{(a)}~Left-hemisphere sagittal maximum projection of the Allen CCFv3 annotation, showing the five regions at their AP--DV footprints with the colors used in (b),(c).
    \textbf{(b)}~Third-cumulant residual $\Delta\kappa_3[K]$ against the variance-matched pairwise model. Violins show the distribution across subsamples, points mark session medians, and bars mark the median across sessions. Stars indicate significance in two-sided, equal-session-weighted cluster sign-flip permutation tests with Benjamini--Hochberg correction across the five pre-specified regions ($q<0.05$ for all but MRN); Fig.~\ref{fig:effscore_kappa3}b shows the $39$-region screen.
    \textbf{(c)}~Standardized efficient score $\langle s_0^\perp\rangle_{\mathrm{data}}/\sqrt{I^\perp_{\gamma\gamma}}$~\eqref{eq:efficient_score_estimate} per subsample, with violins, points, and bars as in (b). Stars mark two-sided, equal-session-weighted cluster sign-flip permutation tests with the same five-region correction ($q<0.05$ for all five).
    \textbf{(d)}~Population spike-count distribution $P(K)$: data (points, mean $\pm$ s.e.m.\ across sessions), variance-matched pairwise model (gray mean), selected deformed model at $\gamma'^\star_{\mathrm{mm}}$ (colored mean), and sagittal atlas locators (insets).
    \textbf{(e)}~Selected $\gamma'^\star_{\mathrm{mm}}$ by region. Black lines mark medians of session medians, and dotted lines mark $\gamma'=0$.
    \textbf{(f)}~Log ratio $\ln[P_{\mathrm{model}}(K)/P_{\mathrm{data}}(K)]$ for the pairwise and selected deformed models, omitting $K=0$. Lines and bands, mean $\pm$ s.e.m.\ across sessions; positive and negative values denote over- and under-prediction.
    \textbf{(g)}~Mean exact silent-reference triplet coefficient $\langle{\tilde\theta_{ijk}^{\mathrm{eff}}}\rangle$ versus pairwise coefficient $\langle{\tilde\theta_{ij}^{\mathrm{eff}}}\rangle$ of the fitted escort distribution for each subsample. Error bars, within-subsample s.d.\ across the $\binom{N}{3}$ triplets or $\binom{N}{2}$ pairs.
    }
    \label{fig:hoi_survey_effscore}
\end{figure*}

Having established that deformation is identifiable through the efficient score, we mapped it across the brain. Using spontaneous Neuropixels recordings from~\cite{steinmetz2019distributed}, we tested random $N=30$ subpopulations from $39$ cortical and subcortical mouse-brain regions in $20$-ms bins for higher-order structure beyond a variance-matched pairwise model (\hyperref[sec:regional_selection]{Methods}).
Without fitting the deformed model, we quantified higher-order structure by comparing the third cumulant of the population spike count, $K=\sum_i x_i$, with its pairwise prediction. For binary activity, matching of all firing rates and pairwise moments gives
\begin{align}
    \Delta\kappa_3[K]
    &\equiv \kappa_3^{\mathrm{data}}[K]
    -\kappa_3^{\mathrm{null}}[K] \nonumber\\
    &=6\sum_{i<j<k}
    \left(\kappa_{ijk}^{\mathrm{data}}
    -\kappa_{ijk}^{\mathrm{null}}\right),
    \label{eq:kappa3_residual}\\
    \kappa_{ijk}
    &=\left\langle
    (x_i-\eta_i)(x_j-\eta_j)(x_k-\eta_k)
    \right\rangle. \nonumber
\end{align}
Here $\eta_i$ is the matched firing probability, and $\kappa_{ijk}$ is the connected triplet cumulant evaluated
separately for the data and null. A negative residual therefore indicates net suppression of triplet correlations relative to the pairwise prediction.
The residual $\Delta\kappa_3[K]$~\eqref{eq:kappa3_residual} had a negative median of per-session medians in $37$ of the $39$ regions, the two exceptions, both in the midbrain, had medians just above zero and $10$ passed brain-wide Benjamini--Hochberg false-discovery-rate (BH-FDR) correction (Fig.~\ref{fig:effscore_kappa3}a,b). Across anatomical groups, the residual was clearly negative in visual, hippocampal, and thalamic populations and weakly negative in frontal and midbrain populations.
At the level of individual subpopulations, $87\%$ of the $2271$ residuals were negative. Moving-block bootstrap tests accounting for temporal dependence identified significant suppression in $61\%$ of subpopulations after BH-FDR correction across all subsamples (one-sided tests; Supplementary Fig.~\ref{fig:supp_perreg_kappa3}).

The moment-matching estimate $\gamma'^\star_{\mathrm{mm}}$ provided both the brain-wide map (Fig.~\ref{fig:effscore_kappa3}c) and the within-region distributions (Fig.~\ref{fig:effscore_kappa3}d). Across anatomical groups, the median of the regional values in Fig.~\ref{fig:effscore_kappa3}d ranged from $-2.8$ (visual) through $-1.9$ (sensorimotor), $-1.8$ (hippocampal), $-1.6$ (thalamic), $-1.3$ (striatal) and $-1.3$ (frontal) to $-1.0$ (midbrain).
At every bin size ($5$, $10$, $20$, and $40$~ms), the moment-matching estimate had negative regional medians in all $39$ regions, and session-level sign-flip tests identified the same $13$ significant regions after BH-FDR correction. These were all the regions with enough sessions for the test to survive correction.
The third-cumulant residual was more sensitive to bin size, with the median of per-session medians negative in $16$, $33$, $37$, and $38$ regions, with $0$, $0$, $10$, and $10$ surviving correction, respectively (Supplementary Figs.~\ref{fig:supp_kappa3_bins_violin} and \ref{fig:supp_gammamap_bins}).
Regional medians of the moment-matching estimate covaried with the third-cumulant residual (Spearman $\rho=0.75$, $p < 0.001$; Fig.~\ref{fig:effscore_kappa3}e), linking the curvature coordinate to the fit-independent departure. More negative deformation was associated with higher mean pairwise correlations across regions (Fig.~\ref{fig:effscore_kappa3}f).

The model at finite $\gamma$ has an all-order hierarchy, whereas the efficient score at $\gamma = 0$ is obtained by projecting the first-order perturbation proportional to $-u^2/2$~\eqref{eq:deformation_score_null} with $u=-\beta E$. Because $u$ is quadratic, the efficient score contains only projected third- and fourth-order components, $s_0^\perp=q_3^\perp+q_4^\perp$ (\hyperref[sec:efficient_score]{Methods}). Across the same $20$~ms subsamples, the third-order component had the larger norm, with median $I_3/(I_3+I_4)=0.89$, where the squared norm $I_j$ is the variance of $q_j^\perp$ under the fitted pairwise model, while the third- and fourth-order components were strongly anti-correlated (median correlation $\rho_{34}=-0.93$; Supplementary Fig.~\ref{fig:supp_effscore_order_fraction}). The third-order contribution therefore dominates the score in norm, although finite $\gamma'$ still encodes all orders.

To examine regional structure, we selected, independently of effect size, the most-sampled region in each anatomical group with more than $50$ pooled $N=30$ subpopulations (\hyperref[sec:regional_selection]{Methods}). This yielded the anteromedial visual area (VISam), secondary motor area (MOs), hippocampal field CA1, dorsal part of the lateral geniculate complex (LGd), and midbrain reticular nucleus (MRN) (Fig.~\ref{fig:hoi_survey_effscore}a). Four of the five (all but MRN) had negative third-cumulant residuals and passed BH-FDR correction both within this coverage-selected set ($q < 0.05$; Fig.~\ref{fig:hoi_survey_effscore}b) and in the $39$-region screen (Fig.~\ref{fig:effscore_kappa3}b). MRN, whose residual was just above zero, passed neither.

All five regions had significantly negative projected scores (Fig.~\ref{fig:hoi_survey_effscore}c). We compared the observed population-count distributions with predictions from the variance-matched pairwise and selected curved models (Fig.~\ref{fig:hoi_survey_effscore}d). The corresponding moment-matching profiles selected predominantly negative deformations (Fig.~\ref{fig:hoi_survey_effscore}e). The curved models better reproduced the central and intermediate population-count range and reduced the pairwise excess high-synchrony tail, although they oversuppressed the extreme tail in MOs and CA1 (Fig.~\ref{fig:hoi_survey_effscore}f). Because selection targets a projected energy-square statistic rather than the full count hierarchy or tail likelihood, one curvature captures the leading suppression but not all subset-specific synchrony. Across $594$ subsamples, mean escort triplet and pairwise coefficients were strongly inversely associated (pooled Spearman $\rho=-0.84$), consistent with the leading-order prediction~\eqref{eq:three_body_coupling} (Fig.~\ref{fig:hoi_survey_effscore}g).

\subsection*{Deformed mean-field theory and the TAP equation}

\begin{figure*}[!t]
    \centering
    \includegraphics[width=0.95\linewidth]{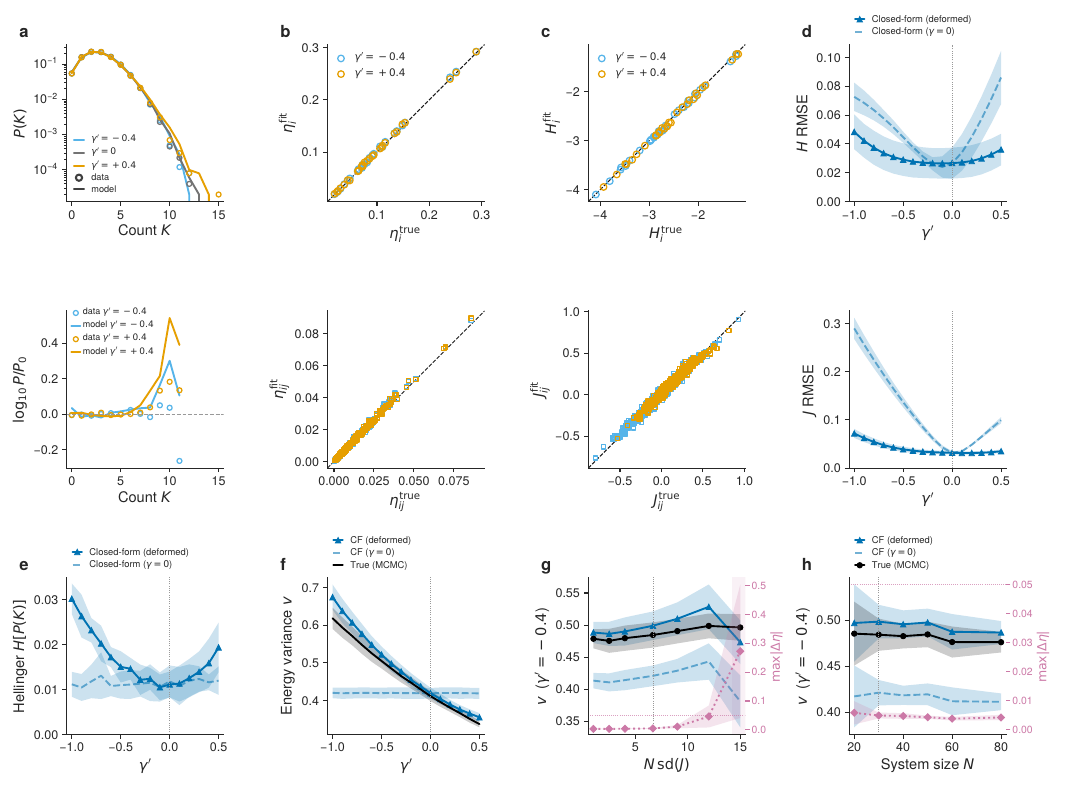}
    \caption{\textbf{Finite-sample inverse recovery at $N=30$ with LGd-matched disorder.} Heterogeneous $N=30$ weak-correlation models use LGd-matched field and coupling disorder while holding target escort moments $\bv{\eta}^{*}$ across deformations. Escort count distributions and moments are estimated from $5\times10^{4}$ draws per model; energy variances use $2\times10^{5}$ escort draws per model (\hyperref[sec:thermo_observables]{Methods}). At each $\gamma'$, the sample-based (MCMC) deformed inverse of $\bv{\eta}^{*}$ is the reference. Blue traces show the closed-form inverse~\eqref{eq:Jij_inverse_main} in the deformed family (solid, triangles) and at $\gamma=0$ (dashed). Panels (a)--(c) show one network and sample at $\gamma'=-0.4$ (sky blue) and $+0.4$ (orange), (d)--(h) show mean $\pm1$ s.d. across $n=10$ independent LGd-matched network draws, each with an independent finite sample.
    \textbf{(a)}~Escort count distribution $P(K)$ for $K=0$--$15$. Top, log-scaled $P(K)$ at $\gamma'=-0.4$, $0$ (gray), and $+0.4$: true-model data (open circles) and closed-form fits (lines). Bottom, $\log_{10}[P_\gamma(K)/P_0(K)]$ relative to $P_0=P(K\mid\gamma'=0)$ for data (open circles) and models (lines) at $\gamma'=\pm0.4$.
    \textbf{(b)}~Recovered dual coordinates $\eta_i$ (top) and $\eta_{ij}$ (bottom) versus truth at $\gamma'=\pm0.4$, dashed, identity.
    \textbf{(c)}~Recovered fields $H_i$ (top) and couplings $J_{ij}$ (bottom) versus truth at $\gamma'=\pm0.4$, dashed, identity.
    \textbf{(d)}~Root-mean-square recovery error across the $N$ fields $H_i$ (top) and $\binom{N}{2}$ couplings $J_{ij}$ (bottom), versus $\gamma'$ for the deformed and $\gamma=0$ closed-form inverses.
    \textbf{(e)}~Hellinger distance $d_{\mathrm{H}}(P,Q)=[1-\sum_{K=0}^{N}\sqrt{P(K)Q(K)}]^{1/2}$ between the true escort count distribution $P$ and each fitted distribution $Q$, both estimated from independent Monte-Carlo samples, versus $\gamma'$.
    \textbf{(f)}~Energy variance per neuron $v=\mathrm{Var}_{\tilde p_\gamma}(E)/N$ at $\beta=1$, using each model's original fields and couplings to define $E$~\eqref{eq:energy}, versus $\gamma'$ for deformed and $\gamma=0$ fits and Monte-Carlo truth (black).
    \textbf{(g)}~Energy variance $v$ versus coupling disorder $N\,\mathrm{sd}(J)$ at $\gamma'=-0.4$; gray dotted, LGd operating point ($N\,\mathrm{sd}(J)=6.68$). \textbf{(h)}~Energy variance $v$ versus $N$ at $\gamma'=-0.4$; gray dotted, LGd operating point ($N=30$). In (g) and (h), the purple right axis shows the deformed closed-form fit's escort-moment residual $\max|\Delta\eta|$, shading in (g) marks coupling values where the across-draw mean exceeds the $0.05$ reliability gate.}
    \label{fig:sim_invising_weakcorr_exact_lgd}
\end{figure*}

Mapping deformation across the brain leaves open which populations admit a simple effective description. Recovering the fields and couplings that fix the induced higher-order hierarchy has so far required a sample-based fit at each finite $\gamma$. We therefore derive a deformed mean-field equation with an iteration-free closed-form inverse for these parameters.

Setting $\beta=1$ and expanding the escort moments, the R\'enyi dual coordinates~\cite{wong2022tsallis,morales2023geometric}, jointly in coupling strength and deformation $\gamma$ yields a closed-form deformed Thouless--Anderson--Palmer (TAP) equation~\eqref{eq:tap_deformed_main}: the classical naive mean-field and Onsager terms~\cite{plefka_convergence_1982,kappen_efficient_1998,roudi2009ising} plus a deformation correction $\gamma\Delta_i(\bv{\eta},\bv{H},\bv{J})$ (\hyperref[sec:methods_tap]{Methods}; Supplementary Notes~\ref{supp_sec:small_deformation_expansion}, \ref{supp_sec:escort_coords_independent} and~\ref{supp_sec:cumulant_deformed_tap}). Adding the deformation response at fixed singleton and pair escort moments to the local $\gamma=0$ inverse gives the iteration-free field and coupling estimators~\eqref{eq:Jij_inverse_main} (\hyperref[sec:methods_tap]{Methods}).

We tested the closed-form inverse under data-realistic disorder and finite sampling in a heterogeneous $N=30$ population matched to the LGd flat-model field and coupling disorder, near the weak-disorder end of the recorded populations (Fig.~\ref{fig:sim_invising_weakcorr_exact_lgd}). Across deformations, we held the target escort moments $\bv{\eta}^{*}$ fixed and replaced the intractable $2^{30}$ enumeration with Monte-Carlo estimates of moments and count distributions from $5\times10^{4}$ draws. At each $\gamma'$, the sample-based deformed inverse of $\bv{\eta}^{*}$ served as a reference for the closed-form inverse~\eqref{eq:Jij_inverse_main}, applied in the deformed family and at $\gamma=0$.
To isolate approximation error from sampling noise, we also tested the inverse by exact enumeration at $N=10$, holding the complete escort vector $(\eta_i,\eta_{ij})$ fixed while re-solving every true $H_i$ and $J_{ij}$ at each deformation (Supplementary Fig.~\ref{fig:sim_invising_weakcorr_exact} and Note~\ref{supp_sec:exact_inversion}).

The closed-form inverse recovered the escort count distribution, first- and second-order moments, and fields and couplings (Fig.~\ref{fig:sim_invising_weakcorr_exact_lgd}a--c). Across deformations, the deformed inverse kept field and coupling errors well below those of the $\gamma=0$ inverse, with its errors lowest near $\gamma'=0$ and increasing with $|\gamma'|$ (Fig.~\ref{fig:sim_invising_weakcorr_exact_lgd}d). Both inverses kept the escort count distribution close to the truth, although the deformed inverse deviated more at strongly negative deformation (Fig.~\ref{fig:sim_invising_weakcorr_exact_lgd}e). The deformed fit tracked the variance of $E$ under the escort distribution, whereas
the $\gamma=0$ inverse missed the deformation-dependent trend because it returns the same parameters at every $\gamma'$ for fixed target moments $\bv{\eta}^{*}$ (Fig.~\ref{fig:sim_invising_weakcorr_exact_lgd}f).
{At $\gamma'=-0.4$, the mean energy-variance curves of the deformed fit and Monte-Carlo reference ($v\approx0.5$) differed by less than $0.017$,} both at the LGd operating point (Fig.~\ref{fig:sim_invising_weakcorr_exact_lgd}g) and for $N=20$--$80$ (Fig.~\ref{fig:sim_invising_weakcorr_exact_lgd}h), and the deformed fit's largest escort-moment mismatch, $\max|\Delta\eta|$, stayed near $0.005$ (Fig.~\ref{fig:sim_invising_weakcorr_exact_lgd}h). This mismatch exceeded the $0.05$ reliability threshold only at the largest coupling disorder, $N\,\mathrm{sd}(J)=15$, and approached it at $N\,\mathrm{sd}(J)=12$ ($0.046$), well beyond the LGd value (gray dotted line in Fig.~\ref{fig:sim_invising_weakcorr_exact_lgd}g).

The closed-form inverse is reliable in the weak-correlation regime but degrades as coupling disorder increases. The moment-matching estimates of the brain-wide survey therefore used sample-based inner fits throughout, which recovered deformation also under strong coupling (Fig.~\ref{fig:efficient_score_recovery}e--h).
 We now ask which recorded populations fall within this weak-correlation regime, using observed correlation statistics rather than coupling disorder.

\subsection*{Applicability of the closed-form theory across the brain}

\begin{figure*}[!t]
    \centering
    \includegraphics[width=0.95\linewidth]{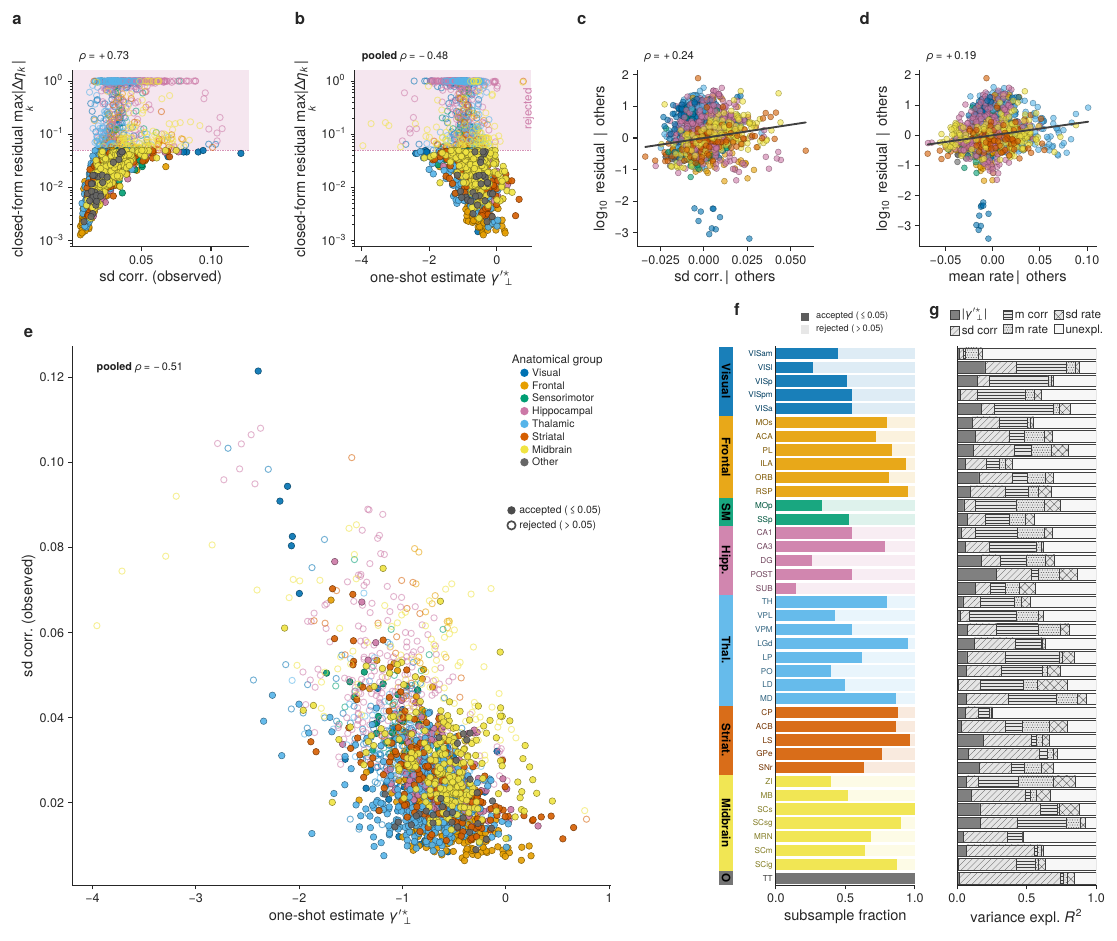}
    \caption[Where the closed-form inverse applies across brain regions]{\textbf{The closed-form inverse's reach is set by correlation structure and deformation.} We evaluate the closed-form inverse at the one-shot efficient-score estimate $\gamma'^\star_\perp$ and quantify its accuracy by the maximum escort-moment residual, $r=\max_k|\Delta\eta_k|$. Across $2271$ random $N=30$ subsamples from $39$ regions ($20$~ms bins), $r$ is plotted against statistics derived from $\bv{\eta}$: the mean and across-unit standard deviation of per-bin firing probability, and the mean and across-pair standard deviation of pairwise Pearson correlation $\rho_{ij}$ (\hyperref[sec:cf_applicability]{Methods}). In (a), (b) and (e), filled points have $r\leq0.05$ and open points $r>0.05$; color denotes anatomical group. \textbf{(a)} Residual (log $y$-axis) versus correlation heterogeneity $\mathrm{sd}(\rho_{ij})$; shading marks $r>0.05$ (two-sided descriptive pooled Spearman $\rho=0.73$, $p<0.001$). \textbf{(b)} Residual (log $y$-axis) versus the one-shot efficient-score estimate $\gamma'^{\star}_\perp$, with the same shading (two-sided descriptive pooled Spearman $\rho=-0.48$, $p<0.001$). \textbf{(c)} Added-variable residuals for correlation heterogeneity across all $n=2271$ subsamples, controlling for the other predictors (two-sided descriptive pooled Spearman $\rho=0.24$, $p<0.001$). \textbf{(d)} Corresponding residuals for mean firing probability ($n=2271$; two-sided descriptive pooled Spearman $\rho=0.19$, $p<0.001$). \textbf{(e)} Correlation heterogeneity versus $\gamma'^{\star}_\perp$ for all $n=2271$ subsamples (two-sided descriptive pooled Spearman $\rho=-0.51$, $p<0.001$). \textbf{(f)} Per-region fractions of accepted ($r\leq0.05$) and rejected ($r>0.05$) subsamples, arranged by anatomical group (SM, sensorimotor; O, other). \textbf{(g)} Per-region Shapley decomposition of linear-model $R^2$ for $\log_{10}r$ among $|\gamma'^\star_\perp|$ and the four firing-probability and correlation statistics; unfilled segments are unexplained. Analysis details and thresholds are given in \hyperref[sec:cf_applicability]{Methods}.}
    \label{fig:cf_obs}
\end{figure*}

We applied the closed-form inverse at the one-shot estimate $\gamma'^\star_\perp$ to all $2271$ subpopulations ($N=30$) and compared its predicted escort moments with the observed moments. The one-shot estimate agreed in sign with the moment-matching estimate in $99\%$ of subpopulations and was rank-correlated with it ($\rho=0.73$; Supplementary Fig.~\ref{fig:tap_mcmc_validation}). We quantified the mismatch by the largest absolute difference across first- and second-order moments, $r=\max_k|\Delta\eta_k|$ (\hyperref[sec:cf_applicability]{Methods}). This mismatch increased with pairwise-correlation heterogeneity, $\mathrm{sd}(\rho_{ij})$ (pooled Spearman $\rho=0.73$; Fig.~\ref{fig:cf_obs}a), and increasingly negative deformation ($\rho=-0.48$; Fig.~\ref{fig:cf_obs}b), but showed weaker association with per-bin firing-probability statistics.
After adjustment for the other population statistics and deformation, pooled associations with mismatch remained positive for correlation heterogeneity (Fig.~\ref{fig:cf_obs}c) and mean firing probability (Fig.~\ref{fig:cf_obs}, with the former slightly stronger. Across the $39$ regions, the median partial Spearman $\rho$ was $0.39$ for correlation heterogeneity and $0.09$ for mean firing probability. The inverse loses accuracy with more negative deformation or greater correlation heterogeneity.

Jointly plotting these limits separated rejected subpopulations ($r>0.05$) from the weakly deformed, weakly heterogeneous applicable regime ($r\leq0.05$; Fig.~\ref{fig:cf_obs}e). Although correlated across pooled subsamples (Spearman $\rho=-0.51$), the factors defined distinct failure modes. Rejection was most prevalent in hippocampal, sensorimotor, and visual populations ($58\%$, $56\%$, and $53\%$), reaching $86\%$ in SUB (subiculum),
$74\%$ in DG (dentate gyrus), and $73\%$ in VISl (lateral visual area), with median $\gamma'^{\star}_\perp$ values of $-1.26$, $-1.10$, and $-1.16$, respectively, while it was rare in frontal populations ($17\%$; Fig.~\ref{fig:cf_obs}f).
Midbrain rejection ($32\%$) arose mainly from correlation heterogeneity, whose equal-region Shapley contribution was $0.34$, compared with $0.09$ for deformation (Fig.~\ref{fig:cf_obs}g). In visual populations, mean correlation contributed most ($0.31$), followed by correlation heterogeneity and deformation ($0.11$ each). These failures identify populations requiring sample-based fitting because the closed-form theory is insufficient.

\section*{Discussion}

Across 39 cortical and subcortical regions of the mouse brain, we found widespread net-suppressive departures from pairwise models. Third-cumulant residuals revealed this structure, while curved-model fits identified predominantly negative R\'enyi curvature. Our closed-form mean-field inverse further delineated the regional limits of a simple analytical description.

The Amari--Hopfield framework provides a standard connection between recurrent neural networks and statistical descriptions of population activity~\cite{amari1972learning,hopfield1982neural}. In its stochastic binary form, each neuron becomes active with probability $p_0(x_i=1\mid\bv{x}_{-i})=\sigma(\beta h_i)$, where $\sigma(z)=(1+e^{-z})^{-1}$ is the logistic function and $h_i=H_i+\sum_{j\ne i}J_{ij}x_j$ is a linear weighted sum of the other neurons' activities plus a bias. With symmetric couplings, updating neurons one at a time according to these probabilities implements Gibbs sampling of the Ising distribution~\cite{ackley_learning_1985}. This logistic input--output form has recently been shown to describe neuronal activity across multiple brain regions and species~\cite{lynn2026simple}.

In curved networks~\cite{aguilera2025explosive}, the activation probability can instead be written as $p_\gamma(x_i=1\mid\bv{x}_{-i})=\sigma(h_i^{\mathrm{eff}})$, with an effective input
\[
h_i^{\mathrm{eff}}=\theta_i^{\mathrm{eff}}+\sum_{j\ne i}\theta_{ij}^{\mathrm{eff}}x_j+\sum_{\substack{j<k\\j,k\ne i}}\theta_{ijk}^{\mathrm{eff}}x_jx_k+\cdots .
\]
The triplet term~\eqref{eq:three_body_coupling} introduces products of two other neurons' activities into this input, and higher-order interactions introduce further combinations, all determined in closed form by the fields, couplings and a single curvature parameter rather than by fitting a free nonlinear function~\cite{gardella2016tractable,humplik2017probabilistic,dicarlo2026extended}. Curvature thus connects effective nonlinear input integration, higher-order population structure and the geometry of the statistical model, allowing all three to be compared across the brain.

In our silence-dominated homogeneous network with weak positive couplings, negative curvature generates an alternating, shrinking hierarchy of higher-order interactions (Fig.~\ref{fig:curvature_orderparameter}d) that reduces the third spike-count cumulant (Fig.~\ref{fig:curvature_orderparameter}b) and increases population silence (Fig.~\ref{fig:curvature_orderparameter}c; see also~\cite{Shimazaki_etal_2015,Tkacik_etal_2014}). In the recurrent-network interpretation, the suppressive triplet component of this hierarchy contributes to sublinear input integration. A similar statistical pattern arises in the dichotomized Gaussian model~\cite{Amari_etal_2003}, in which thresholding correlated Gaussian inputs generates an alternating, shrinking hierarchy of higher-order interactions and excess simultaneous silence~\cite{Shimazaki_etal_2015}. Recurrent models incorporating such hierarchies exhibit threshold-like responses that suppress activation at low input levels, helping to keep the population largely silent~\cite{rodriguez2025higher}. Circuit interactions can also contribute to suppressive nonlinearities in neuronal responses~\cite{carandini2012normalization,kapfer2007supralinear}. Together, these connections suggest that the widespread suppressive structure found here may reflect the combined effects of neuronal thresholds and circuit interactions, with regional variation in the resulting nonlinearity.

Negative curvature was strongest in visual populations and weaker in frontal and midbrain populations (Fig.~\ref{fig:effscore_kappa3}c,d), while the mean-field approximation was broadly applicable to frontal populations (Fig.~\ref{fig:cf_obs}f). Area-specific firing patterns have long been associated with functional specialization~\cite{Shinomoto09,maimon2009beyond,mochizuki2016similarity}. In rodents, dendritic architecture and local connectivity vary across cortical areas~\cite{benavidespiccione2006dendritic,hooks2011laminar}, while membrane physiology and ion-channel function differ across regions and neuronal types~\cite{miller2008region,sheets2011corticospinal,phillips2019repeated}. These cellular and circuit specializations provide distinct substrates for input processing~\cite{harris2015neocortical,baker2018specialized}: neuronal morphology, electrical properties, and cell and branch type shape how dendrites combine and amplify inputs~\cite{fletcher2019neocortical,dembrow2022input}. Regional curvature may therefore capture population-level signatures of how cellular and circuit specializations shape nonlinear input integration and, in turn, regional coding, such as sparse coding of visual inputs~\cite{Ohiorhenuan_etal_2010,Ohiorhenuan10b,ZylberbergSheaBrown_2015}.

In the curved model, nonlinear input integration can also be viewed as a state-dependent modulation of neuronal activation. In this complementary view, the conditional update probability is $p_\gamma(x_i\mid\bv{x}_{-i})={\sigma_\gamma[(2x_i-1)\beta'(\bv{x})h_i]}$, where $\sigma_\gamma(z)=[1+e_\gamma(-z)]^{-1}$ is the deformed logistic function and $e_\gamma(z)=(1+\gamma z)^{1/\gamma}$ is the deformed exponential~\cite{aguilera2025explosive}. The effective inverse temperature $\beta'(\bv{x})=\beta/[1-\gamma\beta E(\bv{x})]$ scales the input according to the population state. In the curved Hebbian associative-memory model with zero external fields and coupling scale $J>0$, positive curvature lowers $\beta'$ for negative-energy memory states, a regime associated with suppression of spurious memories and more robust retrieval~\cite{aguilera2025explosive}. Our silence-dominated model can produce the same direction of modulation with negative curvature. In the homogeneous case, $H<0<J$ and $|H|\ge J$ ensure that the field cost of activation outweighs the coupling gain, so every additional spike raises the energy (Supplementary Note~\ref{supp_sec:meanfield_phase_calibration}). Active states therefore have positive energy relative to silence, and negative curvature lowers $\beta'$ (Supplementary Note~\ref{supp_sec:deformation_sign_invariance}). Thus, under an associative-memory interpretation, the regime supporting the alternating, shrinking hierarchy found here aligns with the temperature modulation associated with robust retrieval.

In the homogeneous model, negative curvature increases population silence but suppresses large synchronous events at fixed firing rates and pairwise correlations (Fig.~\ref{fig:curvature_orderparameter}a,c). The fitted models likewise oversuppress the extreme tail in MOs and CA1 (Fig.~\ref{fig:hoi_survey_effscore}f), although this may partly reflect a limitation of the moment-matching estimate of curvature, which uses a statistic involving moments only up to fourth order. Capturing both frequent silence and heavy-tailed activity requires extending the model. One construction combines an entropy-canceling base measure with an alternating interaction hierarchy to support both features, with a recurrent-network interpretation involving thresholding followed by supralinear activation~\cite{rodriguez2025higher}. Rate-dependent supralinear dendritic integration provides a physiological example of such amplification~\cite{dembrow2022input}. Together, these findings motivate models that suppress weak inputs and amplify stronger drive, potentially allowing sparse activity to coexist with large collective responses. A distinct route is to generalize the deformation itself using two-parameter deformed exponentials within an information-geometric framework~\cite{kaniadakis2005two,newton2024two}.

Our results identify a widespread suppressive component of higher-order neural activity across the mouse brain. R\'enyi curvature summarizes this component and links it to the nonlinearities underlying neural dynamics. This provides a model-based framework for relating regional differences in collective activity to neuronal and circuit specializations.

\section*{Methods}

\subsection*{Curved statistical manifolds}
\label{sec:curved_manifold}

For $\gamma<1$, the curved neural network~\eqref{eq:def_exp_energy} is a R\'enyi maximum-entropy model under escort-moment constraints. A divergence induces a geometry on the space of probability distributions~\cite{amari2000methods}; here we use the R\'enyi divergence
\begin{align}\label{eq:Renyi_div}
    \mathcal{D}_\gamma(p \| q) = -\frac{1}{\gamma} \ln \sum_{\bv{x}} p(\bv{x})^{1-\gamma} q(\bv{x})^{\gamma},
\end{align}
for $N$-dimensional discrete variables $\bv{x}=(x_1,\ldots,x_N)$. Unlike the $\alpha$-divergence~\cite{amarialpha}, it preserves the correspondence between orthogonality in the Fisher metric and a Pythagorean relation, and thereby induces a foliation extending Amari's hierarchical decomposition of higher-order interactions~\cite{amari2001information,morales2021generalization}. For the uniform distribution $q(\bv{x})=2^{-N}$, $\mathcal{D}_\gamma(p\|q)=N\ln 2-H_\gamma[p]$, where~\cite{wong2022tsallis,wong2018logarithmic,morales2021generalization}
\begin{equation}\label{eq:renyi_entropy}
    H_\gamma[p] = \frac{1}{\gamma} \ln \sum_{\bv{x}} p(\bv{x})^{1-\gamma},
\end{equation}
approaching Shannon entropy as $\gamma\to0$. Maximizing it under the escort-moment constraints~\eqref{eq:eta_escort} yields the deformed exponential family~\cite{wong2022tsallis} (Supplementary Note~\ref{supp_sec:max_ent_principle}),
\begin{equation}\label{eq:def_exp}
    p_{\gamma}(\bv{x} | \bv{\theta})
    = \left[1 + \gamma \theta^i F_i(\bv{x}) \right]^{1/\gamma} e^{-\psi_{\gamma}(\bv{\theta})},
\end{equation}
where $\bv{\theta}=(\theta^1,\ldots,\theta^L)$ are the natural parameters (Einstein summation over repeated upper--lower index pairs). We require $1+\gamma\theta^iF_i(\bv{x})>0$ for every configuration, keeping the support fixed. The log partition function is
\begin{align}\label{eq:psi_gamma}
    \psi_{\gamma}(\bv{\theta})
    &= \ln \sum_{\bv{x}} [1+\gamma \theta^i F_i(\bv{x})]^{1/\gamma}.
\end{align}
Curvature replaces the classical Legendre transform with the generalized duality~\cite{wong2022tsallis,morales2023geometric}
\begin{align}
    \phi_\gamma(\bv{\eta}) = \sup_{\bv{\theta}} \left\{\frac{1}{\gamma} \ln(1 + \gamma \theta^i \eta_i) - \psi_\gamma(\bv{\theta}) \right\},
    \label{eq:dual_potential}
\end{align}
over the positivity domain. At an interior maximizer, the dual coordinates satisfy
\begin{subequations}\label{eq:dual_coordinates}
    \begin{align}
        \theta^i &= \frac{\partial^i \phi_{\gamma}}{1-\gamma \partial^k \phi_{\gamma} \eta_k}, \\
        \eta_i &= \frac{\partial_i \psi_{\gamma}}{1-\gamma \theta^k \partial_k \psi_{\gamma}}.
    \end{align}
\end{subequations}
Here $\partial_i=\partial/\partial\theta^i$ and $\partial^i=\partial/\partial\eta_i$. The second relation gives the escort expectations~\eqref{eq:eta_escort} (Supplementary Note~\ref{supp_sec:escort_dual_coordinates}). Equivalently, for $-1<\gamma<1$ (satisfied throughout our search range $\gamma'\in[-3,1]$, i.e.\ $|\gamma|\le0.1$), the same family follows from R\'enyi entropy with deformation $-\gamma$ under ordinary constraints~\cite{aguilera2025explosive}.

\subsection*{Escort reparameterization and moment matching}
\label{sec:escort_fitting}

\begin{table}[t]
    \centering
    \begin{tabular}{lcc}
        \hline
        & Original model $p_\gamma$ & Escort model $\tilde{p}_\gamma = p_{\tilde{\gamma}}$ \\
        \hline
        Deformation & $\gamma$ & $\tilde{\gamma} = \gamma/(1-\gamma)$ \\
        Fields       & $H_i$   & $\tilde{H}_i = (1-\gamma)H_i$ \\
        Couplings    & $J_{ij}$ & $\tilde{J}_{ij} = (1-\gamma)J_{ij}$ \\
        \hline
    \end{tabular}
    \caption{\textbf{Escort-model reparameterization.} The escort distribution of a $\gamma$-deformed model is itself a deformed exponential-family member with shifted deformation and uniformly rescaled parameters.}
    \label{tab:escort_reparam}
\end{table}

Empirical constraints are ordinary sample means $\langle x_i \rangle$ and pairwise moments $\langle x_i x_j \rangle$. For each candidate $\gamma$, we fit $\bv{\theta}$ so the escort expectations~\eqref{eq:eta_escort} match these constraints~\cite{guilmeau2024variational}. The fitted escort then has an explicit reparameterization (Table~\ref{tab:escort_reparam}).

Raising~\eqref{eq:def_exp} to the power $(1-\gamma)$ gives
\begin{align}
    p_\gamma(\bv{x}|\bv{\theta})^{1-\gamma}
    &= \bigl[1 + \gamma\theta^i F_i(\bv{x})\bigr]^{(1-\gamma)/\gamma}
       e^{-(1-\gamma)\psi_\gamma(\bv{\theta})}.
\end{align}
Set $\tilde{\gamma} = \gamma/(1-\gamma)$ and $\tilde{\theta}^{i} = (1-\gamma)\theta^i$. Then $(1-\gamma)/\gamma = 1/\tilde{\gamma}$ and $\tilde{\gamma}{\tilde{\theta}^{i}} = \gamma\theta^i$, so
\begin{equation}
    \bigl[1 + \gamma\theta^i F_i(\bv{x})\bigr]^{(1-\gamma)/\gamma}
    = \bigl[1 + \tilde{\gamma}{\tilde{\theta}^{i}} F_i(\bv{x})\bigr]^{1/\tilde{\gamma}}.
\end{equation}
After normalization, $\tilde{p}_\gamma(\bv{x}|\bv{\theta}) = p_{\tilde{\gamma}}(\bv{x}|\tilde{\bv{\theta}})$. The uniform rescaling preserves coupling-to-field ratios; only the overall scale and effective deformation change. The two distributions coincide as $\gamma\to0$.

To see which effects of deformation can be absorbed by refitting the fields and couplings, we expand the escort log-density to first order in $\gamma$
\begin{equation}
\begin{split}
    \ln p_{\tilde{\gamma}}(\bv{x}\mid{\tilde{\bv{\theta}}})
    &\mathrel{=} \mathrm{const}
    +(1-\gamma)\theta^iF_i(\bv{x}) \\
    &\quad-\frac{\gamma}{2}\bigl[\theta^iF_i(\bv{x})\bigr]^2
    \mathbin{+}{\pazocal{O}(\gamma^2)}.
\end{split}
    \label{eq:escort_factored}
\end{equation}
Rescaling to $\tilde{\theta}^i=(1-\gamma)\theta^i$ absorbs the linear term into a flat family (Table~\ref{tab:escort_reparam}). Part of the centered quadratic term can also be reproduced by refitting the fields and couplings (Supplementary Note~\ref{supp_sec:small_deformation_expansion}). The efficient score derived below~\eqref{eq:efficient_score} removes this pairwise component and isolates the remaining higher-order signal.

\subsection*{Deformation-induced higher-order interactions}
\label{sec:effective_interactions}

Write the natural linear--quadratic form
\begin{equation}\label{eq:u_natural}
\begin{gathered}
    u(\bv{x}) = -\beta E(\bv{x}) = \sum_i \theta_i x_i + \sum_{i<j}\theta_{ij} x_i x_j, \\
    \theta_i=\beta H_i,\;\; \theta_{ij}=\beta J_{ij}.
\end{gathered}
\end{equation}
Then $\ln p_\gamma=\gamma^{-1}\ln(1+\gamma u)-\psi_\gamma$. For $|\gamma u|<1$, $\gamma^{-1}\ln(1+\gamma u)=\sum_{n\ge1}(-1)^{n+1}\gamma^{n-1}u^n/n$ converges; otherwise it is a formal expansion about $\gamma=0$. Through third order,
\begin{equation}\label{eq:logq_nonlin}
\begin{split}
    \ln p_\gamma(\bv{x}) &= \frac{1}{\gamma}\ln (1+\gamma u) - \psi_\gamma \\
    &= u - \tfrac{\gamma}{2} u^2 + \tfrac{\gamma^2}{3} u^3 + \pazocal{O}(\gamma^3) - \psi_\gamma .
\end{split}
\end{equation}
Because $x_i^2=x_i$ and $u$ is quadratic, each power $u^n$ reduces to distinct-index monomials among up to $2n$ neurons. Collecting terms by interaction order recasts the deformed model as an effective higher-order exponential family,
\begin{equation}\label{eq:effective_hamiltonian}
\begin{split}
    \ln p_\gamma = \mathrm{const} &+ \sum_i \theta_i^{\mathrm{eff}} x_i + \sum_{i<j} \theta_{ij}^{\mathrm{eff}} x_i x_j \\
    &+ \sum_{i<j<k} \theta_{ijk}^{\mathrm{eff}} x_i x_j x_k \\
    &+ \sum_{i<j<k<l} \theta_{ijkl}^{\mathrm{eff}} x_i x_j x_k x_l \\
    &+ \text{5-body and higher terms}.
\end{split}
\end{equation}
For any nonempty unit set $A$ whose configurations $\bv 1_B$, $B\subseteq A$, all lie in the positive support, the exact silent-reference coefficient of $\ln p_\gamma$ is the M\"obius transform
\begin{equation}\label{eq:mobius_effective_interaction}
    \theta_A^{\mathrm{eff}}
    =\sum_{B\subseteq A}(-1)^{|A|-|B|}
    \frac{1}{\gamma}\ln [1+\gamma u(\bv 1_B)].
\end{equation}
Here $\bv 1_B$ activates only the units in $B$; at $\gamma=0$, each logarithmic term is interpreted by continuity as $u(\bv 1_B)$. Because $\ln\tilde p_\gamma=(1-\gamma)\ln p_\gamma+\mathrm{const}$, the fitted escort distribution~\eqref{eq:escort_dist} has coefficients $\tilde\theta_A^{\mathrm{eff}}=(1-\gamma)\theta_A^{\mathrm{eff}}$, which Figs.~\ref{fig:curvature_orderparameter}d and~\ref{fig:hoi_survey_effscore}g show. We use this exact finite-$\gamma$ definition for the plotted pair and triplet coefficients, whose configurations all passed the positive-support check.

At $\pazocal{O}(\gamma)$, the quadratic term $-\tfrac{\gamma}{2}u^2$ renormalizes the pairwise interaction,
\begin{equation}\label{eq:two_body_eff}
    \theta_{ij}^{\mathrm{eff}} \simeq \beta J_{ij} - \gamma\beta^2\bigl[H_iH_j + J_{ij}(H_i+H_j) \\
    + \tfrac12 J_{ij}^2\bigr]
\end{equation}
and generates three- and four-body interactions. Three-distinct-index terms give~\eqref{eq:three_body_coupling}, products of disjoint couplings give $\theta_{ijkl}^{\mathrm{eff}}=-\gamma\beta^2[\,J_{ij}J_{kl}+J_{ik}J_{jl}+J_{il}J_{jk}\,]+\pazocal{O}(\gamma^2)$.

\subsection*{Specific heat and energy variance}
\label{sec:thermo_observables}

For the homogeneous temperature sweeps in Fig.~\ref{fig:curvature_orderparameter}e, we first calibrate the model at $\beta=1$. We then vary $\beta$ by raising the calibrated escort configuration probabilities to the power $\beta$ and renormalizing. Its effective energy $\tilde E_\gamma(\bv{x})=-(1-\gamma)\gamma^{-1}\ln[1-\gamma E(\bv{x})]$ gives $\tilde p_{\gamma,\beta}(\bv{x})\propto\tilde p_\gamma(\bv{x})^\beta\propto e^{-\beta\tilde E_\gamma(\bv{x})}$ and $c(\beta)=\beta^2\mathrm{Var}_{\tilde p_{\gamma,\beta}}(\tilde E_\gamma)/N$. The label $\gamma'=-0.4$ identifies the model calibrated at $\beta=1$. The calibrated fields and couplings, raw $\gamma$, and $\tilde E_\gamma$ remain fixed during the sweep. In the homogeneous model, we calculate the specific heat exactly by summing over spike counts $K=0,\ldots,N$ with multiplicities $\binom{N}{K}$.

For the heterogeneous recovery benchmarks, we evaluate the energy variance per neuron $v=\mathrm{Var}_{\tilde p_\gamma}(E)/N$ at $\beta=1$. Each reference or fitted model uses its own original fields and couplings to define $E$~\eqref{eq:energy}. For the finite-sample benchmark (Fig.~\ref{fig:sim_invising_weakcorr_exact_lgd}f--h), we pool four independently seeded escort Gibbs chains, each retaining $5\times10^4$ draws after $5{,}000$ burn-in sweeps, and estimate $\hat v=(NT)^{-1}\sum_{t=1}^{T}(E_t-\bar E)^2$ with $T=2\times10^5$. For $N=10$, we evaluate the same variance by exact enumeration (Supplementary Fig.~\ref{fig:sim_invising_weakcorr_exact}h).

\subsection*{Efficient-score deformation selection}
\label{sec:efficient_score}

The efficient score~\cite{vandervaart1998asymptotic} isolates an infinitesimal deformation after pairwise refitting. Let $p_0$ be the fitted pairwise full-Boltzmann model, that is, the flat model at the fitted parameters, with sufficient statistics $\bv{F}=(x_i,x_ix_j)$, fitted by Boltzmann learning with Gibbs sampling rather than by the closed-form inverse, and let $p_\gamma$ be the deformation path through it, with $p_{\gamma=0}=p_0$. Along this path, $p_\gamma(\bv{x})=(1+\gamma u(\bv{x}))^{1/\gamma}e^{-\psi_\gamma}$, where $u=-\beta E$ uses the fitted pairwise parameters. Expanding about $\gamma=0$ gives
\begin{equation}
    \ln p_\gamma(\bv{x}) = \ln p_0(\bv{x})+\gamma s_0(\bv{x})
    +\pazocal{O}(\gamma^2),
\end{equation}
with
\begin{equation}\label{eq:deformation_score_null}
    s_0(\bv{x})
    \equiv
    \left.\partial_\gamma\ln p_\gamma(\bv{x})\right|_{\gamma=0}
    =-\tfrac12\bigl[u(\bv{x})^2-\langle u^2\rangle_0\bigr].
\end{equation}
By normalization, $\langle s_0\rangle_0=0$, so the Fisher information at $\gamma=0$ is simply
\begin{equation}
    I_{\gamma\gamma} = \langle s_0^2\rangle_0 =\tfrac14 \mathrm{Var}_0(u^2).
\end{equation}

The corresponding pairwise score is $s_\theta\equiv\partial_{\bv{\theta}}\ln p_0 =\bv{F}-\langle\bv{F}{\rangle_0}$. To allow deformation and pairwise refitting simultaneously, define $\delta\vartheta=(\gamma,\delta\theta)^{\!\top}$ and
$S=(s_0,s_\theta)^{\!\top}$. 

Evaluated at $p_0$, the Fisher matrix is
\begin{equation}
    \mathcal{I}=\langle SS^{\!\top}\rangle_0
    =
    \begin{pmatrix}
        I_{\gamma\gamma} & I_{\gamma\theta}\\
        I_{\theta\gamma} & G
    \end{pmatrix},
\end{equation}
where $G=\mathrm{Cov}_0(F,F)$. Writing $b=\mathrm{Cov}_0(u^2,F)$, the mixed block is
    $I_{\theta\gamma}=I_{\gamma\theta}^{\!\top}
    =\mathrm{Cov}_0(F,s_0)=-\tfrac12 b$. 

Let $\ell_0=\langle\ln p_0\rangle_{\mathrm{data}}$ be the mean data log-likelihood at the pairwise model. The Fisher-scoring quadratic for the joint displacement is
\begin{equation} \label{eq:fisher_scoring_quadratic}
    \ell_{\mathrm{quad}}(\delta\vartheta)
    \equiv
    \ell_0 +\delta\vartheta^{\!\top}\langle S\rangle_{\mathrm{data}}
    -\tfrac12\delta\vartheta^{\!\top}\mathcal{I}\delta\vartheta.
\end{equation}

Because the pairwise model matches the data moments, $\langle s_\theta\rangle_{\mathrm{data}}=0$. Stationarity of~\eqref{eq:fisher_scoring_quadratic} with respect to the pairwise parameters therefore gives 
\begin{equation} \label{eq:stationary_eq}
    G\delta\theta = -\gamma I_{\theta\gamma}.
\end{equation}
In the empirical calculation, $G$ is estimated from finite Gibbs samples, so very small Fisher modes are unstable to invert. We therefore compute a regularized Moore--Penrose pseudoinverse $G^+$ from the eigendecomposition of $G$, retaining only eigenmodes whose eigenvalues exceed $10^{-6}$ times the largest eigenvalue. Restricting the refit to these modes gives $\delta\theta^\star(\gamma)=-\gamma G^{+}I_{\theta\gamma}$.

In score space, this compensating refit is the orthogonal projection derived in Supplementary Note~\ref{supp_sec:score_geometry}. Using $P_{\rm pair}s =\operatorname{Cov}_0(s,\bv{F})^\top G^+ \bigl(\bv{F}-\langle \bv{F}{\rangle_0}\bigr)$, the component of the deformation score left after pairwise refitting is
\begin{align}
    s_0^\perp
    &= (1-P_{\rm pair})s_0 =
    s_0-I_{\gamma\theta}G^{+}(\bv{F}-\langle \bv{F}{\rangle_0}) \nonumber\\
    &=
    -\tfrac12\bigl[(u^2-\langle u^2\rangle_0)
    -b^{\!\top}G^{+}(\bv{F}-\langle \bv{F}{\rangle_0})\bigr],
    \label{eq:efficient_score}
\end{align}
with Fisher information
\begin{align}
    I_{\gamma\gamma}^{\perp}
    &=
    I_{\gamma\gamma}
    -I_{\gamma\theta}G^{+}I_{\theta\gamma}
    =\langle(s_0^\perp)^2\rangle_0 \nonumber\\
    &=
    \tfrac14\bigl[\mathrm{Var}_0(u^2)-\bv{b}^{\!\top}G^{+}\bv{b}\bigr],
\end{align}
which removes from $s_0$ the component that can be reproduced by refitting the pairwise parameters. Since the pairwise moments are matched, $\langle s_0^\perp\rangle_{\mathrm{data}} =\langle s_0\rangle_{\mathrm{data}}$.

We fit the model by matching its escort moments to the observed moments. At $\gamma=0$, the deformation score of the escort distribution is
$\tilde{s}_0=s_0-(u-\langle u\rangle_0) =s_0-\bv{\theta}^{\!\top}s_\theta$. Hence $\tilde{s}_0^\perp=s_0^\perp$ and $\tilde I_{\gamma\gamma}^\perp=I_{\gamma\gamma}^\perp$, so the efficient-score estimate is unchanged.
The raw escort information at fixed original fields and couplings is $\tilde I_{\gamma\gamma}=\mathrm{Var}_0(\tilde s_0)$. Figure~\ref{fig:curvature_orderparameter}f shows $\tilde I_{\gamma'\gamma'}=\tilde I_{\gamma\gamma}/(N\beta)^2$ and the common efficient information $\tilde I^\perp_{\gamma'\gamma'}=I^\perp_{\gamma'\gamma'}$.
Substituting the optimal pairwise refit $\delta\theta^\star(\gamma)=-\gamma G^+ I_{\theta\gamma}$ back into~\eqref{eq:fisher_scoring_quadratic} gives a quadratic depending only on $\gamma$ 
\begin{equation}
    \ell_{\mathrm{quad}}^{\mathrm{prof}}(\gamma)
    = \ell_0 + \gamma \langle s_0^\perp\rangle_{\mathrm{data}}
    -\frac{\gamma^2}{2}I_{\gamma\gamma}^{\perp},
\end{equation}
and maximizing it yields
\begin{equation}
    \gamma^\star_\perp
    =
    \frac{\langle s_0^\perp\rangle_{\mathrm{data}}}
    {I^{\perp}_{\gamma\gamma}},
    \qquad
    \gamma'^\star_\perp=N\beta\gamma^\star_\perp .
    \label{eq:efficient_score_estimate}
\end{equation}
For $\gamma'=N\beta\gamma$, $I_{\gamma'\gamma'}=I_{\gamma\gamma}/(N\beta)^2$ and $I^\perp_{\gamma'\gamma'}=I^\perp_{\gamma\gamma}/(N\beta)^2$, so their ratio is invariant under this rescaling. For the empirical fits, we set $\beta=1$, so $\gamma'=N\gamma$.
For the empirical analyses, we used $2\times10^5$ retained Gibbs samples per fitted pairwise model to evaluate the efficient score and its information, and the same number of escort samples per candidate deformation to evaluate the profile statistic.

To estimate finite values of $\gamma'$, we refit the fields and couplings at each candidate deformation by matching the observed moments with escort expectations.
After dropping terms independent of $\gamma'$ from $-2s_0^\perp$, we evaluate $T(\gamma')=\langle u^2\rangle-b^{\!\top}G^{+}\langle F\rangle$ using escort samples from the model refitted at that candidate deformation. The quantities $u$, $F$, and $G^{+}b$ remain fixed at their values in the fitted pairwise model. $T_{\mathrm{data}}$ is the same statistic averaged over the data samples. In Fig.~\ref{fig:efficient_score_recovery}, score estimation and profiling replace $u^2$ with $u^2+2u$. The synthetic recovery test, the representative cost curves, and Fig.~\ref{fig:hoi_survey_effscore} use $\gamma'=-3$ to $1$ in steps of $0.05$. The value
minimizing $|T_{\mathrm{data}}-T(\gamma')|$
defines the moment-matching estimate $\gamma'^\star_{\mathrm{mm}}$; the squared cost in Fig.~\ref{fig:efficient_score_recovery}a,e has the same minimizer. The brain-wide atlas and bin-width survey in Fig.~\ref{fig:effscore_kappa3}c--f and Supplementary Fig.~\ref{fig:supp_gammamap_bins} used nine nodes, followed by a parabolic refinement (linear interpolation in an end interval). Both protocols discard nodes with maximum moment mismatch at least $0.02$. For the empirical analyses, we also reject a selected fit if its refitted moment mismatch or sampled escort-marginal mismatch exceeds $0.05$. Inner fits use sample-based Boltzmann learning because the closed-form inverse is limited to weak correlations.

For the empirical decomposition by interaction order, write the unprojected quadratic term in the deformation score as
\begin{equation}
    q(\bv{x})=-\tfrac12 u(\bv{x})^2
    = q_{\le2}(\bv{x}) + q_3(\bv{x}) + q_4(\bv{x}),
\end{equation}
where $q_3$ and $q_4$ collect the distinct three- and four-body monomials after the binary reduction $x_i^2=x_i$. We apply the same pairwise orthogonal projection at~\eqref{eq:efficient_score} to the $j\in\{3,4\}$ component,
\begin{equation}
    {q_j^\perp}
    = q_j-\langle q_j\rangle_0 
    -\mathrm{Cov}_0(q_j,\bv{F})^{\!\top}G^{+}(\bv{F}-\langle \bv{F}{\rangle_0}).
    \label{eq:score_order_components}
\end{equation}
The centered component $q_{\leq2}-\langle q_{\leq2}\rangle_0$ lies in the pairwise score space and therefore vanishes under this projection. Linearity then results in $s_0^\perp=q_3^\perp+q_4^\perp$, exact in theory and to numerical precision in the calculations. Consequently, 
\begin{equation} 
    I_{\gamma\gamma}^\perp =\mathrm{Var}_0(s_0^\perp) =I_3+I_4 +2 \mathrm{Cov}_0(q_3^\perp,q_4^\perp), 
\end{equation}
where $I_j=\mathrm{Var}_0(q_j^\perp)$.
We compare the squared norms of the two components using
\begin{equation}
    f_3^{\mathrm{norm}} = \frac{I_3}{I_3+I_4},
\end{equation}
and quantify their correlation by
\begin{equation}
    \rho_{34}=
    \frac{\mathrm{Cov}_0(q_3^\perp,q_4^\perp)}
    {\sqrt{I_3I_4}}.
\end{equation}
Supplementary Fig.~\ref{fig:supp_effscore_order_fraction} reports both quantities.

\subsection*{Recording data and regional selection}
\label{sec:regional_selection}

We analyzed spontaneous Neuropixels recordings from~\cite{steinmetz2019distributed} ($39$ sessions from $10$ mice), using the dataset's fine-grained Allen Common Coordinate Framework~\cite{wang2020allen} labels rather than the original report's curated $42$-region grouping. Spontaneous activity was taken from the four intervals labelled ``spontaneous'' in each session of the dataset (Data availability), which its authors define as intervals of sufficient duration with no task or stimulus presentation. Each interval was binned separately and the bins inside the probes' recording ranges were concatenated ($3.9$--$21.0$~min per session, median $14.6$~min). Bins started at the onset of each interval, and we set $x_i=1$ if unit $i$ fired at least one spike in the bin and $x_i=0$ otherwise. We excluded units labelled ``root'' (no assigned region). For each remaining region, we required at least $30$ units with spontaneous-block firing rates above $0.5$~Hz in each of at least three sessions. We included units annotated as multi-unit activity (MUA; $26.5\%$ of units above the rate criterion). The resulting $39$ regions overlap with but are not a subset of the published $42$. Each eligible session contributed random $N=30$ subpopulations ($2271$ total). For a region with $n$ qualifying units in a session, we drew $\min(10,\max(1,n-30))$ subpopulations independently, each without replacement, so subpopulations from one session can share units.

We grouped the $39$ fine-grained Allen labels into eight study-defined anatomical groups (session counts in parentheses): Visual (VISa (4), VISam (11), VISl (3), VISp (10), VISpm (4)); Frontal (ACA (9), ILA (3), MOs (15), ORB (5), PL (8), RSP (9)); Sensorimotor (MOp (3), SSp (4)); Hippocampal (CA1 (18), CA3 (6), DG (11), POST (3), SUB (9)); Thalamic (LD (5), LGd (8), LP (8), MD (3), PO (5), TH (8), VPL (4), VPM (4)); Striatal (ACB (3), CP (5), GPe (3), LS (3), SNr (3)); Midbrain (MB (6), MRN (11), SCig (4), SCm (6), SCs (3), SCsg (3), ZI (3)); and Other (TT (3)).

We defined the third-cumulant residual $\Delta\kappa_3[K]=\kappa_3^{\mathrm{data}}[K]-\kappa_3^{\mathrm{null}}[K]$ against the variance-matched pairwise model. For binary units, $x_i^2=x_i$ reduces every repeated-index term of the count cumulant to firing rates and pairwise covariances,
\begin{align}
    \kappa_3[K]&=\sum_i \eta_i(1-\eta_i)(1-2\eta_i) \nonumber \\
    &\hphantom{=} +3\sum_{i\neq l}(1-2\eta_i)\mathrm{Cov}(x_i,x_l) +6\sum_{i<j<k}\kappa_{ijk},
    \label{eq:kappa3_decomp}
\end{align}
with $\kappa_{ijk}$ the connected triplet cumulant defined in~\eqref{eq:kappa3_residual}, so the raw $\kappa_3[K]$ inherits rate and correlation dependence. Matching of all firing rates and pairwise moments fixes the mean and variance of the population count $K$ and every repeated-index term in $\kappa_3[K]$, leaving the pooled distinct-triplet residual~\eqref{eq:kappa3_residual}. Because the pairwise null itself carries connected triplet correlations induced by its couplings, $\Delta\kappa_3[K]$ measures triplet structure in excess of the fitted null rather than the data's triplet cumulant alone. 

For the brain-wide tests coloring Fig.~\ref{fig:effscore_kappa3}b,d, we separately ranked the absolute nonzero third-cumulant residuals and moment-matching estimates within each region, then restored their signs. We averaged these signed ranks within each session and summed the session averages. We obtained two-sided $p$-values from all possible session-level sign flips, then applied Benjamini--Hochberg correction across $39$ regions. Fig.~\ref{fig:effscore_kappa3}e,f used the $39$ regional medians of per-session medians shown in Fig.~\ref{fig:effscore_kappa3}b,d, as did the shading in Fig.~\ref{fig:effscore_kappa3}a,c, in descriptive two-sided Spearman correlations; the nominal $p$-values do not model repeated subjects or sessions shared across regions. For the exploratory comparison in Fig.~\ref{fig:effscore_kappa3}f, we selected the statistic with the largest absolute regional Spearman correlation with the moment-matching estimate among the means and standard deviations of per-bin firing probability and pairwise correlation. The moving-block bootstrap test of individual subsamples is described in Supplementary Fig.~\ref{fig:supp_perreg_kappa3}.

For the detailed $20$~ms analysis (Fig.~\ref{fig:hoi_survey_effscore}), we selected the region with the most pooled subpopulations in each group, retaining representatives with more than $50$. Because each session contributed at most ten subpopulations, this required at least six sessions. Five groups qualified, represented by VISam (Visual), MOs (Frontal), CA1 (Hippocampal), LGd (Thalamic), and MRN (Midbrain). TH and LGd tied at $80$, so we retained the more specific LGd label. Selection used coverage only, not the residual, inferred curvature, effect size, or significance. For Fig.~\ref{fig:hoi_survey_effscore}b,c, we pooled subsample signed ranks, averaged them within sessions, and obtained two-sided permutation $p$-values by flipping session-level signs. Both panels used Benjamini--Hochberg correction across the five pre-specified regions.

Null third cumulants were evaluated using $3.84\times10^5$ retained Gibbs samples per fitted pairwise model. The count-distribution comparisons in Fig.~\ref{fig:hoi_survey_effscore} used $6.4\times10^5$ samples from the pairwise model and $10^6$ escort samples from the selected curved model.

\subsection*{Derivation of the deformed TAP equation}
\label{sec:methods_tap}

Our mean-field construction uses the R\'enyi dual coordinates, the escort expectations~\eqref{eq:eta_escort}, related to the natural parameters by a generalized Legendre duality that becomes classical as $\gamma\to0$~\cite{wong2022tsallis,morales2023geometric}. At finite $\gamma$, zero pairwise couplings do not produce a factorized reference model, so the standard mean-field construction~\cite{plefka_convergence_1982,kappen_efficient_1998,tanaka_mean-field_1998,roudi2009ising,tanaka_theory_1999} fails. We therefore expand the escort moments directly in connected cumulants of the pair energy, jointly in the coupling scaling parameter $\alpha$ and deformation $\gamma$. Setting $\beta=1$, so $H_i$ and $J_{ij}$ are dimensionless, and inverting this expansion order by order yields a closed-form deformed Thouless--Anderson--Palmer (TAP) equation,
\begin{align}\label{eq:tap_deformed_main}
    \mathrm{logit}(\eta_i)
    &= H_i + \sum_{j\neq i}J_{ij} \eta_j \nonumber\\
    &\quad + \frac{1-2\eta_i}{2}\sum_{j\neq i}J_{ij}^2 \eta_j(1-\eta_j) \nonumber\\
    &\quad + \gamma \Delta_i(\bv{\eta},\bv{H},\bv{J})
    + \pazocal{O}(\alpha^3,\gamma\alpha^2,\gamma^2).
\end{align}
Here $\mathrm{logit}(\eta_i)=\ln[\eta_i/(1-\eta_i)]$. The first two lines are the $\gamma=0$ naive mean-field and Onsager terms, $\gamma \Delta_i(\bv{\eta},\bv{H},\bv{J})$ is the deformation correction, \eqref{eq:tap_deformed_main} is derived below and in Supplementary Note~\ref{supp_sec:cumulant_deformed_tap}.

Let $V(\bv{x})\coloneqq\sum_{a<b}J_{ab}x_ax_b$ and replace $J_{ij}\to\alpha J_{ij}$ to grade coupling order. At $\gamma=0$, $p\propto p_{\mathrm{ind}}(\bv{x})e^{\alpha V(\bv{x})}$, where the independent base has $p_i=\sigma(H_i)=(1+e^{-H_i})^{-1}$. Thus, for any observable $g$ to order $\alpha^2$,
\begin{equation}\label{eq:cumulant_expansion_main}
    \mathbb{E}_\alpha[g] \simeq \langle g\rangle_{\mathrm{ind}}
    + \alpha \kappa_{\mathrm{ind}}(g,V)
    + \frac{\alpha^2}{2} \kappa_{\mathrm{ind}}(g,V,V),
\end{equation}
where $\langle\cdot\rangle_{\mathrm{ind}}$ and $\kappa_{\mathrm{ind}}$ are expectations and joint cumulants under $p_{\mathrm{ind}}$. Setting $g=x_i$ and inverting for $H_i$ through $\pazocal{O}(\alpha^2)$ gives the classical $\{0,1\}$ TAP equation, the $\gamma=0$ part of~\eqref{eq:tap_deformed_main}. The first cumulant, $\kappa_{\mathrm{ind}}(x_i,V)=\chi_i\sum_{j\neq i}J_{ij}p_j$ with $\chi_i=p_i(1-p_i)$, gives the naive mean-field term. The second cumulant sums the three connected edge topologies and gives the Onsager correction.

The small-deformation expansion of Supplementary Note~\ref{supp_sec:small_deformation_expansion} relates the escort coordinate to the ordinary undeformed moment at the same natural parameters, $\eta_i=m_i-\gamma A_i+\pazocal{O}(\gamma^2)$. Transporting classical TAP through this map and linearizing the logit and interaction correction about $\bv{m}=\bv{\eta}$ yields the $\pazocal{O}(\gamma)$ stationarity bracket $\Xi_i$. Its $\alpha^0$ and $\alpha^1$ sectors give the pure deformed-Bernoulli shift and leading coupling correction. Retaining both gives $\Delta_i=\Xi_i^{(0)}+\Xi_i^{(1)}$ in~\eqref{eq:tap_deformed_main}.

For the closed-form inverse from observed escort moments, write
$h_i=\mathrm{logit}(\eta_i)$, $\chi_i^\eta=\eta_i(1-\eta_i)$,
$y_{ij}=(\eta_{ij}-\eta_i\eta_j)/(\chi_i^\eta\chi_j^\eta)$ for $i\ne j$,
$y_{ii}=0$, and $s=h^k\eta_k$.
We retain the local $\gamma=0$ estimates $(H_i^{\rm flat},J_{ij}^{\rm flat})$~\cite{roudi2009ising,roudi2009statistical,tanaka_mean-field_1998,sessak_small-correlation_2009}
and add the deformation response at fixed singleton and pair escort moments
\begin{equation}\label{eq:Jij_inverse_main}
    \begin{aligned}
    \widehat J_{ij}
     &= J_{ij}^{\rm flat}+\gamma h_i h_j\\
     &\quad+\gamma y_{ij}\{1+s+h_i(1-2\eta_i)+h_j(1-2\eta_j)\},\\
    \widehat H_i &= H_i^{\rm flat}+\gamma G_{H,i}.
    \end{aligned}
\end{equation}
Here $J_{ij}^{\rm flat}=y_{ij}-c_{ij}^{\eta}y_{ij}^2$,
$c_{ij}^{\eta}=\tfrac12(1-2\eta_i)(1-2\eta_j)$; the corresponding
$H_i^{\rm flat}$ and the explicit $G_{H,i}$ are given in Supplementary
Note~\ref{supp_sec:cumulant_deformed_tap},
equations~\eqref{eq:wc_fixed_full_eta_flat} and \eqref{eq:wc_fixed_full_eta_response}.
This iteration-free estimator is used in Figs.~\ref{fig:sim_invising_weakcorr_exact_lgd}
and~\ref{fig:sim_invising_weakcorr_exact} and in the empirical applicability
analysis. Its deformation response is complete through linear coupling
at fixed singleton and pair escort moments; the local flat-closure error, $\pazocal{O}(\alpha^2)$, remains.
The truncation order of the forward TAP equation~\eqref{eq:tap_deformed_main}
is therefore not the overall accuracy order of this inverse.

\subsection*{Closed-form applicability analysis}
\label{sec:cf_applicability}

We evaluate the closed-form inverse at the one-shot efficient-score estimate $\gamma'^\star_\perp$ and quantify its accuracy by the maximum escort-moment residual, $r=\max_k|\Delta\eta_k|$. Direct deformation associations used the signed one-shot estimate $\gamma'^{\star}_\perp$. Partial-regression and Shapley predictors used its magnitude $|\gamma'^{\star}_\perp|$ and the mean and standard deviation of per-bin firing probabilities $\eta_i$ and pairwise Pearson correlations $\rho_{ij}=(\eta_{ij}-\eta_i\eta_j)/\sqrt{\eta_i(1-\eta_i)\eta_j(1-\eta_j)}$. Descriptive pooled Spearman analyses included all $2271$ subpopulations. These two-sided $p$-values treat subsamples as independent and do not account for dependence within sessions.

For each partial-regression plot, we separately fitted $\log_{10}r$ and the predictor plotted in that panel by ordinary least squares using the other four predictors and an intercept. We then computed the Spearman correlation between their residuals; for the within-region medians, we repeated both steps within each region.

Within each region, we fitted $\log_{10}r$ by linear regression using the five standardized predictors. To obtain each predictor's Shapley contribution~\cite{lipovetsky2001analysis}, we averaged the increase in $R^2$ when adding it to the model over all possible orders of adding the predictors. Group-level values are unweighted regional means. We included subsamples in the correlation, partial regression, and Shapley analyses even when the closed-form approximation was inaccurate. The categorical summary classified $r\leq0.05$ as accepted and $r>0.05$ as rejected; group-level rejection fractions pool subsamples within each group.

Because sessions differ in length ($T=11{,}560$--$63{,}054$ bins), a fixed
residual threshold sits at a session-dependent multiple of the sampling-noise
floor.
For firing rates and pairwise moments separately, we divided the maximum absolute residual by the mean of the sampling standard errors $\sqrt{\eta_k(1-\eta_k)/T}$. We used the larger of these two standardized residuals. We set the standardized cutoff so that the same number of subsamples was classified as rejected before and after standardization. This kept $92\%$ of the $2271$ subsamples in the same accepted/rejected class. The standardized pairwise-moment residual was larger in $96\%$ of subsamples. Standardization increased rejection in SUB ($86\%\to92\%$). In the midbrain group, correlation heterogeneity contributed more than deformation under both criteria. We therefore report the residual in moment
units.

\section*{Data availability}
The Neuropixels recordings analyzed in this study are publicly available as the Steinmetz et al.\ dataset, \emph{Distributed coding of choice, action and engagement across the mouse brain}, from the DANDI Archive under Dandiset~000017, version 0.240329.1926: \url{https://doi.org/10.48324/dandi.000017/0.240329.1926}, and from figshare: \url{https://doi.org/10.6084/m9.figshare.9974357}. Both datasets are distributed under a CC-BY-4.0 license.

\section*{Code availability}
The analysis code used in this study is available in the GitHub repository:\\
\url{https://github.com/shimazaki/deftap_paper}. 

\bibliographystyle{naturemag}
\bibliography{ref}

\begin{acknowledgments}
The authors thank Miguel Aguilera, Fernando Rosas, and Henrik Jeldtoft Jensen for helpful comments and discussions. P.A.M. acknowledges support by JSPS KAKENHI Grant Numbers 23K16855 and 24K21518. H.S. is supported by JSPS KAKENHI Grant Numbers JP 20K11709, 21H05246, 24K21518, and 25K03085.
\end{acknowledgments}

\section*{Author contributions}
Conceptualization: H.S.;
Methodology: P.A.M., T.M.; 
Formal analysis: P.A.M., T.M., H.S.;
Software: P.A.M., T.M., H.S.; 
Visualization: P.A.M., H.S.;
Funding acquisition: P.A.M., H.S.; 
Supervision: H.S.; 
Writing -- original draft: T.M., P.A.M., H.S.;
Writing -- review \& editing: P.A.M., T.M., H.S.

\section*{USE OF LARGE LANGUAGE MODELS}
Claude (Anthropic) and Codex (OpenAI) assisted with code and text editing; the authors verified all outputs and take full responsibility for the content.

\section*{ETHICS STATEMENT}

No new animal experiments were performed; the original experiments were conducted according to the UK Animals (Scientific Procedures) Act 1986 under personal and project licenses issued by the Home Office following appropriate ethics review~\cite{steinmetz2019distributed}.

\section*{Competing interests}
The authors declare no competing interests.

\needspace{4\baselineskip}
\section*{Materials \& Correspondence}
Correspondence and requests for materials should be addressed to P.A.M. or H.S.

\clearpage
\clearpage

\appendix
\onecolumngrid

\begin{center}

{\large Curved geometry of collective neural activity across the mouse brain}
\vspace{1em} \\
{\large \textbf{Supplementary Information}}
\vspace{1em} \\
Pablo A. Morales \\
    \textit{Research Division, Araya Inc., Tokyo, Japan \\
    Centre for Complexity Science, Imperial College London, London, UK}
\vspace{1em} \\
Th\'eo Marchal \\
\textit{LMPS, ENS Paris-Saclay, Gif-sur-Yvette, France}
\vspace{1em} \\
Hideaki Shimazaki \\
    \textit{Graduate School of Informatics, Kyoto University, Kyoto, Japan\\
    }
\end{center}

\renewcommand\appendixname{Supplementary Note}
\setcounter{figure}{0}
\renewcommand{\thefigure}{S\arabic{figure}}
\renewcommand{\theHfigure}{S\arabic{figure}}
\renewcommand{\thesection}{\arabic{section}}
\renewcommand{\theequation}{S\thesection.\arabic{equation}}
\setcounter{page}{1}

\makeatletter
\renewcommand\addcontentsline[3]{\def\@tempa{toc}\def\@tempb{#1}\def\@tempc{section}\def\@tempd{#2}\ifx\@tempa\@tempb\ifx\@tempc\@tempd\theo@acl{#1}{#2}{#3}\fi\else\theo@acl{#1}{#2}{#3}\fi}
\makeatother
\tableofcontents
\vspace{1em}

\newpage

\section{R\'enyi maximum-entropy construction of the curved exponential family}
\label{supp_sec:max_ent_principle}

We derive the deformed exponential family by maximizing R\'enyi entropy under escort moment constraints, then state the corresponding result under standard constraints.

The maximum R\'enyi entropy principle seeks a distribution $p$ maximizing $H_\gamma [p]$
\begin{equation}\label{eq:renyi_entropy_supp}
    H_\gamma[p] = \frac{1}{\gamma} \ln \sum_{\bv{x}} p(\bv{x})^{1-\gamma},
\end{equation}
subject to normalization and escort-moment constraints $\langle F_i \rangle_{\tilde{p}} = c_i$ with $c_i$ constants, on sufficient statistics $F_i (\bv{x})$.

We first establish existence and global optimality on the finite state space used here. Assume $\gamma<1$ and let $\alpha=1-\gamma>0$. The escort map $p\mapsto\tilde p$, defined by $\tilde p(\bv{x})=p(\bv{x})^\alpha/\sum_{\bv{y}}p(\bv{y})^\alpha$, is a bijection of the closed probability simplex $\Delta$, with inverse
\begin{equation}\label{eq:escort_inverse_maxent}
    p(\bv{x})
    =\frac{\tilde p(\bv{x})^{1/\alpha}}
    {\sum_{\bv{y}}\tilde p(\bv{y})^{1/\alpha}}.
\end{equation}
Thus the escort constraints define a compact convex polytope $\pazocal{R}_c=\{\tilde p\in\Delta:\sum_{\bv{x}}\tilde p(\bv{x})F_i(\bv{x})=c_i\}$, and direct substitution gives
\begin{equation}\label{eq:renyi_escort_objective}
    H_\gamma[p]
    =-\frac{\alpha}{\gamma}
    \ln\sum_{\bv{x}}\tilde p(\bv{x})^{1/\alpha}.
\end{equation}
If $0<\gamma<1$, maximizing~\eqref{eq:renyi_escort_objective} is equivalent to minimizing the strictly convex function $\sum_{\bv{x}}\tilde p(\bv{x})^{1/\alpha}$ on $\pazocal{R}_c$. If $\gamma<0$, this function is strictly concave and must instead be maximized. Hence, a unique global maximizer exists whenever $\pazocal{R}_c$ is nonempty. For $\gamma \ne 0$, the first-order equations below are necessary and sufficient for global optimality at an interior solution. At $\gamma=0$, the same existence, uniqueness, and optimality conclusions follow from strict concavity of Shannon entropy. The multipliers need not be unique when the sufficient statistics are redundant.

For $\gamma<0$, the maximizer is positive on every state not forced to zero by the constraints. If it vanished on such a state, mixing its escort distribution with a feasible one positive there, with a small weight $t>0$, would produce a gain of order $t^{1/\alpha}$. This gain dominates the order-$t$ change on the existing support because $1/\alpha<1$. Thus the maximizer has full support whenever $\pazocal{R}_c$ contains a full-support distribution. For $0<\gamma<1$, the maximizer can lie on the boundary even when full-support feasible distributions exist. The corresponding Karush--Kuhn--Tucker conditions give $p(\bv{x})\propto[a+b^iF_i(\bv{x})]_+^{1/\gamma}$, where $[u]_+=\max(u,0)$. We study the full-support case, in which $a+b^iF_i(\bv{x})>0$ for every configuration.

We now derive the interior solution. The escort distribution is $\tilde{p} \coloneqq p^{1-\gamma}/Z$ with partition function $Z=\sum_{\bv{x}}p(\bv{x})^{1-\gamma} = e^{\gamma H_\gamma}$, and the constraints read
\begin{equation}
    \langle F_i \rangle_{\tilde{p}} \coloneqq \sum_{\bv{x}} \tilde{p}(\bv{x}) F_i(\bv{x}) = c_i.
\end{equation}
We define the Lagrangian with multipliers $\lambda_0$ and $\lambda^i$
\begin{equation}
    \mathcal{L}[p] = H_\gamma[p] + \lambda_0 \Big(1 - \sum_{\bv{x}} p(\bv{x})\Big) + \lambda^i \Big( c_i - \langle F_i \rangle_{\tilde{p}} \Big).
\end{equation}
At the stationary point,
\begin{equation}
    \frac{1-\gamma}{\gamma} \frac{p(\bv{x})^{-\gamma}}{Z} - \lambda^i \frac{1-\gamma}{Z} p(\bv{x})^{-\gamma} \big( F_i(\bv{x}) - c_i \big) = \lambda_0.
\end{equation}
Factoring out $(1-\gamma)p^{-\gamma}/Z$ from the left-hand side yields
\begin{equation}
    \frac{(1-\gamma)\,p(\bv{x})^{-\gamma}}{Z}\left[\frac{1}{\gamma} - \lambda^i\!\big(F_i(\bv{x}) - c_i\big)\right] = \lambda_0.
\end{equation}
Solving for $p(\bv{x})^{-\gamma}$ and taking the $(-1/\gamma)$-th power gives
\begin{equation}
    p(\bv{x})
    = \left(\frac{\gamma \lambda_0 Z}{1-\gamma}\right)^{-\frac{1}{\gamma}} \left[1 - \gamma \lambda^i \big(F_i(\bv{x}) - c_i\big) \right]^{1/\gamma},
\end{equation}
where the bracket must be strictly positive for every configuration in the interior solution.
Let $A=1+\gamma{\lambda^i}c_i$. For the binary sufficient statistics used here, $F_i(\bv 0)=0$, so $A$ is the bracket evaluated at the all-zero configuration and is strictly positive. The bracket then factors as $A[1-\gamma(\lambda^i/A)F_i]$. Thus the factor $A^{1/\gamma}$ and the $Z$-dependent prefactor are absorbed into the normalization, while the natural parameters are defined as the rescaled, sign-reversed multipliers, $\theta^i=-\lambda^i/A$. This yields $1+\gamma\theta^iF_i$ inside the bracket.
This leads to the deformed exponential family~\cite{wong2018logarithmic,wong2022tsallis}
\begin{equation}
    \label{eq:def_exp_supp_escort}
    p_{\gamma}(\bv{x} | \bv{\theta})
    = \left[1 + \gamma \theta^i F_i(\bv{x}) \right]^{1/\gamma} e^{-\psi_{\gamma}(\bv{\theta})},
\end{equation}
where $\psi_{\gamma}(\bv{\theta})$ is a log partition function
\begin{align}\label{eq:psi_gamma_supp}
    \psi_{\gamma}(\bv{\theta})
    &= \ln \sum_{\bv{x}} [1+\gamma \theta^i F_i(\bv{x})]^{1/\gamma}.
\end{align}

Under the standard constraints $\sum_{\bv{x}} p(\bv{x}) F_i(\bv{x}) = c_i$, the feasible set is also a compact convex polytope. For $0<\gamma<1$, maximizing R\'enyi entropy is equivalent to maximizing the strictly concave function $\sum_{\bv{x}}p(\bv{x})^{1-\gamma}$, whereas for $\gamma<0$, it is equivalent to minimizing the same strictly convex function. Thus, for the standard case, a unique global optimizer exists whenever the constraints are feasible. At $\gamma=0$, the same conclusion follows from the strict concavity of Shannon entropy.
Repeating the interior variational argument, the constraint derivative no longer carries the $p^{-\gamma}$ factor, which changes the exponent from $+1/\gamma$ to $-1/\gamma$
\begin{equation}
    \label{eq:def_exp_supp_original}
    p_{\gamma}^{\mathrm{std}}(\bv{x} | \bv{\theta})
    = \left[1 + \gamma \theta^i F_i(\bv{x}) \right]^{-1/\gamma} e^{-\psi_{\gamma}^{\mathrm{std}}(\bv\theta)},
\end{equation}
where the bracket must be strictly positive for every configuration in a full-support interior solution and this family has its own log partition function,
\begin{equation}
    \psi_{\gamma}^{\mathrm{std}}(\bv\theta)
    =\ln\sum_{\bv{x}}
    \left[1 + \gamma \theta^i F_i(\bv{x}) \right]^{-1/\gamma}.
\end{equation}
In the notation of~\cite{wong2022tsallis}, the $\pazocal{F}^{(\pm a)}$ families with $a>0$ have densities proportional to $(1+a\,\vartheta^iF_i)^{\mp1/a}$. The standard-constraint model is therefore the $\pazocal{F}^{(a)}$ family with $a=\gamma$ if $\gamma>0$, and the $\pazocal{F}^{(-a)}$ family with $a=-\gamma$ and $\bv\vartheta=-\bv\theta$ if $\gamma<0$. The escort-constraint family~\eqref{eq:def_exp_supp_escort}, which is the model used throughout this paper, is their $\lambda$-exponential family with $\lambda=\gamma$: the $\pazocal{F}^{(-a)}$ family with $a=\gamma$ if $\gamma>0$, and the $\pazocal{F}^{(a)}$ family with $a=-\gamma$ and $\bv\vartheta=-\bv\theta$ if $\gamma<0$.

\newpage

\section{Escort dual coordinates and derivatives of the curved log partition function}
\label{supp_sec:escort_dual_coordinates}

In the classical setting ($\gamma=0$), the dual coordinates are gradients of a scalar potential, $\eta_i = \partial_i \psi_0(\bv{\theta})$, as guaranteed by the dually flat geometry of the exponential family. At finite $\gamma$, the derivatives $\partial_i\eta_j$ and $\partial_j\eta_i$ generally differ~\eqref{eq:escort_jacobian_exact}, so the dual coordinates follow the generalized relation~\eqref{eq:dual_coordinates} rather than the gradient relation above. We now evaluate that relation explicitly and derive the first derivative of $\psi_\gamma$ and the exact escort Jacobian needed for the small-deformation expansion.
Assume throughout that $1+\gamma\theta^kF_k(\bv{x})>0$ for every configuration, so the model remains in the interior of its positivity domain and its support is fixed under parameter differentiation. The derivative of $\psi_{\gamma}(\bv{\theta})$ with respect to $\theta^i$ is
\begin{align}
\label{eq:psi_derivative}
    \partial_i \psi_\gamma(\bv{\theta}) &= \frac{\sum_{\bv{x}} (1+\gamma\theta^k F_k(\bv{x}))^{1/\gamma-1} F_i(\bv{x})}{\sum_{\bv{x}} (1+\gamma \theta^k F_k(\bv{x}) )^{1/\gamma}} \nonumber\\
    &= \sum_{\bv{x}} \frac{(1+\gamma \theta^k F_k(\bv{x}))^{1/\gamma}}{1+\gamma \theta^k F_k(\bv{x})} e^{-\psi_{\gamma}(\bv{\theta})} F_i(\bv{x}) \nonumber\\
    &= \sum_{\bv{x}} \hat{p}_{\gamma}(\bv{x} | \bv{\theta}) F_i(\bv{x}),
\end{align}
with $\hat{p}_\gamma (\bv{x} | \bv{\theta})$ denoting the unnormalized escort weight $\hat{p}_\gamma (\bv{x}|\bv{\theta}) = p_{\gamma}(\bv{x}|\bv{\theta})/(1+\gamma \theta^k F_k(\bv{x}))$.
This recovers the expression $\partial_i \psi_\gamma(\bv{\theta}) = \mathbb{E}_{p_\gamma}[Z_{\theta i}]$ with $Z_{\theta i} \coloneqq F_i(\bv{x}) / (1+\gamma \theta^k F_k(\bv{x}))$. Using $\partial_i \psi_\gamma(\bv{\theta})$, the dual coordinate $\eta_i$ given by~\eqref{eq:dual_coordinates} is computed as
\begin{align}
    \eta_i
    &= \frac{\sum_{\bv{x}} \hat{p}_\gamma (\bv{x} | \bv{\theta})F_i(\bv{x})}{1-\gamma \sum_{\bv{x}} \hat{p}_\gamma (\bv{x} | \bv{\theta}) \theta^l F_l(\bv{x})} \nonumber\\
    &=\frac{\sum_{\bv{x}} \hat{p}_\gamma (\bv{x} | \bv{\theta}) F_i (\bv{x})}{1- \sum_{\bv{x}}\hat{p}_\gamma (\bv{x} | \bv{\theta})(\gamma \theta^j F_j(\bv{x}) + 1) + \sum_{\bv{x}} \hat{p}_\gamma (\bv{x} | \bv{\theta})} \nonumber\\
    &= \frac{\sum_{\bv{x}} \hat{p}_\gamma (\bv{x} | \bv{\theta}) F_i(\bv{x})}{\sum_{\bv{x}} \hat{p}_\gamma (\bv{x} | \bv{\theta})},
\end{align}
where the last step uses $\hat{p}_\gamma(\bv{x}|\bv{\theta})(1+\gamma \theta^k F_k(\bv{x})) = p_\gamma(\bv{x}|\bv{\theta})$ and $\sum_{\bv{x}} p_\gamma(\bv{x}|\bv{\theta}) = 1$, so that $\sum_{\bv{x}} \hat{p}_\gamma (1+\gamma \theta^k F_k) = 1$ and the denominator collapses.
Thus, the dual coordinate reduces to the escort expectation of $F_i$:
\begin{align}
    \eta_i = \sum_{\bv{x}} \tilde {p}_{\gamma}(\bv{x}|\bv{\theta}) F_i(\bv{x}).
    \label{eq:eta_escort_supp}
\end{align}
Indeed, the normalized weight is exactly the escort distribution in~\eqref{eq:escort_dist}:
\begin{equation}
    \tilde {p}_{\gamma}(\bv{x}|\bv{\theta})
    = \frac{\hat{p}_{\gamma}(\bv{x}|\bv{\theta})}{\sum_{\bv{y}} \hat{p}_{\gamma}(\bv{y}|\bv{\theta})}
    = \frac{p_{\gamma}(\bv{x}|\bv{\theta})^{1-\gamma}}{\sum_{\bv{y}}p_{\gamma}(\bv{y}|\bv{\theta})^{1-\gamma}},
\end{equation}
because $\hat p_\gamma=e^{-\gamma\psi_\gamma}p_\gamma^{1-\gamma}$ and the common factor cancels on normalization.
This representation also gives the exact Jacobian directly. Since
$\partial_i\log\hat p_\gamma=(1-\gamma)Z_{\theta i}-\partial_i\psi_\gamma$, differentiating the normalizing sum gives
\begin{equation}
    \partial_i\log\tilde p_\gamma(\bv{x})
    =(1-\gamma)\left\{Z_{\theta i}(\bv{x})
    -\mathbb E_{\tilde p_\gamma}[Z_{\theta i}]\right\}.
\end{equation}
Differentiating the escort expectation in~\eqref{eq:eta_escort_supp} therefore yields
\begin{equation}
    \partial_i\eta_j
    =(1-\gamma)\,\mathrm{Cov}_{\tilde p_\gamma}\!\left(F_j,Z_{\theta i}\right).
    \label{eq:escort_jacobian_exact}
\end{equation}
At $\gamma=0$, $Z_{\theta i}=F_i$ and this reduces to the symmetric Fisher covariance. At finite deformation, the unequal arguments $F_j$ and $Z_{\theta i}$ make the Jacobian generally asymmetric.

\newpage

\section{R\'enyi projection and escort-moment matching}
\label{supp_sec:projection}

We consider a nested setting with two parameterizations of different dimensionality. Let $\bv{\theta}^{\mathrm{full}} = (\theta_{\mathrm{full}}^1,\ldots,\theta_{\mathrm{full}}^{L})$ denote a larger parameter vector defining a deformed exponential distribution $p_\gamma(\bv{x}\mid \bv{\theta}^{\mathrm{full}})$. We introduce a submanifold of this family by constraining a subset of parameters to zero. Denoting the reduced parameter vector by $\bv{\theta} = (\theta^1,\ldots,\theta^{M}, 0, \ldots, 0)$ with $M < L$, the corresponding distribution $p_\gamma(\bv{x}\mid \bv{\theta})$ lies on a lower-dimensional submanifold $\mathcal{E}_\gamma \subset \mathcal{E}_\gamma^{\mathrm{full}}$. Let $\Theta_\gamma^{(M)}$ be the admissible set of these reduced parameter vectors, for which $1+\gamma\theta^iF_i(\bv{x})>0$ for every configuration.
Given a target distribution parameterized by $\bv{\theta}^{\mathrm{full}}$, the R\'enyi divergence between the two nested distributions can be written in the generalized Bregman form
\begin{equation}\label{eq:Bregman_Renyidiv_nested}
    \mathcal{D}_{\gamma}[\bv{\theta}^{\mathrm{full}};\bv{\theta}]
    = \psi_{\gamma}(\bv{\theta}) + \phi_{\gamma}(\bv{\eta}^{\mathrm{full}})
    - \frac{1}{\gamma} \ln \bigl(1 + \gamma \theta^{i}\eta^{\mathrm{full}}_{i}\bigr),
\end{equation}
where $\bv{\eta}^{\mathrm{full}}$ are the escort expectations generated by $\bv{\theta}^{\mathrm{full}}$ and are fixed while the reduced parameter varies. At $\gamma=0$, equation~\eqref{eq:Bregman_Renyidiv_nested} is understood by continuity and reduces to the ordinary Bregman representation of the Kullback--Leibler divergence. For finite $\gamma$, the logarithm is well defined throughout $\Theta_\gamma^{(M)}$ because
\begin{equation}
    1+\gamma\theta^i\eta_i^{\mathrm{full}}
    = \mathbb{E}_{\tilde p_\gamma}\!\left[1+\gamma\theta^iF_i\right] > 0,
\end{equation}
where $\tilde p_\gamma$ is the target's escort distribution and the strict inequality follows from admissibility of $\bv{\theta}$. Since $\theta^i = 0$ for $i > M$, the sum $\theta^i\eta_i^{\mathrm{full}}$ involves only $i = 1,\ldots,M$. Minimizing this divergence over the free parameters $\theta^i$ yields the R\'enyi projection of the full distribution onto the submanifold:
\begin{equation}
    \bv{\theta}^\star
    \in
    \arg\min_{\bv{\theta}\in \Theta_\gamma^{(M)}}
    \mathcal{D}_\gamma[\bv{\theta}^{\mathrm{full}};\bv{\theta}].
\end{equation}

Assuming a minimizer is in the interior of $\Theta_\gamma^{(M)}$, we show below that its stationarity conditions reduce to $\eta_i(\bv{\theta}^\star) = \eta_i^{\mathrm{full}}$ for $i = 1,\ldots,M$, i.e.\ the R\'enyi projection matches the escort expectations in the free coordinates of the reduced model.

We start from the generalized Bregman representation~\eqref{eq:Bregman_Renyidiv_nested}; note that $\bv{\eta}^{\mathrm{full}}$ does not depend on $\bv{\theta}$. Differentiating with respect to $\theta^i$ gives
\begin{equation}
    \partial_i \mathcal{D}_{\gamma}
    =
    \partial_i \psi_\gamma(\bv{\theta})
    -
    \frac{\eta_i^{\mathrm{full}}}{1+\gamma \theta^{k}\eta_k^{\mathrm{full}}}.
\end{equation}
Setting this to zero at an interior minimizer $\bv{\theta}^\star$ yields
\begin{equation}\label{eq:stat_partialpsi}
    \partial_i \psi_\gamma(\bv{\theta}^\star)
    =
    \frac{\eta_i^{\mathrm{full}}}{1+\gamma \theta^{\star k}\eta_k^{\mathrm{full}}},
    \qquad i=1,\ldots,M.
\end{equation}

Multiplying~\eqref{eq:stat_partialpsi} by $\theta^{\star i}$, summing over $i$, and rearranging gives the direct chain
\begin{align}
    \theta^{\star i}\partial_i \psi_\gamma(\bv{\theta}^\star)
    &= \frac{\theta^{\star i}\eta_i^{\mathrm{full}}}
    {1+\gamma \theta^{\star k}\eta_k^{\mathrm{full}}}, \nonumber\\
    1-\gamma\theta^{\star i}\partial_i \psi_\gamma(\bv{\theta}^\star)
    &= \frac{1}{1+\gamma\theta^{\star k}\eta_k^{\mathrm{full}}}, \nonumber\\
    1+\gamma\theta^{\star i}\eta_i^{\mathrm{full}}
    &= \frac{1}{1-\gamma\theta^{\star k}\partial_k \psi_\gamma(\bv{\theta}^\star)}.
    \label{eq:1plusA_identity}
\end{align}

The generalized dual-coordinate relation~\eqref{eq:dual_coordinates} gives
\begin{equation}\label{eq:dual_eta_general}
    \eta_i(\bv{\theta})
    =
    \frac{1}{1-\gamma \theta^{k}\partial_k \psi_\gamma(\bv{\theta})}
    \,\partial_i \psi_\gamma(\bv{\theta}).
\end{equation}
Evaluating at $\bv{\theta}^\star$ and substituting~\eqref{eq:stat_partialpsi} and~\eqref{eq:1plusA_identity}:
\begin{equation}
\eta_i(\bv{\theta}^\star) = \Bigl(1+\gamma \theta^{\star k}\eta_k^{\mathrm{full}}\Bigr)\,
    \frac{\eta_i^{\mathrm{full}}}{1+\gamma \theta^{\star k}\eta_k^{\mathrm{full}}} = \eta_i^{\mathrm{full}}.
\end{equation}

Thus every interior minimizer satisfies
\begin{equation}
    \eta_i(\bv{\theta}^\star)
    =
    \eta_i^{\mathrm{full}},
    \qquad i = 1,\ldots,M,
\end{equation}
so any interior R\'enyi projection onto the submanifold matches the escort expectations in the free coordinates. This stationarity argument does not by itself establish existence, uniqueness, or an interior solution.

\newpage

\section{Mean-field theory and rate calibration of curved networks}
\label{supp_sec:meanfield_phase_calibration}

We derive the mean-field equation in the thermodynamic limit for a homogeneous curved neural network of binary units $x_i\in\{0,1\}$, with uniform field $H$ and a common coupling $J/N$. Unlike the small-deformation expansion used in the main text, this saddle-point treatment retains the full dependence on the rescaled deformation $\gamma'=\beta N\gamma$.

\subsection*{Homogeneous model}

With energy~\eqref{eq:energy} specialized to the interaction parameters above
\begin{equation}\label{eq:supp_homog_energy}
    E(\bv{x}) = -H\sum_{i} x_i - \frac{J}{N}\sum_{i<j} x_i x_j,
    \qquad
    p_\gamma(\bv{x} | \bv{\theta}) \propto \big[ 1-\gamma\beta E(\bv{x}) \big]^{1/\gamma},
    \qquad 1-\gamma\beta E(\bv{x})>0\quad \forall\bv{x},
\end{equation}
the distribution depends on $\bv{x}$ only through the population count $S=\sum_i x_i$. For a configuration with activity $m=S/N$, the energy per neuron in the thermodynamic limit is
\begin{equation}\label{eq:supp_homog_eofm}
    \frac{E}{N} \xrightarrow[N\to\infty]{} e(m) = -\Big(Hm + \tfrac12 Jm^2\Big),
\end{equation}
using $\sum_{i<j}x_i x_j = \tfrac12 S(S-1)\simeq \tfrac12 N^2 m^2$.

\subsection*{Intensive deformation and saddle point}

Following~\cite{aguilera2025explosive} we keep the deformation intensive by setting $\gamma=\gamma'/(N\beta)$, so that the system-size factors cancel,
\begin{equation}\label{eq:supp_homog_cancel}
    \gamma\beta E = \frac{\gamma'}{N\beta}\beta\big(N e(m)\big) = \gamma' e(m),
    \qquad
    \frac{1}{\gamma} = \frac{N\beta}{\gamma'}.
\end{equation}
A configuration of activity $m$ then carries weight $[1-\gamma' e(m)]^{N\beta/\gamma'}$, and there are $\binom{N}{Nm}\simeq e^{N H_2(m)}$ such configurations, with the binary entropy $H_2(m)=-m\ln m-(1-m)\ln(1-m)$. By Laplace's method the normalization is dominated by the maximum of the free-energy density
\begin{equation}\label{eq:supp_homog_freeenergy}
    f(m) = \frac{\beta}{\gamma'}\ln\big(1-\gamma' e(m)\big) + H_2(m).
\end{equation}

\subsection*{Mean-field equation}

Write the effective inverse temperature as
\begin{equation}\label{eq:supp_inv_temp_mf}
    \beta'=\frac{\beta}{D(m)}, \qquad D(m)\coloneqq 1-\gamma'e(m)
    =1+\gamma'\left(Hm+\tfrac12 Jm^2\right).
\end{equation} 
Using $H_2'(m)=-\mathrm{logit}(m)$ and $-e'(m)=H+Jm$, the stationarity condition becomes
\begin{equation}\label{eq:supp_homog_mf}
    \mathrm{logit}(m) = \beta'(m)(H+Jm)
    \; \Longleftrightarrow\; m=\sigma\big(\beta'(m)(H+Jm)\big)
\end{equation}
where $\sigma(z)=(1+e^{-z})^{-1}$. The deformation acts as a state-dependent rescaling of the inverse temperature, $\beta\to\beta'$, by the same mechanism as in the $\{-1,+1\}$ analysis of~\cite{aguilera2025explosive}. The exact encoding map also generates an additive energy constant, which must be retained or absorbed into the inverse temperature as described below.

\subsection*{Activation cost, effective temperature and the majority-active branch}

Adding one active neuron to a state of activity $m$ changes the energy by
$-(H+Jm)$~\eqref{eq:supp_homog_eofm}: the field cost $-H$ minus the coupling
gain $Jm$. For $\gamma'<0$, $\beta'=\beta/D(m)$ therefore falls with activity
where $H+Jm<0$ and rises where $H+Jm>0$. The same quantity constrains the fixed
points. Because $\beta'>0$, the two sides of~\eqref{eq:supp_homog_mf} have the
same sign, so a majority-active fixed point ($m>\tfrac12$) requires $H+Jm>0$.
Since $H+Jm\le H+J$, this requires $H>-J$; the deformation rescales $\beta$ but
cannot change this sign.

For $H<0<J$ two regimes follow. If the fields are strongly inhibitory,
$|H|\ge J$, every additional spike raises the energy, negative deformation
lowers $\beta'$ at every activity level, and every fixed point has
$m<\tfrac12$; for example, $H=-2$ and $J=1$ give $H+Jm\le-1$ for all $m$.
A strongly state-dependent $\beta'$ can still produce several fixed points
with $m<\tfrac12$. If the couplings are strong, $|H|<J$, additional spikes
beyond $m=|H|/J$ lower the energy, negative deformation raises $\beta'$ as
activity grows, and a majority-active branch can coexist with the silent one.
This is the regime of accelerated retrieval and explosive
transitions~\cite{aguilera2025explosive}.

\begin{figure}[p]
    \centering
    \includegraphics[width=0.95\linewidth]{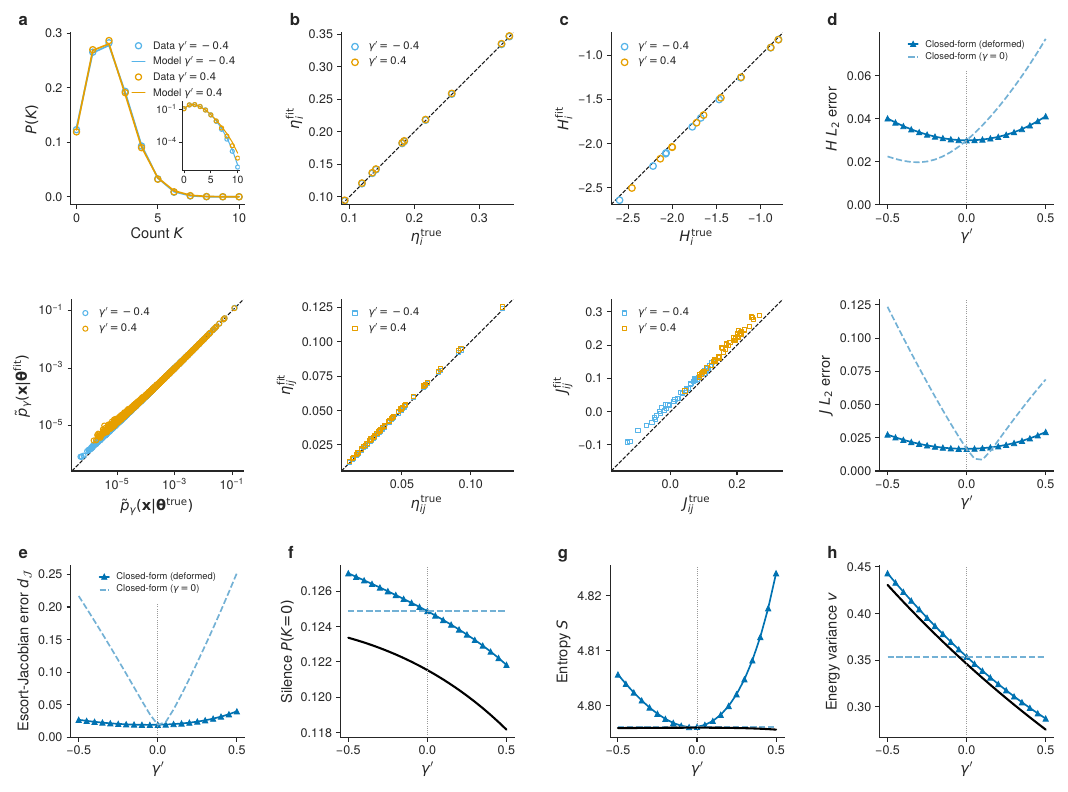}
    \caption{\textbf{Closed-form mean-field inverse compared with the exact solution ($N=10$).} Recovery of a synthetic heterogeneous $N=10$ model in the weak-correlation regime by exact enumeration. All displayed probability distributions are escort distributions; the fields, couplings, and $\gamma'=N\beta\gamma$ retain their original coordinates. The complete escort-coordinate vector $\bv{\eta}^{*}=(\eta_i,\eta_{ij})$ is defined by the seeded reference model at $\gamma'=0$ and held fixed at every deformation: for each $\gamma'$ all 55 true natural parameters $(H_i,J_{ij})$ are re-solved so that the exact escort moments equal the same $\bv{\eta}^{*}$. Scatter panels (a--c) use rescaled deformations $\gamma'\in\{-0.4,0.4\}$ and sweeps (d--h) cover a 21-point grid of $\gamma'\in[-0.5,0.5]$ in steps of $0.05$. Throughout (a--c) color encodes the deformation ($\gamma'=-0.4$ sky-blue, $\gamma'=+0.4$ orange); throughout (d--h) line style encodes the family (deformed solid with markers, $\gamma=0$ lighter dashed). The true reference, where shown, is black. \textbf{(a)}~Population spike-count distribution $P(K)$, data (open circles) vs.\ fitted model (line), linear scale (inset: log); lower, log--log scatter of the fitted distribution $\tilde p_\gamma(\bv{x}|\bv{\theta}^{\mathrm{fit}})$ against the true $\tilde p_\gamma(\bv{x}|\bv{\theta}^{\mathrm{true}})$ over all $2^N$ configurations. \textbf{(b)}~Recovered dual coordinates $\eta_i$ (top) and $\eta_{ij}$ (bottom) vs.\ true. \textbf{(c)}~Recovered fields $H_i$ (top) and couplings $J_{ij}$ (bottom) vs.\ true; dashed lines, perfect recovery. \textbf{(d)}~Parameter-recovery error in the \emph{flat parameter geometry}: $L_2$ (root-mean-square) deviation of recovered from true parameters, for fields $H_i$ (top) and couplings $J_{ij}$ (bottom) vs.\ $\gamma'$, for the closed-form inverse~\eqref{eq:Jij_inverse_main} applied to the same exact escort moments, at the true $\gamma$ (deformed) or at $\gamma=0$. \textbf{(e)}~The same parameter error weighted by the symmetric part of the local escort-coordinate Jacobian, $d_{\mathcal J}=\sqrt{\Delta\bv{\theta}^{\!\top}G_{\rm sym}\Delta\bv{\theta}}$, where $\Delta\bv{\theta}=\bv{\theta}^{\mathrm{fit}}-\bv{\theta}^{\mathrm{true}}$, $\mathcal{J}=\partial\bv{\eta}/\partial\bv{\theta}$, and $G_{\rm sym}=(\mathcal{J}+\mathcal{J}^{\!\top})/2$. At $\gamma=0$, $G_{\rm sym}$ is the Fisher matrix and $d_{\mathcal J}=\sqrt{2\,\mathrm{KL}}$ to leading order. At finite $\gamma$, $G_{\rm sym}$ need not be positive-semidefinite; the displayed nonnegative quadratic values are therefore escort-coordinate sensitivity diagnostics, not a norm or Fisher distance. \textbf{(f)}~Silence probability $P(K\!=\!0)=\tilde p_\gamma(\bv{0})$; \textbf{(g)}~Shannon entropy $S=-\sum_{\bv{x}}\tilde p_\gamma(\bv{x})\ln\tilde p_\gamma(\bv{x})$; \textbf{(h)}~Energy variance per neuron $v=\mathrm{Var}_{\tilde p_\gamma}(E)/N$ at $\beta=1$, where $E$ is defined by each model's original fields and couplings~\eqref{eq:energy} (\hyperref[sec:thermo_observables]{Methods}). Panels (f)--(h) evaluate each escort model at the deformation used in its inversion.}
    \label{fig:sim_invising_weakcorr_exact}
\end{figure}

\subsection*{Heterogeneous (SK-type) networks and the calibration of Fig.~\ref{fig:sim_invising_weakcorr_exact}}

The synthetic network of Fig.~\ref{fig:sim_invising_weakcorr_exact} replaces the common $(H,J)$ by quenched random fields $H_i$ and couplings $J_{ij}$, giving a Sherrington--Kirkpatrick-type disordered version of the curved network. With $J_{ii}=0$, the naive mean-field equation becomes
\begin{equation}\label{eq:supp_sk_mf}
    m_i = \sigma\Big(\beta'\big(H_i + \sum_{j} J_{ij} m_j\big)\Big),
    \qquad
    \beta' = \frac{\beta}{1-\gamma' e(\bv{m})},
\end{equation}
where $e(\bv{m})$ is the mean-field energy per neuron. This is the $\{0,1\}$ image of the spin self-consistency $m_i^{s}=\tanh\big(\beta'(h_i+\sum_j J^{s}_{ij}m_j^{s})\big)$, whose homogeneous and replica forms were analyzed in~\cite{aguilera2025explosive}. Substituting $s_i=2x_i-1$, $m_i^{s}=2m_i-1$, and $\tanh u = 2\sigma(2u)-1$ converts the local field term by term, with $J_{ij}=4J^{s}_{ij}$ and $H_i=2h_i-\tfrac12\sum_j J_{ij}$. Writing the energy in spin variables gives $E(\bv{x})=E_s(\bv{s})+c$, with $c$ defined in~\eqref{eq:supp_encoding_constant}. Because a deformed family is not invariant under energy shifts, an exact correspondence must retain $c$ inside $1-\gamma\beta(E_s+c)$ or, when $1-\gamma\beta c>0$, absorb it through $\beta_{\rm eff}=\beta/(1-\gamma\beta c)$ as in~\eqref{eq:supp_shift_beta_eff}. Thus the encoding change preserves $\gamma$ but can rescale $\beta$ and hence the magnitude of $\gamma'=N\beta\gamma$.

\paragraph*{Rate calibration.} In the undeformed reference model of Fig.~\ref{fig:sim_invising_weakcorr_exact}, $\bar H$ and $\bar J$ denote the disorder means of the fields and off-diagonal couplings. For the scaling $\bar J=J/N$, the expected total coupling to a neuron is
\begin{equation}\label{eq:supp_JN_step}
    J_N=(N-1)\bar J=J\Big(1-\frac1N\Big).
\end{equation}
As a first approximation, replace $H_i$ by $\bar H$, $m_i$ by $m$, and $\sum_{j\ne i}J_{ij}$ by $J_N$. Then~\eqref{eq:supp_sk_mf} reduces to the homogeneous form~\eqref{eq:supp_homog_mf} with total coupling $J_N$.
At $\gamma'=0$, the field required for a target activity $m$ is
\begin{equation}\label{eq:supp_sk_calib}
    \mathrm{logit}(m)=\beta (\bar H + J_N m)
    \Longrightarrow
    \bar H = \frac{\mathrm{logit}(m)}{\beta} - J_N m.
\end{equation}
For $\beta=1$, $J=1$, and $m=0.2$, this gives $\bar H=-1.566$ at $N=10$ and $\bar H\to-1.586$ as $N\to\infty$. These estimates neglect the variation of the local field across neurons and the fluctuations of their binary activities.
We include both contributions in a Gaussian approximation to estimate the field center for the undeformed reference model in Fig.~\ref{fig:sim_invising_weakcorr_exact}.

\paragraph*{Disorder-smeared calibration.}
In the undeformed reference of Fig.~\ref{fig:sim_invising_weakcorr_exact}, quenched disorder produces a spread of local fields across neurons, while fluctuations of the binary activities broaden their distribution further. The nonlinear response $\sigma$ makes the mean activity sensitive to this spread, which is neglected in the homogeneous estimate~\eqref{eq:supp_sk_calib}. In the Gaussian independent-site approximation with common mean activity $m$, the local field has mean $\bar H+J_Nm$, with~\eqref{eq:supp_JN_step} 
for $J_{ii}=0$, and variance
\begin{equation}\label{eq:supp_sk_Delta}
    \Delta \simeq \sigma_H^2 + m^2\sum_{j \neq i} \mathrm{Var}(J_{ij}) + \sum_{j \neq i} \overline{J_{ij}^2} m(1-m),
    \qquad \sigma_H^2=\mathrm{Var}(H_i),
\end{equation}
where the overbar denotes a disorder average. The first two terms describe the variation of the fields and coupling sums across neurons, and the third arises from fluctuations of binary activity. 
Averaging the nonlinear response over the local-field distribution gives the mean activity at $\gamma'=0$,
\begin{equation}\label{eq:supp_sk_calib_noisy}
    m = \int \mathcal{D}z\sigma\Big(\beta'\big(\bar H + J_N m + \sqrt{\Delta} z\big)\Big),
    \qquad \mathcal{D}z = \frac{e^{-z^2/2}}{\sqrt{2\pi}} dz,
\end{equation}
with $\beta'$ in~\eqref{eq:supp_inv_temp_mf}. At fixed $m$ and $\Delta$, the right-hand side increases strictly with $\bar H$, so each prescribed activity determines a unique field center within this approximation.
At finite $N$, the site-wise mean-field rates depend on the quenched fields and coupling sums,
\begin{equation}
    m_i \simeq \sigma\Big(\beta'\Big[H_i+m\sum_jJ_{ij}\Big]\Big),\qquad H_i\sim\mathcal{N}(\bar H,\sigma_H^2).
\end{equation}
For $\bar J=J/N$ and $\mathrm{Var}(J_{ij})=\pazocal{O}(N^{-2})$, both coupling contributions in~\eqref{eq:supp_sk_Delta} vanish as $1/N$, and the coupling sums approach $J$. Thus $\Delta\to\sigma_H^2$ as $N\to\infty$, while the quenched fields sustain a spread of firing rates.
For the undeformed calibration with $\beta'=\beta=1$, $J=1$, $m=0.2$, and $\sigma_H^2=0.25$, the variance estimate $\Delta \simeq 0.25+0.019$ and total coupling $J_N=0.9$ at $N=10$ give $\bar H\simeq-1.644$ by Gaussian quadrature, compared with $-1.566$ in the homogeneous approximation. In the thermodynamic limit, $J_N\to1$ and $\Delta \to 0.25$ give $\bar H\to-1.658$, compared with the homogeneous value $-1.586$. More negative fields compensate for the increase in mean activity caused by local-field fluctuations.
The dependence on deformation in Fig.~\ref{fig:sim_invising_weakcorr_exact} is studied at fixed first- and second-order escort moments $\eta^*=(\eta_i^*,\eta_{ij}^*)$, set by the undeformed reference. The fields and couplings required to preserve these moments are determined from the exact distribution over all $2^N$ configurations. The inferred fields and couplings are compared with these values at each $\gamma'$.

\clearpage

\section{Pairwise projection and the efficient deformation score}
\label{supp_sec:score_geometry}

We derive the deformation score after pairwise refitting, its relation to energy fluctuations, and its order decomposition.
Unless stated otherwise, expectations and scores in this note
are defined under the original distribution $p_\gamma$.
 Let $p_0=p_{\gamma=0}$ be the fitted pairwise model, that is, the flat model at the fitted parameters, with natural parameters $\bv{\theta}=(\theta^i)$, sufficient statistics $F=(F_i)=(x_a,x_ax_b)$, and expectations $\langle\cdot{\rangle_0}$.

\subsection*{Projection onto the pairwise tangent space}

At the fitted pairwise reference $p_0$, an infinitesimal change of the natural parameters gives $\delta\ln p_0(\bv{x})=\delta\theta^i g_i(\bv{x})$ where $g_i(\bv{x}) \coloneqq \partial_{\theta^i}\ln p_0(\bv{x})= F_i(\bv{x})-\langle F_i\rangle_0$. Thus the score functions $g_i$ span the pairwise tangent space $\mathcal{T}_{\rm pair}=\mathrm{span}\{g_i\}$, whose vectors are generally not orthogonal. With the inner product $(f,h)_{0} \equiv \langle fh\rangle_0$, the mutual overlap of the $g_i$ vectors is captured by the Gram matrix $G_{ij}=(g_i,g_j)_{0}=\mathrm{Cov}_0(\bv{F}_i,\bv{F}_j)$, the Fisher metric of the pairwise family at $p_0$. To project a centered statistic $\bar f=f-\langle f\rangle_0$, onto their span, write its projected component as $P_{\rm pair} \bar{f} = a^i g_i$, to be defined below. Orthogonal projection requires the residual $\bar f - P_{\rm pair}\bar{f}$ to be orthogonal to every $g_j$,
\begin{equation}
    \bigl( g_j, \bar{f} - a^i g_i \bigr)_{0} = 0 .
\end{equation}
Using $(g_i , g_j)_{0} = G_{ij}$ gives,
\begin{equation}\label{eq:supp_proj_coefficients}
    G_{ji} a^i = (g_j , \bar f)_{0} = \mathrm{Cov}_0 (\bv{F}_j, f) .
\end{equation}
Thus, when the pairwise scores are linearly independent, $a^i = (G^{-1})^{ij}\mathrm{Cov}_0 (\bv{F}_j, f)$. As discussed in the Methods, $G$ is difficult to invert so more generally, we may take the Moore-Penrose pseudoinverse $G^+$.
This operation gives the same orthogonal projection onto the span of independent pairwise scores,
\begin{equation}\label{eq:supp_pairwise_projector}
    P_{\rm pair} f = \mathrm{Cov}_0(f,\bv{F})^{\!\top} G^+ \bigl(\bv{F}-\langle\bv{F}{\rangle_0}\bigr),
\end{equation}
as used in the Methods. From~\eqref{eq:supp_proj_coefficients}, it directly follows that
\begin{equation}
    \mathrm{Cov}_0\bigl(\bv{F},(1-P_{\rm pair})\bar{f}\bigr) = 0.
\end{equation}
Moreover, the pseudoinverse is defined to satisfy $G^+GG^+=G^+$, which implies $P_{\rm pair}^2=P_{\rm pair}$. Thus $P_{\rm pair}$ is the orthogonal projector onto $\mathcal{T}_{\mathrm{pair}}$ in $L^2(p_0)$, and $(1-P_{\mathrm{pair}})\bar f$ is the component that cannot be produced by an infinitesimal pairwise refit. In the empirical calculation, the
eigenmode cutoff specified in Methods restricts this projection to the well-resolved pairwise modes.

\subsection*{Deformation score and specific heat}
\label{supp_subsec:score_heat}

The score associated with the deformation parameter is
$s_\gamma(\bv{x})=\partial_\gamma\ln p_\gamma(\bv{x}|\bv{\theta})$. Define 
$u(\bv{x})=\theta^iF_i(\bv{x})$, which in the field/coupling parametrization becomes $u=\beta\big(\sum_iH_ix_i+\sum_{i<j}J_{ij}x_ix_j\big)=-\beta E$ with energy
$E$ given by~\eqref{eq:energy}. Differentiating $\ln p_\gamma=\gamma^{-1}\ln(1+\gamma u)-\psi_\gamma$ and expanding about $\gamma=0$,
\begin{equation}\label{eq:supp_score}
    s_\gamma = -\frac{\ln(1+\gamma u)}{\gamma^2}
    + \frac{u}{\gamma(1+\gamma u)} - \partial_\gamma\psi_\gamma
    \xrightarrow[\gamma\to0]{}
    -\tfrac12\big(u^2-\langle u^2\rangle_0\big).
\end{equation}
Here $\partial_\gamma\psi_\gamma|_0=-\tfrac12\langle u^2\rangle_0$, 
so that $\langle s_0\rangle_0=0$. More generally, $\langle s_\gamma\rangle_{p_\gamma}=0$ as long as the support of $p_\gamma$ does not change with $\gamma$.
For an empirical fit, let the undeformed reference be the pairwise model $p_0$.
The ordinary data moments are matched to moments of the escort distribution $\tilde p_\gamma$, so the corresponding escort score at $\gamma=0$ is
\begin{equation}
    \tilde s_0
    \equiv \left.\partial_\gamma\log\tilde p_\gamma\right|_{\gamma=0}
    =s_0-\bigl(u-\langle u\rangle_0\bigr).
\end{equation}
Since $u=\theta^iF_i$, we have $u-\langle u\rangle_0=\theta^i g_i\in\mathcal{T}_{\mathrm{pair}}$. This term is therefore removed by the projection in~\eqref{eq:supp_pairwise_projector}, giving $\tilde{s}_0^\perp=s_0^\perp$. Thus the fixed-parameter deformation path and the escort-moment path have the same efficient score at $\gamma=0$.
At fixed natural parameters, the deformation score at $\gamma=0$ is $-\tfrac12\big(u^2-\langle u^2\rangle_0\big)$. If the ordinary means of the sufficient statistics $F_i$ are matched, then $\langle u\rangle=\theta^i\langle F_i\rangle$ is matched as well. Since $\mathrm{Var}(u)=\langle u^2\rangle-\langle u\rangle^2$, differences in $\langle u^2\rangle$ then coincide with differences in $\mathrm{Var}(u)$. At $\gamma=0$, this variance determines the specific heat,
\begin{equation}
    c=\frac{1}{N}\mathrm{Var}(u)
     =\frac{\beta^2}{N}\mathrm{Var}(E).
\end{equation}
At finite deformation, the power-tempered specific heat instead
uses the effective escort energy $\tilde E_\gamma$.
The heterogeneous benchmarks report the energy variance
$v=\mathrm{Var}_{\tilde p_\gamma}(E)/N$ at $\beta=1$,
with each model's own fields and couplings
(\hyperref[sec:thermo_observables]{Methods}).

Before refitting the pairwise parameters,
\begin{equation}
    I_{\gamma\gamma}=\mathrm{Var}_0(s_0)=\tfrac14\mathrm{Var}_0(u^2).
\end{equation}
Refitting removes the component of $s_0$ contained in the pairwise score
subspace, leaving the efficient score $s_0^\perp$.
In the synthetic tests, 
only $0.4$--$2.3\%$ of the raw escort deformation information remains after removing the component absorbed by pairwise refitting, with the larger fractions in the strong-coupling populations (Fig.~\ref{fig:efficient_score_recovery}). Energy variance provides a complementary diagnostic of the fitted
models, whereas deformation selection uses the projected score.
In Fig.~\ref{fig:sim_invising_weakcorr_exact}(h) the energy variance separates the deformed and $\gamma=0$ fits. 
Fig.~\ref{fig:sim_invising_weakcorr_exact}(e) plots the parameter error weighted by the symmetric part of the escort-coordinate Jacobian,  Fig.~\ref{fig:sim_invising_weakcorr_exact}(f) the silence probability,
and Fig.~\ref{fig:sim_invising_weakcorr_exact}(g) the entropy $S$.
The observables in panels (f)--(h) are evaluated under the
escort distribution.

At $\gamma=0$, the response of a fixed observable $A$ at fixed
pairwise parameters is
$\left.\partial_\gamma\langle A\rangle\right|_0
=\mathrm{Cov}_0(A,s_0)$.
With the full pairwise projection, an infinitesimal refit that
preserves ordinary moments gives the response
$\mathrm{Cov}_0(A,s_0^\perp)$.
Under an exact escort-moment refit, it is instead
$\mathrm{Cov}_0(A,s_0^\perp+u-\langle u\rangle_0)$.
At finite $\gamma$, let
$r_i=\partial_{\theta^i}\log p_\gamma=F_i/(1+\gamma\theta^kF_k)-\langle F_i/(1+\gamma\theta^kF_k)\rangle_{p_\gamma}$ (on fixed positive support). The efficient deformation score tangent is
\begin{equation}\label{eq:supp_finite_effscore}
    s_\gamma^\perp
    =s_\gamma
    -\mathrm{Cov}_{p_\gamma}(s_\gamma,r)^{\!\top}
     \mathcal I(\gamma)^{+}r,
\end{equation}
where $\mathcal I(\gamma)=\mathrm{Cov}_{p_\gamma}(r,r)$.
If the pairwise parameter scores are linearly dependent, $\mathcal{I}(\gamma)^+$ gives the minimum norm projection. For a fixed observable $A$, the local Fisher-orthogonal path
under $p_\gamma$ satisfies
\begin{equation}
\frac{d}{d\gamma}\langle A\rangle_{p_\gamma}
=\mathrm{Cov}_{p_\gamma}(A,s_\gamma^\perp).
\end{equation}
In the empirical analysis, at each candidate $\gamma$, the model is refit by matching ordinary data moments to escort moments, while the projection defining the statistic is kept fixed at the fitted pairwise model (\hyperref[sec:efficient_score]{Methods}).
Therefore, for $\gamma\neq0$, the profiled empirical statistic is not the finite-$\gamma$ efficient score in equation~\eqref{eq:supp_finite_effscore}.
Deformation selection uses this fixed projection evaluated over
the escort-fitted candidates.
The third-cumulant residual $\Delta\kappa_3[K]$ defined in the
\hyperref[sec:regional_selection]{recording-data Methods} probes higher-order
structure outside the fitted first- and second-order moments independently of the
deformed fit. In the present analysis this residual validates the selected
curvature but does not itself define $\gamma'^\star_{\mathrm{mm}}$.

\subsection*{Order decomposition of the local deformation score}
\label{supp_subsec:score_order_decomposition}

The same projection geometry clarifies what can, and cannot, be meant by assigning
the deformation score to third- or fourth-order structure. At the fitted pairwise
null $p_0$, write
\begin{equation}\label{eq:supp_q_decomp}
    q(\bv{x})=-\tfrac12 u(\bv{x})^2,
    \qquad
    u(\bv{x})=\widehat{\bv{\theta}}\!\cdot\!F(\bv{x}).
\end{equation}
Because $F$ contains one- and two-body binary features, $u^2$ contains monomials up
to fourth order after the reduction $x_i^2=x_i$. Grouping distinct monomials by
their order gives
\begin{equation}
    q=q_1+q_2+q_3+q_4
      \equiv q_{\le2}+q_3+q_4,
\end{equation}
where $q_r$ collects all distinct $r$-body monomials and $q_{\leq 2}=q_1+q_2$. Because $u$ contains no constant feature, $q$ has no zeroth-order term; the constant term in the likelihood score comes only from centering $q$. The $q_r$ group the deformation score by order, with coefficients fixed by the original fields and couplings.
The projected components $q_3^\perp$ and $q_4^\perp$~\eqref{eq:score_order_components}, with $s_0^\perp=q_3^\perp+q_4^\perp$, are defined in the \hyperref[sec:efficient_score]{Methods}.
For each subsample, the covariances entering $P_{\mathrm{pair}}$ were estimated from $2\times10^5$ Gibbs samples of the fitted pairwise model (eight chains, $2.5\times10^4$ samples per chain, after $8000$ burn-in sweeps). In the empirical calculation, $G^+$ retains covariance eigenmodes larger than $10^{-6}$ times the
largest eigenvalue.
The truncation at fourth order concerns only the derivative at $\gamma=0$. At finite deformation,
expansion of $\gamma^{-1}\log(1+\gamma u)$ about $\gamma=0$ contains all powers $u^n$ and can therefore generate a constrained interaction hierarchy of every order up to $N$.

The norm fraction $f_3^{\mathrm{norm}}=I_3/(I_3+I_4)$ with $I_j=\mathrm{Var}_0(q_j^\perp)$, the correlation $\rho_{34}$ and the decomposition $I^\perp=I_3+I_4+2\mathrm{Cov}_0(q_3^\perp,q_4^\perp)$ are defined in the \hyperref[sec:efficient_score]{Methods} and shown in Supplementary Fig.~\ref{fig:supp_effscore_order_fraction}.
Because the covariance contributes to $I^\perp$, a covariance-based allocation such as 
$[I_3+\mathrm{Cov}_0(q_3^\perp,q_4^\perp)]/I^\perp$
can fall outside $[0,1]$ when the two components are strongly anti-correlated. The quantity
$f_3^{\mathrm{norm}}$ therefore compares $I_3$ and $I_4$ directly; $f_3^{\mathrm{norm}}>1/2$ corresponds to $I_3>I_4$, while the value of $I^\perp$ also depends on their covariance.

\clearpage

\section{First-order deformation expansion of the distribution and escort coordinates}
\label{supp_sec:small_deformation_expansion}

\subsection*{Expansion of the deformed distribution}
\label{supp_subsec:expansion_deformed_pmf}

To recover the undeformed reference distribution from the curved model, whose deformation breaks the dual-flat structure of ordinary exponential families, we expand the deformed exponential distribution $p_{\gamma}(\bv{x}|\bv{\theta})$ for small $\gamma$. We then use this expansion to approximate the escort distribution and the dual coordinates needed for minimizing the R\'enyi divergence. The validity of the approximation is confirmed numerically in later sections. To linear order, $p_{\gamma}(\bv{x}|\bv{\theta})$ is
\begin{align}\label{eq:Pablo_expansion}
    p_{\gamma} (\bv{x} | \bv{\theta}) &\simeq p_0(\bv{x}|\bv{\theta})\left(1 + \frac{1}{2}\left[\mathbb{E}_{p_0}[(\theta^{i} F_{i})^2]-(\theta^{i} F_{i})^2 \right] \gamma \right).
\end{align}

Using this approximation, we can also approximate the escort distribution $\tilde p_{\gamma}(\bv{x}|\bv{\theta})$. Writing $\tilde p_{\gamma}(\bv{x}|\bv{\theta}) \simeq c_0 + c_1 \gamma$ and using the escort definition~\eqref{eq:escort_dist}, we have
\begin{equation}\label{eq:phat_approx}
    \frac{p_{\gamma} (\bv{x} | \bv{\theta})}{1+\gamma \theta^k F_k(\bv{x})} \simeq \left( c_0 + c_1 \gamma \right) \sum_{\bv{y}} \frac{p_{\gamma} (\bv{y} | \bv{\theta})}{1+\gamma \theta^k F_k(\bv{y})},
\end{equation}
where both sides correspond to the unnormalized escort distribution $\hat{p}_\gamma$. We match orders to determine $c_0$ and $c_1$. The LHS of~\eqref{eq:phat_approx} up to the linear term with respect to $\gamma$ is
\begin{equation}
    p_{0} \left(1 + \frac{\gamma}{2}\left( \theta^k \theta^\ell \mathbb{E}_{p_0} \left[ F_k F_\ell \right] - (\theta^k F_k )^2 - 2\theta^k F_k \right) \right)
    \label{eq:pgamma_exp}
\end{equation}
and its normalization up to $\pazocal{O}(\gamma)$ reads
\begin{equation}
    \sum_{\bv{y}} \frac{p_{\gamma} (\bv{y} | \bv{\theta})}{1+\gamma \theta^k F_k(\bv{y})} \simeq 1 - \gamma \theta^k \mathbb{E}_{p_0}[F_k ].
\end{equation}
The RHS of~\eqref{eq:phat_approx} up to $\pazocal{O}(\gamma)$ is thus approximated as $c_0 + ( c_1 - c_0 \theta^{k} \mathbb{E}_{p_0} [ F_{k}]) \gamma$.
Matching both sides of~\eqref{eq:phat_approx} results in
\begin{align}
    c_0 &= p_{0}(\bv{x} | \bv{\theta}) \nonumber \\
    c_1 - c_0 \theta^k \mathbb{E}_{p_0}[F_k] &= \frac{p_{0}(\bv{x} | \bv{\theta})}{2}\big\{\theta^k \theta^\ell (\mathbb{E}_{p_0}[F_k F_\ell] - F_k F_\ell) - 2\theta^k F_k\big\}
\end{align}
allowing us to identify $c_0$ and $c_1$. The escort distribution $\tilde p_{\gamma}(\bv{x} | \bv{\theta})$ is therefore approximated as
\begin{equation}\label{eq:supp_ptil_approximation}
    \tilde p_{\gamma}(\bv{x} | \bv{\theta}) \simeq p_0(\bv{x}|\bv{\theta})\Big(1 + \frac{\gamma}{2}\big[\theta^k \theta^\ell (\mathbb{E}_{p_0}[F_k F_\ell] - F_k F_\ell) - 2\theta^k (F_k - \mathbb{E}_{p_0}[F_k])\big]\Big).
\end{equation}
The escort correction adds $-2\theta^k(F_k-\mathbb{E}_{p_0}[F_k])$ to the deformed-density correction~\eqref{eq:Pablo_expansion}; both are centered under $p_0$ and preserve first-order normalization.

\subsection*{Approximate dual coordinates}
\label{supp_subsec:approx_dual_coordinates}

Finite curvature shifts the relevant sampling measure from $p_0$ to the escort distribution, so that expectations are taken with respect to that escort~\eqref{eq:eta_escort_supp}. The deformation therefore shifts $\eta_i$ away from its baseline $\mathbb{E}_{p_0}[F_i]$, with sign and magnitude set by covariance corrections.
Although the connection to the potential variables in~\eqref{eq:dual_coordinates} is not immediately transparent, its $\gamma=0$ limit is well understood. We therefore expand $\eta_i$ perturbatively around $\gamma = 0$.

The dual variable $\eta_i$ to first order in $\gamma$ can be obtained from~\eqref{eq:dual_coordinates}. Expanding the log partition function gives
\begin{equation}
    \psi_\gamma \simeq \psi_0-\frac{\gamma}{2}\theta^k\theta^\ell\mathbb{E}_{p_0}[F_kF_\ell].
\end{equation}
Differentiation uses $\partial_i\mathbb{E}_{p_0}[g]=\mathbb{E}_{p_0}[\partial_i g]+\mathrm{Cov}_0(g,F_i)$ because both $p_0$ and $g=(\theta^kF_k)^2$ depend on $\bv\theta$. Hence
\begin{equation}\label{eq:1st_deriv}
    \partial_i \psi_\gamma \simeq \mathbb{E}_{p_0}[F_i] + \frac{\gamma}{2}\big(\theta^k \theta^\ell \mathbb{E}_{p_0}[F_k F_\ell] \mathbb{E}_{p_0}[F_i] - \theta^k \theta^\ell \mathbb{E}_{p_0}[F_k F_\ell F_i] - 2\theta^k \mathbb{E}_{p_0}[F_k F_i]\big).
\end{equation}
Direct substitution into~\eqref{eq:dual_coordinates} yields
\begin{equation}\label{eq:eta_expanded}
    \eta_i = \frac{1}{1-\gamma \theta^k \partial_k \psi_{\gamma}} \partial_i \psi_\gamma
    \simeq \mathbb{E}_{p_0}[F_i] - \frac{\gamma}{2} \{2\theta^k \mathrm{Cov}_{0}(F_k,\!F_i) + \theta^k \theta^\ell \mathrm{Cov}_{0}(F_k F_\ell,\!F_i)\}.
\end{equation}
Thus, finite deformation modifies escort expectations through covariance corrections computed under $p_0$. The linear correction changes sign with $\gamma$ with even-order contributions entering beyond this approximation. Depending on the sign of $\gamma$ and the braced term, the deformation raises or lowers the corresponding escort expectation.

\newpage
\section{Explicit escort coordinates at the independent reference}
\label{supp_sec:escort_coords_independent}

Here we evaluate the first-order dual coordinates~\eqref{eq:eta_expanded} explicitly for the binary model, treating the singleton and pair blocks in turn. At the independent reference, we partition the natural parameters into one- and two-body blocks and set the latter to zero,
$\bv{\theta}=(\bv{\theta}^{(1)},\bv{\theta}^{(2)})=(\bv{\theta}^{(1)},\bv{0})$.
For all $i \in \{1, \dots, N \}$ and $j \in \{i+1, \dots, N\}$ we define
\begin{align}
    f_{i} &\coloneqq x_i, \\
    f_{ij} &\coloneqq x_{i} x_{j} ,
\end{align}
so that the feature vector is $F=(F^{(1)},F^{(2)})\coloneqq(\{f_i\},\{f_{ij}\})$.
The pairwise features $f_{ij}$ remain random variables even though their natural
parameters vanish. Furthermore, we denote $p_i=1/(1+e^{-\theta_i})$.

We begin with the singleton block $\bv{\eta}_1$ in~\eqref{eq:eta_expanded}, taking
$f_i=x_i$ with the pairwise natural parameters set to zero. In that case,
$\mathbb{E}_{p_0}[f_i]=p_i$, and
\begin{align}
    \begin{aligned}
        \theta^j \mathrm{Cov}_{0}(F_j,\!f_i) 
        &= \theta^j (\mathbb{E}_{p_0}[F_j f_i] - p_j p_i ) \\ 
        &= \theta^j (\delta_{ji} (\mathbb{E}_{p_0}[F_j f_i] - p_j p_i)+ (1-\delta_{ji})(\mathbb{E}_{p_0}[F_j f_i]  - p_j p_i ) )\\ 
        &= \theta^j \delta_{ji} (\mathbb{E}_{p_0}[F_j f_i] - p_j p_i)\\  
        &= \theta_i (\mathbb{E}_{p_0}[f_i^2]  -  p_i^2 ) \\ 
        &= \theta_i (\mathbb{E}_{p_0}[f_i] - p_i^2 )  = p_i (1-p_i) \theta_i\\
    \end{aligned}
\end{align}
For the quadratic contraction, only singleton indices contribute. The off-diagonal summand is symmetric under
$k\leftrightarrow\ell$ and vanishes unless one index equals $i$. We may therefore fix $\ell=i$ and multiply that contribution by two,
\begin{align}
    \begin{aligned}
        \theta^k \theta^\ell \mathrm{Cov}_{0}(F_k F_\ell,\!f_{i}) 
        &= \theta^k \theta^\ell [\delta_{kl}+(1-\delta_{kl})]\mathrm{Cov}_{0}(F_k F_\ell,\!f_{i}) \nonumber \\
        &= \theta^\ell \theta^\ell \mathrm{Cov}_{0}(f_\ell,\!f_{i}) +\theta^k \theta^\ell (1-\delta_{kl})\mathrm{Cov}_{0}(F_k F_\ell,\!f_{i})\nonumber \\
        &=p_i (1-p_i)(\theta_i)^2 +\theta^k \theta^\ell (1-\delta_{kl})\mathrm{Cov}_{0}(F_k F_\ell,\!f_{i})\nonumber \\
        &=p_i (1-p_i)(\theta_i)^2
        +2 \theta^k \theta^\ell (1-\delta_{kl})\delta_{\ell i}
        \mathrm{Cov}_{0}(F_k F_\ell,\!f_i)\nonumber \\
        &=p_i (1-p_i)(\theta_i)^2 +2 \theta^k \theta^i (1-\delta_{ki}) \mathrm{Cov}_{0}(F_k F_i,\!f_{i}) \nonumber \\
        &=p_i (1-p_i)(\theta_i)^2 +2 \theta^k \theta^i (1-\delta_{ki})  p_k p_i (1-p_i)\\
    \end{aligned}
\end{align}
Therefore, by arranging the members according to \eqref{eq:eta_expanded}, we obtain
\begin{equation}\label{eq:eta1_closed}
    (\bv{\eta}_1)_{i} \simeq p_i - \gamma p_i (1-p_i)\theta_i\{1 + \tfrac{1}{2} \theta_i - \theta_i p_i+ \bar u_0  \}.
\end{equation}
Here we denote the mean singleton potential by $\bar u_0\coloneqq\mathbb E_{p_0}[u]=\sum_{k=1}^{N}\theta_kp_k$.

Secondly, we suppose that $f_{i_1 i_2} = x_{i_1} x_{i_2}$ and $F^{(1)}$ is as shown above. The independence of the neurons under $p_0$ is a consequence of the null pairwise interaction, which leads to $\mathbb{E}_{p_0}[f_{i_1 i_2}] = p_{i_1} p_{i_2}$.
In $\theta^k \theta^\ell \mathrm{Cov}_{0}(F_k F_\ell,\!f_{i_1 i_2})$ the contractions allow for the indices to match or not those at the second argument of the covariance. We must therefore identify the non-vanishing covariances, when both $(k,\ell)$ have overlap with $(i_1,i_2)$
\begin{align}
\mathrm{Cov}_{0}(F_{i_1}F_{i_1},f_{i_1 i_2}) &= \mathrm{Cov}_{0}(F_{i_1},f_{i_1 i_2})
 = p_{i_1}p_{i_2}(1-p_{i_1}),\\
\mathrm{Cov}_{0}(F_{i_2}F_{i_2},f_{i_1 i_2}) &= \mathrm{Cov}_{0}(F_{i_2},f_{i_1 i_2})
 = p_{i_1}p_{i_2}(1-p_{i_2}),\\
\mathrm{Cov}_{0}(F_{i_1}F_{i_2},f_{i_1 i_2})
&= \mathrm{Cov}_{0}(F_{i_2}F_{i_1},f_{i_1 i_2}) = p_{i_1}p_{i_2}(1-p_{i_1}p_{i_2}).
\end{align}
When only one index overlaps ($m\neq i_1,i_2$), then,
\begin{align}
\mathrm{Cov}_{0}(F_{i_1}F_m,f_{i_1 i_2}) &= \mathrm{Cov}_{0}(F_mF_{i_1},f_{i_1 i_2})
= p_{i_1}p_{i_2}p_m(1-p_{i_1}),\\
\mathrm{Cov}_{0}(F_{i_2}F_m,f_{i_1 i_2}) &= \mathrm{Cov}_{0}(F_mF_{i_2},f_{i_1 i_2})
= p_{i_1}p_{i_2}p_m(1-p_{i_2}).
\end{align}
So the full contraction reads,
\begin{align}
    \theta^k \theta^\ell \mathrm{Cov}_{0}(F_k F_\ell,\!f_{i_1 i_2}) &= p_{i_1}p_{i_2}(1-p_{i_1})(\theta^{i_1})^2 + p_{i_1}p_{i_2}(1-p_{i_2})(\theta^{i_2})^2 + 2p_{i_1}p_{i_2}(1-p_{i_1}p_{i_2})\theta^{i_1}\theta^{i_2} \nonumber \\ 
    &\quad + 2 p_{i_1}p_{i_2}(1-p_{i_1})\theta^{i_1}\sum_{m\neq i_1,i_2}\theta^m p_m + 2p_{i_1}p_{i_2}(1-p_{i_2})\theta^{i_2}\sum_{m\neq i_1,i_2}\theta^m p_m.
\end{align}
Equivalently,
\begin{align}
    \theta^k \theta^\ell \mathrm{Cov}_{0}(F_k F_\ell,\!f_{i_1 i_2}) &= p_{i_1}p_{i_2}\Big[(\theta^{i_1})^2(1-p_{i_1}) + (\theta^{i_2})^2(1-p_{i_2}) + 2(1-p_{i_1}p_{i_2})\theta^{i_1}\theta^{i_2} \nonumber\\
    &\qquad\qquad + 2(\bar u_0 -\theta^{i_1}p_{i_1}-\theta^{i_2}p_{i_2})\big((1-p_{i_1})\theta^{i_1}+(1-p_{i_2})\theta^{i_2}\big)\Big].
\end{align}
Replacing in~\eqref{eq:eta_expanded} leads to,
\begin{align}
    (\eta_2)_{i_1 i_2} &\simeq p_{i_1}p_{i_2} - \frac{\gamma}{2} \Big\{
    2p_{i_1}p_{i_2}\big[(1-p_{i_1})\theta^{i_1}+(1-p_{i_2})\theta^{i_2}\big] \nonumber \\
    &\qquad\qquad + p_{i_1}p_{i_2}\Big[ (\theta^{i_1})^2(1-p_{i_1})
    + (\theta^{i_2})^2(1-p_{i_2}) + 2(1-p_{i_1}p_{i_2})\theta^{i_1}\theta^{i_2} \nonumber\\
    &\qquad\qquad\qquad\qquad + 2(\bar u_0 -\theta^{i_1}p_{i_1}-\theta^{i_2}p_{i_2}) \big((1-p_{i_1})\theta^{i_1}+(1-p_{i_2})\theta^{i_2}\big) \Big] \Big\}.\label{eq:eta2_closed}
\end{align}
These are the singleton and pair closed forms~\eqref{eq:eta1_closed} and~\eqref{eq:eta2_closed}. The pair closed form~\eqref{eq:eta2_closed} shows that deformation generates pair structure even at $\bv{\theta}^{(2)}=\bv{0}$.

\clearpage

\section{Cumulant derivation of deformed TAP and its inverse}
\label{supp_sec:cumulant_deformed_tap}

This Note derives the deformed TAP equation~\eqref{eq:tap_deformed_main} and its inverse from a weak-correlation cumulant expansion of the escort moments. We proceed in two stages, first recovering the classical $\{0,1\}$ TAP equation at $\gamma=0$, isolating the $\alpha^2 J^2$ Onsager contribution that the deformed equation carries as its $\gamma^0\alpha^2$ term, and then extending the same expansion to $\gamma$-linear order. We expand the escort moments directly about the independent base and invert the resulting relation order by order, without a Legendre inversion or a variational free energy.

\subsection*{Cumulant expansion of the escort mean}
\label{supp_subsec:weakcorr_cumulant_expansion}

At $\gamma=0$ the deformed exponential reduces to the ordinary exponential, and the pairwise binary model factorizes
\begin{equation}\label{eq:wc_model}
    p(\bv{x}\mid \bv{H}, \alpha\bv{J}) = \frac{1}{Z(\alpha)} p_{\mathrm{ind}}(\bv{x}) e^{\alpha V(\bv{x})},
    \qquad
    V(\bv{x}) = \sum_{a<b} J_{ab} x_a x_b,
\end{equation}
with the independent base distribution
\begin{equation}\label{eq:wc_base}
    p_{\mathrm{ind}}(\bv{x}) = \prod_i p_i^{x_i}(1-p_i)^{1-x_i},
    \qquad
    p_i = \sigma(H_i) = \frac{1}{1+e^{-H_i}}.
\end{equation}
Throughout this section we write $\chi_i \coloneqq p_i(1-p_i)$, and $\langle\cdot\rangle_{\mathrm{ind}}$ and $\kappa_{\mathrm{ind}}(\cdot,\ldots)$ for expectations and joint cumulants under~$p_{\mathrm{ind}}$. For an observable $g(\bv{x})$ the expectation under~\eqref{eq:wc_model} reads
\begin{equation}\label{eq:wc_escort_def}
    \mathbb{E}_\alpha[g] = \frac{\langle g e^{\alpha V}\rangle_{\mathrm{ind}}}{\langle e^{\alpha V}\rangle_{\mathrm{ind}}}.
\end{equation}
Introducing the cumulant-generating function
\begin{equation}\label{eq:wc_cgf}
    G(\alpha,r,s) = \ln\!\left\langle e^{\alpha V + r A + s B}\right\rangle_{\!\mathrm{ind}} - \ln\!\left\langle e^{\alpha V}\right\rangle_{\!\mathrm{ind}},
\end{equation}
mixed derivatives at $(r,s)=(0,0)$ yield the $\alpha$-deformed covariance $\mathrm{Cov}_\alpha(A,B) = \partial_{r,s}G|_{(0,0)}$, and so on for higher-order cumulants. Expanding $G$ in $\alpha$ and setting $g = A$ in~\eqref{eq:wc_escort_def} gives
\begin{equation}\label{eq:wc_g_expansion}
    \mathbb{E}_\alpha[g] \simeq \langle g\rangle_{\mathrm{ind}} + \alpha \kappa_{\mathrm{ind}}(g,V) + \frac{\alpha^2}{2} \kappa_{\mathrm{ind}}(g,V,V) + \pazocal{O}(\alpha^3),
\end{equation}
where we used $(1+\alpha a + \tfrac{\alpha^2}{2} b)^{-1} \simeq 1 - \alpha a + \alpha^2(a^2 - b/2)$ with $a=\langle V\rangle_{\mathrm{ind}}$, $b=\langle V^2\rangle_{\mathrm{ind}}$, and collected the resulting combination of ordinary moments into connected cumulants. The expansion~\eqref{eq:wc_g_expansion} is the weak-correlation analogue of the escort expectation used throughout Supplementary Note~\ref{supp_sec:escort_coords_independent}, refined here to second order so that the Onsager term survives.

\subsection*{Evaluation of the base-point cumulants}
\label{supp_subsec:weakcorr_cumulants}

Specializing $g=x_i$ in~\eqref{eq:wc_g_expansion} requires two cumulants of $p_{\mathrm{ind}}$.

\paragraph*{First cumulant.}  By independence of the base distribution,
\begin{equation}\label{eq:wc_kappa_xi_V}
    \kappa_{\mathrm{ind}}(x_i,V) = \sum_{a<b} J_{ab} \kappa_{\mathrm{ind}}(x_i, x_a x_b) = \chi_i \sum_{j\neq i} J_{ij} p_j \eqqcolon \chi_i S_i,
\end{equation}
since $\kappa_{\mathrm{ind}}(x_i, x_a x_b)$ vanishes unless $i\in\{a,b\}$, and $\kappa_{\mathrm{ind}}(x_i, x_i x_j) = \langle x_i^2 x_j\rangle_{\mathrm{ind}} - \langle x_i\rangle_{\mathrm{ind}}\langle x_i x_j\rangle_{\mathrm{ind}} = \chi_i p_j$.

\paragraph*{Second cumulant by edge enumeration.}  Writing
\begin{equation}\label{eq:wc_kappa_xi_VV_raw}
    \kappa_{\mathrm{ind}}(x_i,V,V) = \sum_{a<b}\sum_{c<d} J_{ab}J_{cd} \kappa_{\mathrm{ind}}(x_i, x_a x_b, x_c x_d),
\end{equation}
the three-point joint cumulant vanishes whenever the random variables can be partitioned into two statistically independent groups under~$p_{\mathrm{ind}}$. Taking the two edges $(a,b)$ and $(c,d)$, the only configurations that survive are those in which index~$i$ and the two edges form a connected set, yielding the three surviving topologies
\begin{equation}\label{eq:wc_three_topologies}
    \begin{aligned}
    \text{(I)}&\quad (a,b)=(c,d)=(i,j), \qquad j\neq i,\\
    \text{(II)}&\quad (a,b)=(i,j),\ (c,d)=(i,k), \qquad k\neq j,  k\neq i,\\
    \text{(III)}&\quad (a,b)=(i,j),\ (c,d)=(j,k), \qquad k\neq i,  k\neq j,
    \end{aligned}
\end{equation}
together with the symmetric counterparts of~(II)--(III) obtained by swapping the two edges. Using $x_i^n = x_i$ on $\{0,1\}$, each topology contributes an elementary base-point moment
\begin{equation}\label{eq:wc_topology_values}
    \begin{aligned}
    \text{(I)}&:\quad \kappa_{\mathrm{ind}}(x_i, x_i x_j, x_i x_j) = \chi_i p_j (1-2 p_i p_j),\\[2pt]
    \text{(II)}&:\quad \kappa_{\mathrm{ind}}(x_i, x_i x_j, x_i x_k) = \chi_i p_j p_k (1-2p_i),\\[2pt]
    \text{(III)}&:\quad \kappa_{\mathrm{ind}}(x_i, x_i x_j, x_j x_k) = \chi_i \chi_j p_k.
    \end{aligned}
\end{equation}
Topology~(III) is recognized
as the three-point cumulant splits into a pair $(x_i,x_j)$ coupled to the marginal $p_k$, giving the $\chi_i\chi_j p_k$ weight. Summing the three topologies with the combinatorial factors in~\eqref{eq:wc_kappa_xi_VV_raw} and collecting powers of $p_i$ yields
\begin{equation}\label{eq:wc_kappa_xi_VV}
    \tfrac{1}{2} \kappa_{\mathrm{ind}}(x_i,V,V)
    = \chi_i \Bigg[\sum_{j\neq i} J_{ij} \chi_j S_j
    + \frac{1-2p_i}{2}\bigg(S_i^2 + \sum_{j\neq i} J_{ij}^2 \chi_j\bigg)\Bigg],
\end{equation}
with $S_i$ defined in~\eqref{eq:wc_kappa_xi_V}. The parenthesized $S_i^2$ piece collects the off-diagonal contributions of~(II) with the diagonal $(j,j)$ part of~(I), the $\sum_j J_{ij}^2\chi_j$ piece and the $k=i$ term of $S_j$ are the residual of~(I).

\paragraph*{Undeformed weak-coupling mean to second order.}  Substituting \eqref{eq:wc_kappa_xi_V} and~\eqref{eq:wc_kappa_xi_VV} into \eqref{eq:wc_g_expansion} with $g=x_i$ gives the mean,
\begin{equation}\label{eq:wc_m_expansion}
    m_i = p_i + \alpha \chi_i S_i
    + \alpha^2 \chi_i\Bigg[\sum_{j\neq i} J_{ij} \chi_j S_j
    + \frac{1-2p_i}{2}\bigg(S_i^2 + \sum_{j\neq i} J_{ij}^2 \chi_j\bigg)\Bigg]
    + \pazocal{O}(\alpha^3).
\end{equation}

\subsection*{Inversion via the logit expansion}
\label{supp_subsec:weakcorr_logit_expansion}

To recover the effective field $H_i$ from the escort mean we invert~\eqref{eq:wc_m_expansion} by expanding $\mathrm{logit}(m_i) = \ln[m_i/(1-m_i)]$ around $m_i=p_i$. Using
\begin{equation}\label{eq:wc_logit_derivs}
    \frac{\mathrm{d}}{\mathrm{d}p}\mathrm{logit}(p) = \frac{1}{\chi(p)},
    \qquad
    \frac{\mathrm{d}^2}{\mathrm{d}p^2}\mathrm{logit}(p) = \frac{2p-1}{\chi(p)^2},
\end{equation}
with $\chi(p)=p(1-p)$, the second-order Taylor expansion gives
\begin{equation}\label{eq:wc_logit_taylor}
    \mathrm{logit}(m_i) \simeq H_i + \frac{\delta_i}{\chi_i} + \frac{2p_i-1}{2\chi_i^2} \delta_i^2,
    \qquad \delta_i = \delta_i^{(1)} + \delta_i^{(2)},
\end{equation}
where $\delta_i^{(1)} = \alpha\chi_i S_i$ and $\delta_i^{(2)}$ is the bracketed $\alpha^2$ term of~\eqref{eq:wc_m_expansion}. Keeping terms through $\alpha^2$,
\begin{equation}\label{eq:wc_delta_sq}
    \delta_i^2 \simeq (\delta_i^{(1)})^2 = \alpha^2\chi_i^2 S_i^2 + \pazocal{O}(\alpha^3),
    \qquad
    \frac{2p_i-1}{2\chi_i^2} \delta_i^2 = -\alpha^2 \frac{1-2p_i}{2} S_i^2.
\end{equation}
Substituting into~\eqref{eq:wc_logit_taylor} and collecting orders of $\alpha$,
\begin{equation}\label{eq:wc_logit_intermediate}
    \mathrm{logit}(m_i)
    \simeq H_i + \alpha S_i
    + \alpha^2 \Bigg[\sum_{j\neq i} J_{ij} \chi_j S_j
    + \frac{1-2p_i}{2}\bigg(S_i^2 + \sum_{j\neq i} J_{ij}^2 \chi_j \bigg)\Bigg]
    -\alpha^2 \frac{1-2p_i}{2} S_i^2,
\end{equation}
and the $S_i^2$ pieces cancel, leaving
\begin{equation}\label{eq:wc_logit_cancelled}
    \mathrm{logit}(m_i)
    \simeq H_i + \alpha S_i + \alpha^2 \Bigg[\sum_{j\neq i} J_{ij} \chi_j S_j + \frac{1-2p_i}{2}\sum_{j\neq i} J_{ij}^2 \chi_j\Bigg].
\end{equation}

\subsection*{The classical TAP equation}
\label{supp_subsec:weakcorr_tap_eq}

The first two terms in~\eqref{eq:wc_logit_cancelled} combine into a single naive mean-field sum by the substitution $p_i\mapsto m_i$, accurate to $\pazocal{O}(\alpha^2)$
\begin{equation}\label{eq:wc_nmf_combine}
    \alpha S_i + \alpha^2\sum_{j\neq i} J_{ij} \chi_j S_j
    = \alpha\sum_{j\neq i} J_{ij}\bigl(p_j + \alpha\chi_j S_j\bigr)
    = \alpha\sum_{j\neq i} J_{ij} m_j + \pazocal{O}(\alpha^3).
\end{equation}
For the Onsager piece it suffices to replace $p_i\to m_i$ and $\chi_j = p_j(1-p_j)\to m_j(1-m_j)$, which induces only $\pazocal{O}(\alpha^3)$ corrections. We arrive at the classical Plefka/TAP equation for the $\{0,1\}$ model,
\begin{equation}\label{eq:wc_tap}
    \mathrm{logit}(m_i) = H_i + \alpha\sum_{j\neq i} J_{ij} m_j
    + \alpha^2 \frac{1-2m_i}{2}\sum_{j\neq i} J_{ij}^2 m_j(1-m_j) + \pazocal{O}(\alpha^3).
\end{equation}
At $\gamma=0$, the correction $\gamma\Delta_i$ vanishes, and~\eqref{eq:tap_deformed_main} reduces to~\eqref{eq:wc_tap} through the second order in $\alpha$. The $\alpha J^1$ and $\alpha^2 J^2$ terms give the naive mean-field contribution and the standard Onsager correction in the forward iteration, respectively. The combinatorial derivation above fixes the coefficient of the Onsager term, $(1-2m_i)/2$, and the $J_{ij}^2 m_j(1-m_j)$ weight inside the sum.

\subsection*{Pair covariance from the single-spin expansion}
\label{supp_subsec:weakcorr_pair_cov_forward}

Differentiating the single-spin result~\eqref{eq:wc_m_expansion} with respect to $H_j$ ($j\neq i$) gives the connected pair covariance $C_{ij} = \mathrm{Cov}_\alpha(x_i,x_j) = \partial m_i/\partial H_j$. The required base-point derivatives are
\begin{equation}\label{eq:wc_base_derivs}
    \frac{\partial p_i}{\partial H_j} = \delta_{ij} \chi_i,
    \qquad
    \frac{\partial\chi_i}{\partial H_j} = \delta_{ij} (1-2p_i) \chi_i,
    \qquad
    \frac{\partial S_i}{\partial H_j} = J_{ij} \chi_j\quad (j\neq i),
\end{equation}
together with the composite identities
\begin{equation}\label{eq:wc_composite_derivs}
    \frac{\partial \mathcal{A}_i}{\partial H_j} = J_{ij} (1-2p_j) \chi_j S_j + \chi_j\sum_{k\neq i,j} J_{ik}J_{kj} \chi_k,
    \qquad
    \frac{\partial \mathcal{B}_i}{\partial H_j} = J_{ij}^2 (1-2p_j) \chi_j,
\end{equation}
where $\mathcal{A}_i = \sum_{k\neq i}J_{ik}\chi_k S_k$ and $\mathcal{B}_i = \sum_{k\neq i}J_{ik}^2\chi_k$ are the composite sums appearing in~\eqref{eq:wc_m_expansion}. Collecting orders of $\alpha$ and extracting the $(i,j)$ edge from $S_i = J_{ij}p_j + S_i'$ and $S_j = J_{ij}p_i + S_j'$,
\begin{equation}\label{eq:wc_cov_forward}
    C_{ij} = \alpha \chi_i\chi_j J_{ij}
    + \alpha^2 \chi_i\chi_j \Bigg[\sum_{k\neq i,j}J_{ik}J_{kj} \chi_k
    + J_{ij} (1-2p_i) S_i'
    + J_{ij} (1-2p_j) S_j'
    + \frac{1-4 p_ip_j}{2} J_{ij}^2\Bigg]
    + \pazocal{O}(\alpha^3),
\end{equation}
with reduced sums $S_i'\coloneqq\sum_{k\neq i,j}J_{ik}p_k$ and $S_j'\coloneqq\sum_{k\neq i,j}J_{jk}p_k$. The $J_{ij}^2$ coefficient $(1-4p_ip_j)/2$ collects $(1-2p_i)p_j + (1-2p_j)p_i + \tfrac12(1-2p_i)(1-2p_j)$ after absorbing the $(i,j)$-edge contributions of $S_i$ and $S_j$. Combinatorially, the leading term is the independent-pair susceptibility, the two-edge piece $\sum_{k\neq i,j} J_{ik}J_{kj}\chi_k$ is the non-local path $i\!\to\!k\!\to\!j$,
and the terms proportional to $S_i'$ and $S_j'$ account for the shifts in the susceptibilities at $i$ and $j$, respectively, due to couplings to sites $k\ne i,j$.

\subsection*{Pair moment to $\pazocal{O}(\alpha^2)$}
\label{supp_subsec:weakcorr_pair_moment}

Since $m_{ij}\coloneqq\mathbb{E}_\alpha[x_i x_j] = m_i m_j + C_{ij}$, \eqref{eq:wc_m_expansion} and~\eqref{eq:wc_cov_forward} combine into the closed-form two-spin expectation
\begin{equation}\label{eq:wc_mij_expansion}
    m_{ij} = m_i m_j + \alpha \chi_i\chi_j J_{ij}
    + \alpha^2 \chi_i\chi_j \Bigg[\sum_{k\neq i,j}J_{ik}J_{kj} \chi_k
    + J_{ij} (1-2p_i) S_i'
    + J_{ij} (1-2p_j) S_j'
    + \frac{1-4 p_ip_j}{2} J_{ij}^2 \Bigg]
    + \pazocal{O}(\alpha^3).
\end{equation}
This is the forward target used in pairwise moment matching at $\gamma=0$, and the off-diagonal companion of~\eqref{eq:wc_m_expansion}.

\subsection*{Pair covariance from the TAP equation}
\label{supp_subsec:weakcorr_pair_cov_tap}

The same covariance is recovered directly from the TAP equation~\eqref{eq:wc_tap} by applying $\partial/\partial H_j$ and using $\partial_{m_i}\mathrm{logit}(m_i) = 1/\chi_i$, so that $\partial_{H_j}\mathrm{logit}(m_i) = C_{ij}/\chi_i$. Through $\pazocal{O}(\alpha^2 J^2)$,
\begin{equation}\label{eq:wc_tap_diff}
    \frac{C_{ij}}{\chi_i}
    = \alpha\sum_{k} J_{ik} C_{kj}
    + \alpha^2 \frac{1-2m_i}{2}\sum_{k}J_{ik}^2 (1-2m_k) C_{kj}
    + \pazocal{O}(\alpha^3),
\end{equation}
where the diamagnetic piece $-C_{ij}\sum_k J_{ik}^2\chi_k$ generated by the $(1-2m_i)$ factor is $\pazocal{O}(\alpha^3)$ and has been dropped. At leading order $C_{jj}\simeq\chi_j$ and $C_{kj}\simeq\alpha\chi_k\chi_j J_{kj}$ for $k\neq j$, so the first sum splits as
\begin{equation}\label{eq:wc_tap_first_sum}
    \alpha\sum_k J_{ik} C_{kj}
    = \alpha J_{ij} \chi_j
    + \alpha^2 \chi_j\sum_{k\neq i,j}J_{ik}J_{kj} \chi_k
    + \pazocal{O}(\alpha^3),
\end{equation}
and the Onsager sum is dominated by $k=j$, contributing $\tfrac{1-2m_i}{2} J_{ij}^2 (1-2m_j) \chi_j$. Multiplying through by $\chi_i$,
\begin{equation}\label{eq:wc_cov_tap}
    C_{ij}
    \simeq \alpha \chi_i\chi_j J_{ij}
    + \alpha^2 \chi_i\chi_j\sum_{k\neq i,j} J_{ik}J_{kj} \chi_k
    + \alpha^2 \tfrac{1}{2} \chi_i\chi_j (1-2m_i)(1-2m_j) J_{ij}^2
\end{equation}
The covariance~\eqref{eq:wc_cov_tap} agrees with the forward result~\eqref{eq:wc_cov_forward} at $\pazocal{O}(J^2)$ once the latter is expressed in the TAP means $m_i$. The local end-point terms $J_{ij}(1-2p_i)S_i'$, $J_{ij}(1-2p_j)S_j'$ are absorbed into the leading $\chi(p_i)\chi(p_j)J_{ij}$ term via the shift $\chi(p_i) = \chi(m_i) - (1-2m_i) \alpha\chi_i S_i + \pazocal{O}(\alpha^2)$, and the residual $J_{ij}^2$ coefficient reduces through
\begin{equation}\label{eq:wc_bracket_equivalence}
    \frac{1-4 p_ip_j}{2} - (1-2p_i) p_j - (1-2p_j) p_i
    = \frac{(1-2m_i)(1-2m_j)}{2} + \pazocal{O}(\alpha).
\end{equation}
The canonical factor $(1-2m_i)(1-2m_j)/2$ is therefore the correct local Onsager weight of the pair covariance, without auxiliary $\chi$ factors at the endpoints. The result~\eqref{eq:wc_cov_tap} supplies the self-contained cumulant-level derivation of this Onsager weight.

\subsection*{Deformed extension: $\gamma$-linear corrections}
\label{supp_subsec:weakcorr_deformed_extension}

The classical TAP derivation above is at $\gamma=0$. We now extend the cumulant route to the $\gamma$-linear regime, obtaining the deformed TAP equation~\eqref{eq:tap_deformed_main} and the leading deformation correction to the pair couplings, $\gamma h_ih_j$ with $h_i=\mathrm{logit}(\eta_i)$, which enters the main-text estimator~\eqref{eq:Jij_inverse_main}.

Recall the small-$\gamma$ expansion of the escort dual coordinates from~\eqref{eq:eta_expanded},
\begin{equation}\label{eq:wc_eta_gamma_expansion}
    \eta_r \simeq \mathbb{E}_{p_0}[F_r] - \gamma  A_r,
    \qquad
    A_r \coloneqq \tfrac{1}{2}\bigl\{2 \theta^k \mathrm{Cov}_0(F_k, F_r) + \theta^k\theta^\ell \mathrm{Cov}_0(F_k F_\ell, F_r)\bigr\}.
\end{equation}
For singletons $F_r = x_i$, write $A_i$ for the $\pazocal{O}(\gamma)$ correction; for pairs $F_r = x_i x_j$, write $A_{ij}$. The corresponding pair-covariance is determined by $Q_{ij}\coloneqq \eta_{ij}-\eta_i \eta_j$, however, $m_r = \mathbb{E}_{p_0}[F_r]$ and so from~\eqref{eq:wc_eta_gamma_expansion}, we obtain $Q_{ij} = m_{ij}-m_i m_j -\gamma B_{ij}$ with the correction $B_{ij}$ defined as,
\begin{equation}\label{eq:wc_Bij_def}
    B_{ij} \coloneqq A_{ij} - m_i A_j - m_j A_i.
\end{equation}
Let us remember that $\partial_{H_j}m_i = \mathbb{E}_{\alpha}[x_i x_j] - \mathbb{E}_{\alpha}[x_i]\mathbb{E}_{\alpha}[x_j] = {\rm Cov}_\alpha (x_i,x_j)$, defined as $C_{ij}$ which implies that $m_{ij}=m_i m_j + C_{ij}$ in the undeformed case and so the connected pair covariance to $\pazocal{O}(\gamma)$ reads $Q_{ij} = C_{ij} - \gamma B_{ij}$ with $C_{ij}$ the $\gamma=0$ pair covariance of~\eqref{eq:wc_cov_forward}.\newline

\paragraph*{Singleton coefficient $A_i$ at $\bv{J}=0$.} Setting $\theta^{ij}=0$ in~\eqref{eq:wc_eta_gamma_expansion} and using $\mathrm{Cov}_0(x_k, x_i) = \delta_{ki}\chi_i$ and $\mathrm{Cov}_0(x_k x_\ell, x_i)$ non-zero only when $k=i$ or $\ell=i$, direct enumeration yields
\begin{equation}\label{eq:wc_Ai_J0}
    A_i\big|_{\bv{J}=0}
    = \tfrac{1}{2} \chi_i H_i\bigl[2 + 2 s + H_i(1-2 p_i)\bigr],
    \qquad
    s \coloneqq H^k p_k.
\end{equation}

\paragraph*{Pair coefficient $A_{ij}$ and $B_{ij}$ at $\bv{J}=0$.} A parallel enumeration for $A_{ij}$ at $\bv{J}=0$, splitting the bilinear cumulant $\mathrm{Cov}_0(x_k x_\ell, x_i x_j)$ into the surviving topologies $(k,\ell)\in\{(i,i),(j,j),(i,j),(j,i),(i,m\neq i,j),(j,m\neq i,j)\}$, gives after collection
\begin{align}\label{eq:wc_Aij_J0}
    A_{ij}\big|_{\bv{J}=0}
    &= p_i \chi_j H_j + p_j \chi_i H_i \nonumber\\
    &\hphantom{=}\;\,+\tfrac{1}{2}\bigl[H_i^2 \chi_i p_j + H_j^2 p_i \chi_j + 2 H_i H_j p_i p_j(1-p_i p_j) \nonumber\\
    &\hphantom{= -\tfrac{1}{2}\bigl[}+ 2 H_i \chi_i p_j\sum_{m\neq i,j}H_m p_m + 2 H_j p_i \chi_j\sum_{m\neq i,j}H_m p_m\bigr].
\end{align}
Substituting~\eqref{eq:wc_Ai_J0} and~\eqref{eq:wc_Aij_J0} into~\eqref{eq:wc_Bij_def} and using the cancellation
\begin{equation}\label{eq:wc_qq_identity}
    1 - p_i p_j - (1-p_i) p_j - (1-p_j) p_i
    = (1-p_i)(1-p_j),
\end{equation}
all terms involving $s$ and the bare $H_i, H_j$ contributions cancel, leaving the compact identity
\begin{equation}\label{eq:wc_Bij_J0}
    B_{ij}\big|_{\bv{J}=0}
    = \chi_i \chi_j H_i H_j.
\end{equation}
This identity yields the coupling term $\gamma h_i h_j$ of the main-text estimator~\eqref{eq:Jij_inverse_main}, as shown next.

\paragraph*{Inverse pair-coupling correction at $\bv{J}=0$.} The $\gamma^1$ shift of the connected pair covariance at $\bv{J}=0$ is therefore
\begin{equation}\label{eq:wc_Qij_gamma_correction}
    Q_{ij}\big|_{\bv{J}=0}
    = C_{ij}\big|_{\bv{J}=0} - \gamma B_{ij}\big|_{\bv{J}=0}
    = -\gamma \chi_i\chi_j H_i H_j
    \qquad (\text{since }C_{ij}\big|_{\bv{J}=0}=0).
\end{equation}
With the normalized covariance $y_{ij}$ of~\eqref{eq:wc_yij_inverse_def}, which equals $Q_{ij}/(\chi_i\chi_j)$ to this order, inversion of~\eqref{eq:wc_Qij_gamma_correction} yields the one-step corrected estimator
\begin{equation}\label{eq:wc_theta_ij_gamma1}
    J_{ij}
    = y_{ij} + \gamma H_i H_j,
\end{equation}
which is the $\bv{J}=0$, $\pazocal{O}(\gamma)$ part of the coupling estimator~\eqref{eq:J_TAP_gamma_inverse_chat} and of the fixed-escort-moment coupling $\widehat J_{ij}$ in~\eqref{eq:wc_fixed_full_eta_response}, since $h_i=H_i+\pazocal{O}(\gamma)$ at $\bv J=0$.

\subsection*{Deformed TAP via the cumulant route}
\label{supp_subsec:weakcorr_deformed_tap}

To assemble the full deformed TAP equation in escort coordinates, we invert the singleton relation~\eqref{eq:wc_eta_gamma_expansion} as $m_i = \eta_i + \gamma A_i + \pazocal{O}(\gamma^2)$ where $m_i=\mathbb E_{p_0}[x_i]$ is the ordinary moment of the undeformed model at the same natural parameters and $\eta_i$ is the finite-$\gamma$ escort moment, so $\eta_i = m_i$ at $\gamma=0$. Let us define the interaction contribution to the classical TAP~\eqref{eq:wc_tap} by
\begin{equation}
R_i(\bv m)\coloneqq \sum_{j\ne i}J_{ij}m_j
+\frac{1-2m_i}{2}\sum_{j\ne i}J_{ij}^2m_j(1-m_j).
\end{equation}
Expanding both $\operatorname{logit}(m_i)$ and $R_i(\bv m)$ around $\bv m=\bv\eta$ gives $\operatorname{logit}(m_i)\simeq\operatorname{logit}(\eta_i)+\gamma A_i/\chi_i$ and $R_i(\bv\eta+\gamma\bv A)\simeq R_i(\bv\eta)+\gamma\sum_j(\partial_{m_j}R_i)_{\bv m=\bv\eta}A_j$. Hence the $\pazocal{O}(\gamma)$ correction is
\begin{equation}\label{eq:wc_def_bracket}
    \Xi_i \coloneqq \sum_j \left.\partial_{m_j} R_i\right|_{\bv m=\bv\eta} A_j - \frac{A_i}{\chi_i}
    = -\frac{A_i}{\chi_i} - A_i \sum_{j\neq i} J_{ij}^2 \chi_j + \sum_{j\neq i} \left[J_{ij} + \frac{1-2\eta_i}{2} J_{ij}^2(1-2\eta_j)\right] A_j,
\end{equation}
where $\chi_i = \eta_i (1-\eta_i)$ and the second equality follows from
$\partial_{m_i}R_i=-\sum_{k\ne i}J_{ik}^2m_k(1-m_k)$ and
$\partial_{m_j}R_i=J_{ij}+\tfrac{1-2m_i}{2}J_{ij}^2(1-2m_j)$ for $j\ne i$.
To organize the mixed $\alpha \gamma$-expansion, we first grade $A_i$~\eqref{eq:wc_eta_gamma_expansion} in coupling strength at fixed field~$\bv H$
\begin{equation}\label{eq:wc_A_fixedH_grading}
    A_i(\bv H,\alpha\bv J)
    =A_i^{(0)}(\bv H)+\alpha A_i^{(1)}(\bv H)+\pazocal O(\alpha^2).
\end{equation}
Thus $A_i^{(r)}$ denotes fixed-$\bv H$ coefficients of order $\alpha^r$. Here $A_i^{(0)}=A_i|_{\bv J=0}$~\eqref{eq:wc_Ai_J0}, and expanding $A_i$~\eqref{eq:wc_eta_gamma_expansion} to first order in $\alpha$, with $\mathrm{Cov}_0$ taken in the undeformed model at the same natural parameters $(\bv H,\alpha\bv J)$, gives
\begin{equation}\label{eq:wc_Ai_alpha1}
\begin{aligned}
    A_i^{(1)} &= (1-2p_i) S_i A_i^{(0)}
    + p_i(1-p_i)\Bigl\{\bigl[1+s+H_i(1-p_i)\bigr]S_i
    + \bigl[1+H_i(1-p_i)\bigr]W_i + H_i T_i + \sum_{j\neq i}J_{ij}A_j^{(0)}\Bigr\},\\
    W_i &\coloneqq \sum_{j\neq i}J_{ij}H_j p_j(1-p_j),
    \qquad
    T_i \coloneqq \sum_{\substack{a<b\\ a,b\neq i}}J_{ab} p_a p_b\bigl[1+H_a(1-p_a)+H_b(1-p_b)\bigr],
\end{aligned}
\end{equation}
where $p_i=\sigma(H_i)$, and $S_i$ and $s$ are defined in~\eqref{eq:wc_kappa_xi_V} and~\eqref{eq:wc_Ai_J0}. The correction~\eqref{eq:wc_def_bracket} then gives
\begin{subequations}\label{eq:wc_def_sectors}
\begin{align}
    \Xi_i^{(0)} &= -\frac{A_i^{(0)}}{\chi_i}, \\
    \Xi_i^{(1)} &= -\frac{A_i^{(1)}}{\chi_i} + \sum_{j\neq i}J_{ij} A_j^{(0)}.
\end{align}
\end{subequations}
The field response required to keep $\bv \eta$ fixed enters only when the implicit TAP equation is inverted. We write
\begin{equation}\label{eq:wc_fixed_eta_field_branch}
    \bv H(\alpha,\gamma)
    =\bv h+\alpha\bv H_{\alpha}
    +\gamma\bv K^{(0)}+\gamma\alpha\bv K^{(1)}+\cdots,
    \qquad
    \bv H_{\alpha}=-\bv J\bv\eta,
    \qquad
    h_i\coloneqq\mathrm{logit}(\eta_i).
\end{equation}
Differentiating the implicit equation at fixed $\bv\eta$ gives
\begin{subequations}\label{eq:wc_fixed_eta_field_response}
\begin{align}
    K_i^{(0)}
    &=\frac{A_i^{(0)}}{\chi_i},\\
    K_i^{(1)}
    &=\frac{A_i^{(1)}}{\chi_i}
      -\sum_{j\ne i}J_{ij}A_j^{(0)}
      +\frac{1}{\chi_i}\sum_k
       \frac{\partial A_i^{(0)}}{\partial H_k}H_{\alpha,k}.
\end{align}
\end{subequations}
The last term results from evaluating the pure $\gamma$ correction at the $\pazocal{O}(\alpha)$-corrected undeformed field~\eqref{eq:wc_fixed_eta_field_branch}.
These two sectors give the deformation correction (sectors at $\pazocal{O}(\gamma\alpha^2)$ and beyond are not retained)
\begin{equation}\label{eq:wc_Delta_NMF_TAP}
    \Delta_i = \Xi_i^{(0)} + \Xi_i^{(1)}.
\end{equation}
Substituting back into the classical TAP \eqref{eq:wc_tap} and collecting orders gives the deformed TAP equation through $\pazocal{O}(\gamma\alpha^1)$
\begin{align}\label{eq:wc_deformed_tap}
    \mathrm{logit}(\eta_i)
    &= H_i + \sum_{j\neq i}J_{ij}\,\eta_j \nonumber\\
    &\quad + \frac{1-2\eta_i}{2}\sum_{j\neq i}J_{ij}^2\,\eta_j(1-\eta_j) \nonumber\\
    &\quad + \gamma\,\Delta_i(\bv{\eta},\bv{H},\bv{J})
    + \pazocal{O}(\alpha^3,\gamma\alpha^2,\gamma^2),
\end{align}
which is the main-text deformed TAP equation~\eqref{eq:tap_deformed_main}. The cumulant route thus reaches the deformed TAP equation without a generalized Legendre construction or a variational free energy, and every coefficient is fixed by connected cumulants at the independent base.

\subsection*{Inverse deformed TAP}
\label{supp_subsec:weakcorr_deformed_inverse}

We derive two closed-form inverses, beginning with the local-closure inverse, in which the deformation correction is also evaluated within the pair-local closure to obtain the coupling estimator~\eqref{eq:J_TAP_gamma_inverse_chat} and the field estimator~\eqref{eq:wc_field_estimator}. The closed-form inverse of the main text~\eqref{eq:Jij_inverse_main}, used in Figs.~\ref{fig:sim_invising_weakcorr_exact_lgd} and~\ref{fig:cf_obs} and in Supplementary Fig.~\ref{fig:sim_invising_weakcorr_exact}, is the fixed-escort-moment inverse~\eqref{eq:wc_fixed_full_eta_flat} and~\eqref{eq:wc_fixed_full_eta_response} at the end of this subsection. The two coupling estimators agree at $\bv J=0$ but differ at order $\gamma\alpha$.

The full second order weak-correlation expansion of $C_{ij}$ in~\eqref{eq:wc_cov_tap} contains an explicit two-step path contribution $\chi_i\chi_j\sum_{k\neq i,j} J_{ik}J_{kj} \chi_k$. Retaining this term leads to a coupled inverse problem. To obtain a closed pairwise estimator, we adopt a local closure in which indirect paths are assumed either small due to weak-common neighbor structure, low susceptibilities, or cancellation of mixed sign couplings. This approximation is controlled for sparse neural activity or locally tree-like weakly correlated systems.
We first define the normalized inverse coordinate in terms of the observed escort moments. Let
\begin{equation}
    Q_{ij}\coloneqq\eta_{ij}-\eta_i\eta_j,
    \qquad
    \chi_i^\eta \coloneqq\eta_i(1-\eta_i).
    \label{eq:wc_Q_chieta_inverse_def}
\end{equation}
At leading weak-correlation order, the connected covariance satisfies $Q_{ij}\simeq \chi_i^\eta \chi_j^\eta J_{ij}$. This motivates the normalized inverse
coordinate
\begin{equation}
    y_{ij} \coloneqq \frac{Q_{ij}}{\chi_i^\eta \chi_j^\eta} = \frac{\eta_{ij}-\eta_i\eta_j}{\eta_i(1-\eta_i)\eta_j(1-\eta_j)} .
    \label{eq:wc_yij_inverse_def}
\end{equation}
At leading order, the escort-normalized covariance equation gives the coupling estimate $y_{ij}$ without a separate fit at $\gamma=0$.

Within the local closure approximation, the classical TAP covariance in
\eqref{eq:wc_cov_tap} reduces to the direct-edge form
\begin{equation}
    C_{ij}^{\rm loc} \simeq \chi_i^\eta\chi_j^\eta \left[ J_{ij} + c_{ij}^{\eta}J_{ij}^2 \right],
    \label{eq:wc_Cij_loc_inverse_eta}
\end{equation}
where $c_{ij}^{\eta} \coloneqq \frac{1}{2}(1-2\eta_i)(1-2\eta_j)$. Here the ordinary means $m_i$ of~\eqref{eq:wc_cov_tap} are set equal to $\eta_i$, which holds at $\gamma=0$.
Including the deformation correction, the connected escort covariance has the local form $Q_{ij} \simeq C_{ij}^{\rm loc} -\gamma B_{ij}^{\rm loc}$, previously determined at~\eqref{eq:wc_Qij_gamma_correction}. For $\gamma\neq0$, the prefactor of $C_{ij}^{\rm loc}$ is $\chi(\eta_i+\gamma A_i)\chi(\eta_j+\gamma A_j)$, the susceptibilities of the ordinary means $\mathbb{E}_{p_0}[x_i]\simeq\eta_i+\gamma A_i$, rather than $\chi_i^\eta\chi_j^\eta$. The term $B_{ij}^{\rm loc}$ is the local closure of $B_{ij}$ in~\eqref{eq:wc_Bij_def}, derived below.
Dividing by the escort-coordinate denominator in \eqref{eq:wc_yij_inverse_def}, the normalized local forward relation can be written as
\begin{equation}
    y_{ij} \simeq J_{ij} + c_{ij}^{\eta}J_{ij}^2 - \gamma \Omega_{ij}^{\rm loc}.
    \label{eq:wc_yij_forward_inverse}
\end{equation}
Here $-\gamma\Omega_{ij}^{\rm loc}$ collects all $\pazocal{O}(\gamma)$ terms of $y_{ij}$. Since $\chi(\eta_i+\gamma A_i)\simeq\chi_i^\eta\bigl[1-\gamma d_i^{{}_{(\gamma)}}\bigr]$, with $d_i^{{}_{(\gamma)}}$ given in~\eqref{eq:di_gamma_inverse_m} below, $\Omega_{ij}^{\rm loc}\simeq B_{ij}^{\rm loc}/(\chi_i^\eta\chi_j^\eta)+\bigl(d_i^{{}_{(\gamma)}}+d_j^{{}_{(\gamma)}}\bigr)J_{ij}$. The remaining task is to compute $\Omega_{ij}^{\rm loc}$ in the same normalized coordinate as $y_{ij}$.

We evaluate $B_{ij}$ in the same local closure and impose locality before the deformation correction is computed. This gives a self-contained approximation for the local contribution to the inverse estimator and avoids an after-the-fact truncation of nonlocal terms. For a fixed pair $(i,j)$, we replace the interaction potential by its direct edge, so $V \mapsto V_{ij}^{\rm loc}(x)=J_{ij}x_i x_j$. The remaining singleton contribution from all sites $k\neq i,j$ is retained only through its independent mean. We introduce an auxiliary independent-base coordinate $m$ for this raw cumulant calculation, with $\chi_i^m \coloneqq m_i(1-m_i)$ and $s_{ij}^m \coloneqq \sum_{k\neq i,j} H_k m_k$. The corresponding local two-site Hamiltonian is
\begin{equation}
    H_{\alpha,ij}^{\rm loc} = s_{ij}^m +H_i x_i+H_j x_j+\alpha J_{ij}x_i x_j .
    \label{eq:H_alpha_local}
\end{equation}
By~\eqref{eq:wc_eta_gamma_expansion}, $A_r=\mathrm{Cov}_0(F_r,u+\tfrac12u^2)$ with $u=\theta^kF_k$, so $B_{ij}$ in~\eqref{eq:wc_Bij_def} is the joint cumulant of $x_i$, $x_j$ and $u+\tfrac12u^2$ under $p_0$. The local closure replaces $u$ by $H_{\alpha,ij}^{\rm loc}$ and $p_0$ by the local two-site model, which gives the local observable
\begin{equation}
    L_{\alpha,ij}^{\rm loc} = H_{\alpha,ij}^{\rm loc}
    +\frac{1}{2}\left(H_{\alpha,ij}^{\rm loc}\right)^2 .
    \label{eq:L_alpha_local}
\end{equation}
Hence, within the local closure approximation,
\begin{equation}
    B_{\alpha, ij}^{\rm loc}
    \coloneqq \kappa_{\alpha}^{\rm loc}
    \left(x_i,x_j,L_{\alpha,ij}^{\rm loc}\right),
    \label{eq:B_Omega_local_def}
\end{equation}
where $\kappa_{\alpha}^{\rm loc}$ denotes cumulants computed in the local two-site model.
We expand $B_{\alpha,ij}^{\rm loc}$ to first order in $\alpha$. At $\alpha=0$, the local measure factorizes over $x_i$ and $x_j$, with means $m_i$ and $m_j$, while $s_{ij}^m$ is a constant background. We write $\kappa_0^{\rm loc}$ for the joint cumulants of this independent two-site Bernoulli measure. In this limit, $H_{0,ij}^{\rm loc} = s_{ij}^m +H_i x_i+H_j x_j$ and so only the coefficient of $x_i x_j$ in $L_{0,ij}^{\rm loc}$ contributes to the three-point cumulant. Since this coefficient is $H_i H_j$, we obtain
\begin{equation}
    B_{ij}^{(0),{\rm loc}}= H_i H_j\, \kappa_0^{\rm loc}(x_i,x_j,x_i x_j) = \chi_i^m \chi_j^m H_i H_j .
    \label{eq:B0_local}
\end{equation}
We next compute the first derivative of $B_{ij}^{\rm loc}(\alpha)$ at $\alpha=0$. There are two contributions, the explicit $\alpha$-dependence of $L_{\alpha,ij}^{\rm loc}$ and the change of the local measure generated by the interaction $J_{ij}x_i x_j$. This gives,
\begin{equation}
    B_{ij}^{(1),\rm loc} = J_{ij} \left[ \kappa_0^{\rm loc} (x_i,x_j,(1+H_{0,ij}^{\rm loc})x_i x_j) + \kappa_0^{\rm loc} (x_i,x_j,L_{0,ij}^{\rm loc},x_i x_j)\right] .
    \label{eq:B1_local_general}
\end{equation}
The first cumulant in~\eqref{eq:B1_local_general} is immediate, because $(1+H_{0,ij}^{\rm loc})x_i x_j = (1+ s_{ij}^m +H_i+H_j)x_i x_j$ and therefore
\begin{equation}
    \kappa_0^{\rm loc} \left(x_i,x_j,(1+H_{0,ij}^{\rm loc})x_i x_j\right)
    = \chi_i^m\chi_j^m(1+s_{ij}^m+H_i+H_j).
    \label{eq:first_cumulant_result}
\end{equation}
For the second cumulant in \eqref{eq:B1_local_general}, we expand
\begin{equation}
    L_{0,ij}^{\rm loc}
    = {\rm const} + \bigg[ H_i(1+s_{ij}^m)+\frac{1}{2}H_i^2 \bigg]x_i + \bigg[H_j(1+s_{ij}^m) + \frac{1}{2}H_j^2 \bigg]x_j + H_i H_j x_i x_j.
\end{equation}
The constant term does not contribute to the cumulant. The required cumulants of the independent Bernoulli measure are
\begin{align}
    \kappa_0^{\rm loc}(x_i,x_j,x_i,x_i x_j) &= \chi_i^m\chi_j^m (1-2m_i)\nonumber \\
    \kappa_0^{\rm loc}(x_i,x_j,x_j,x_i x_j) &= \chi_i^m\chi_j^m(1-2m_j)\nonumber \\
    \kappa_0^{\rm loc}(x_i,x_j,x_i x_j,x_i x_j) &= \chi_i^m\chi_j^m(1-4m_i m_j).
    \label{eq:bernoulli_cumulants}
\end{align}
Combining \eqref{eq:first_cumulant_result} and \eqref{eq:bernoulli_cumulants} and setting $\alpha=1$,
we obtain the raw local expansion
\begin{equation}
    B_{ij}^{\rm loc} = \chi_i^m\chi_j^m \left[ H_i H_j + J_{ij}\mathcal{C}_{ij}^{\rm loc}(m,H) \right] + O(J_{ij}^2),
    \label{eq:B_loc_raw_expansion}
\end{equation}
where
\begin{equation}
    \mathcal{C}_{ij}^{\rm loc}(m,H) = 1 + s_{ij}^m + H_i + H_j + \left[H_i(1+s_{ij}^m) + \frac{1}{2}H_i^2 \right](1-2m_i) + \left[H_j(1+s_{ij}^m) + \frac{1}{2}H_j^2 \right](1-2m_j) + H_i H_j (1-4 m_i m_j).
\end{equation}
The expansion~\eqref{eq:B_loc_raw_expansion} is a statement about the raw covariance numerator, naturally factored by the auxiliary susceptibility $\chi_i^m\chi_j^m$. To insert it into the inverse relation~\eqref{eq:wc_yij_forward_inverse}, it must be converted to the escort-normalized denominator $\chi_i^\eta\chi_j^\eta$. Since $B_{ij}^{\rm loc}$ enters $y_{ij}$ multiplied by $\gamma$, this conversion is needed only at $\gamma=0$, where it uses the shift of $\chi_i^m$ with $J_{ij}$ derived below. The $\pazocal{O}(\gamma)$ shift of $\chi_i^\eta$ in~\eqref{eq:chi_eta_chi_m_inverse} below is needed for the prefactor of $C_{ij}^{\rm loc}$.
At $J=0$, the singleton escort map, to linear order in $\gamma$, has the form $\eta_i \simeq m_i-\gamma A_i^{{}_{(0)}}(m,H)$, with $A_{\vphantom{i}}^{{}_{(0)}}$ previously calculated. Consequently,
\begin{equation}
    \chi_i^\eta = \eta_i(1-\eta_i) \simeq
    \chi_i^m \left[ 1+\gamma d_i^{(\gamma)}(m,H) \right]
    \label{eq:chi_eta_chi_m_inverse}
\end{equation}
with
\begin{equation}
    d_i^{(\gamma)}(m,H) \coloneqq
    -\frac{(1-2m_i)A_i^{(0)}(m,H)}{\chi_i^m} =
    -\frac{1}{2}(1-2m_i)H_i \left[2+2s^m+H_i(1-2m_i)\right].
    \label{eq:di_gamma_inverse_m}
\end{equation}
Here $s^m\coloneqq\sum_{k}H_k m_k$ is $s$ of~\eqref{eq:wc_Ai_J0} with $p_k$ replaced by $m_k$. There is also a local susceptibility-normalization contribution from the direct edge itself. At $\gamma=0$,
\begin{equation}
    \frac{\partial m_i}{\partial J_{ij}}\bigg|_{J_{ij}=0} =
    \operatorname{Cov}_0(x_i,x_ix_j) = \chi_i^m m_j ,
    \label{eq:dm_dJ_inverse}
\end{equation}
so to linear order in $J_{ij}$,
\begin{equation}
    \chi_i^m(J_{ij}) \simeq
    \chi_i^m \left[1+J_{ij}(1-2m_i)m_j\right].
\end{equation}
Therefore
\begin{equation}
    \chi_i^m(J_{ij})\chi_j^m(J_{ij}) = \chi_i^m\chi_j^m \left[1+J_{ij}d_{ij}^{(J)}(m) \right] + \pazocal{O}(J_{ij}^2),
    \label{eq:chi_pair_J_inverse}
\end{equation}
where
\begin{equation}
    d_{ij}^{(J)}(m) \coloneqq (1-2m_i)m_j+(1-2m_j)m_i .
    \label{eq:dJ_inverse_m}
\end{equation}
At $\gamma=0$, $\chi_i^\eta\chi_j^\eta=\chi_i^m(J_{ij})\chi_j^m(J_{ij})$, so dividing~\eqref{eq:B_loc_raw_expansion} by~\eqref{eq:chi_pair_J_inverse} gives $B_{ij}^{\rm loc}/(\chi_i^\eta\chi_j^\eta)\simeq H_iH_j+J_{ij}\bigl(\mathcal C_{ij}^{\rm loc}-H_iH_j\,d_{ij}^{{}_{(J)}}\bigr)$. Combining these normalization effects, the local deformation correction in the escort-normalized inverse coordinate is
\begin{equation}
    \Omega_{ij}^{\rm loc} \simeq H_i H_j + J_{ij}\widehat{\mathcal C}_{ij}^{\rm loc}
    \label{eq:Omega_local_normalized}
\end{equation}
where
\begin{equation}
    \widehat{\mathcal C}_{ij}^{\rm loc} \coloneqq
    \mathcal C_{ij}^{\rm loc}(m,H)
    + d_i^{(\gamma)}(m,H) + d_j^{(\gamma)}(m,H)
    - H_i H_j d_{ij}^{(J)}(m).
    \label{eq:Chat_inverse_m}
\end{equation}
The terms $d_i^{{}_{(\gamma)}}+d_j^{{}_{(\gamma)}}$ convert the $\gamma$-shifted singleton susceptibilities to the escort-coordinate denominator in \eqref{eq:wc_yij_inverse_def}. The term proportional to $d_{ij}^{{}_{(J)}}$ is the corresponding direct-edge susceptibility normalization of the leading deformation contribution $H_i H_j$. Using \eqref{eq:Omega_local_normalized}, the normalized local forward relation becomes
\begin{equation}
    y_{ij} \simeq J_{ij} + c_{ij}^{\eta}J_{ij}^2 - \gamma H_i H_j
    - \gamma J_{ij}\widehat{\mathcal C}_{ij}^{\rm loc}.
    \label{eq:y_forward_normalized_inverse}
\end{equation}
Equivalently,
\begin{equation}
    c_{ij}^{\eta}J_{ij}^2 +
    \left( 1-\gamma\widehat{\mathcal C}_{ij}^{\rm loc}\right)J_{ij}
    -\left(y_{ij}+\gamma H_i H_j\right)=0.
\end{equation}
The small-coupling branch is therefore
\begin{equation}
    J_{ij} =
    \frac{
        -\big(1-\gamma\widehat{\mathcal C}_{ij}^{\rm loc}\big)
        +
        \sqrt{
            \big(1-\gamma\widehat{\mathcal C}_{ij}^{\rm loc}\big)^2
            +
            4c_{ij}^{\eta}
            \big(y_{ij}+\gamma H_i H_j\big)
        }
    }
    {2c_{ij}^{\eta}} .
    \label{eq:quadratic_brancH_inverse_corrected}
\end{equation}
Expanding this branch perturbatively gives
\begin{equation}
    J_{ij}^{\rm TAP,\gamma}
    \simeq y_{ij}
    + \gamma H_i H_j - c_{ij}^{\eta} y_{ij}^2 +
    \gamma \left[ \widehat{\mathcal C}_{ij}^{\rm loc} - 2c_{ij}^{\eta}H_i H_j \right] y_{ij} .
    \label{eq:J_TAP_gamma_inverse_chat}
\end{equation}
It is important to notice that at this point replacing
$m_i$ by $\eta_i$ inside coefficients already multiplied by $\gamma$ changes the estimator only at $O(\gamma^2)$. Hence, in the final inverse formula we write $\mathcal C_{ij}^{\rm loc}$, $d_i^{{}_{(\gamma)}}$, and $d_{ij}^{{}_{(J)}}$ as functions of $\eta$.

In~\eqref{eq:J_TAP_gamma_inverse_chat} we set $H_i=h_i=\mathrm{logit}(\eta_i)$, also inside $\widehat{\mathcal C}_{ij}^{\rm loc}$, so that $s^m= h^k\eta_k$ and $s_{ij}^m=\sum_{k\neq i,j}h_k\eta_k$. The fields then follow from the deformed TAP equation~\eqref{eq:wc_deformed_tap} solved for $H_i$,
\begin{equation}\label{eq:wc_field_estimator}
H_i^{\rm TAP,\gamma}
    = h_i - \sum_{j\neq i}J_{ij}^{\rm TAP,\gamma}\eta_j
    - \frac{1-2\eta_i}{2}\sum_{j\neq i}\bigl(J_{ij}^{\rm TAP,\gamma}\bigr)^2\chi_j^\eta
    - \gamma\Delta_i\bigl(\bv\eta,\bv h,\bv J^{\rm TAP,\gamma}\bigr),
\end{equation}
where $\Delta_i$~\eqref{eq:wc_Delta_NMF_TAP}, with sectors~\eqref{eq:wc_def_sectors}, uses $A_i^{(0)}$~\eqref{eq:wc_Ai_J0} and $A_i^{(1)}$~\eqref{eq:wc_Ai_alpha1} at $\bv H=\bv h$, so that $p_i=\eta_i$. No iteration is performed: the couplings and $\Delta_i$ are not re-evaluated at $H_i^{\rm TAP,\gamma}$. Relative to the fixed-$\bv\eta$ expansion~\eqref{eq:wc_fixed_eta_field_branch}--\eqref{eq:wc_fixed_eta_field_response}, taken at $\alpha=1$ and $\bv J=\bv J^{\rm TAP,\gamma}$ with its coefficients evaluated at $\bv H=\bv h$, equation~\eqref{eq:wc_field_estimator} includes the Onsager term of~\eqref{eq:wc_deformed_tap} and omits the last term of $K_i^{(1)}$, a contribution of order $\gamma\alpha$, because $\Delta_i$ is evaluated at $\bv h$.

\paragraph*{Inverse at fixed escort moments.}
We hold all singleton and pair escort moments fixed as $\gamma$ varies. At $\gamma=0$ we have $m_i=\eta_i$ and $C_{ij}=Q_{ij}$. With $h_i=\mathrm{logit}(\eta_i)$, the local pair covariance~\eqref{eq:wc_Cij_loc_inverse_eta} and the singleton TAP relation~\eqref{eq:wc_tap} give the flat estimates
\begin{equation}\label{eq:wc_fixed_full_eta_flat}
    \begin{aligned} J_{ij}^{\rm flat}&=y_{ij}-c_{ij}^{\eta}y_{ij}^2,\\ H_i^{\rm flat}&=h_i-\sum_{j\ne i}J_{ij}^{\rm flat}\eta_j-\frac{1-2\eta_i}{2}\sum_{j\ne i}(J_{ij}^{\rm flat})^2\chi_j^\eta.
    \end{aligned}
\end{equation}
At fixed parameters, an infinitesimal deformation changes $\eta_r$ by $-A_r d \gamma$ according to~\eqref{eq:wc_eta_gamma_expansion}. Varying the parameters contributes ${\rm Cov}(F_r,F_b) d\theta^b$. Requiring $d\eta_r=0$ for every singleton and pair gives
\begin{equation}
    {\rm Cov}_{\gamma=0}(F_r,F_k)\frac{d\theta^k}{d\gamma}\Big|_{\bv\eta,\gamma=0} =\theta^k\operatorname{Cov}_{\gamma=0}(F_r,F_k) + \tfrac12\theta^k\theta^\ell\operatorname{Cov}_{\gamma=0}(F_r,F_kF_\ell)
\end{equation}
Since $u(\bv{x})=\theta^kF_k(\bv{x})$, the field and coupling derivatives are the coefficients of $x_i$ and $x_ix_j$ in the projection of $u+u^2/2$ onto these features in the covariance inner product.
Since paths $i\to k\to j$ first contribute at order $\alpha^2$ in~\eqref{eq:wc_cov_tap}, the full covariance and its local closure~\eqref{eq:wc_Cij_loc_inverse_eta} give the same first-order couplings, $J_{ij}\simeq y_{ij}$. At this order, the corresponding fields follow from~\eqref{eq:wc_tap} as $H_i\simeq h_i-\sum_{j\ne i}y_{ij}\eta_j$. Substituting into the undeformed potential and collecting the terms for each pair gives,
\begin{equation}
    \theta^r F_r(\bv{x}) \simeq h^i x_i+\sum_{i<j}y_{ij}\big[(x_i-\eta_i)(x_j-\eta_j)-\eta_i\eta_j\big].
\end{equation}
At zeroth order, $x_i^2=x_i$ gives the coefficients $h_i+\tfrac12 h_i^2$ and $h_i h_j$ of $x_i$ and $x_ix_j$ in $u+u^2/2$. The term linear in $y$ follows by multiplying the pair sum above by $1+\sum_k h_kx_k$. Because this term is already of order $\alpha$, evaluating its projection under the independent Bernoulli measure with means $\eta_i$ changes the result only at order $\alpha^2$.

At this reference, $s$ in~\eqref{eq:wc_Ai_J0} is $h^k \eta_k$, so $h^k x_k=s+ h^k(x_k-\eta_k)$. For a given pair $i,j$, the product $(x_i-\eta_i)(x_j-\eta_j)(x_k-\eta_k)$ is orthogonal to every singleton and pair when $k\ne i,j$, since one centered factor averages to zero independently. For the two repeated indices we use $(x_k-\eta_k)^2=\chi_k^\eta+(1-2\eta_k)(x_k-\eta_k)$, in particular, the coefficient of $x_i$ in $(x_i-\eta_i)(x_j-\eta_j)^2$ is $\chi_j^\eta-\eta_j(1-2\eta_j)=\eta_j^2$. Collecting the coefficients of $x_ix_j$ and $x_i$ gives the inverse used in the figures,
\begin{equation} \label{eq:wc_fixed_full_eta_response}
    \begin{aligned} 
        \widehat J_{ij}&=J_{ij}^{\rm flat}+\gamma\Big\{h_i h_j+y_{ij}\big[1+s+h_i(1-2\eta_i)+h_j(1-2\eta_j)\big]\Big\},\qquad i<j \\
        \widehat H_i &=H_i^{\rm flat}+\gamma\bigg\{h_i+\tfrac12 h_i^2-h_i\sum_{k<\ell}y_{k\ell}\eta_k\eta_\ell-\sum_{j\ne i}y_{ij}\eta_j\big[1+s+h_i(1-2\eta_i)-h_j\eta_j\big]\bigg\}.
    \end{aligned}
\end{equation}
At fixed input moments, the deformation correction in~\eqref{eq:wc_fixed_full_eta_response} has error $\pazocal{O}(\gamma\alpha^2,\gamma^2)$, in addition to the $\pazocal{O}(\alpha^2)$ error of the flat closure.

The local-closure coupling estimator~\eqref{eq:J_TAP_gamma_inverse_chat} is not the same as $\widehat J_{ij}$ in~\eqref{eq:wc_fixed_full_eta_response}. With $H_i=h_i$ and $m_i=\eta_i$, the definition~\eqref{eq:Chat_inverse_m} simplifies to $\widehat{\mathcal C}_{ij}^{\rm loc}=1+s_{ij}^m+h_i+h_j+2c_{ij}^{\eta}h_ih_j$, and the two coupling estimators differ by exactly $J_{ij}^{\rm TAP,\gamma}-\widehat J_{ij}=\gamma\,y_{ij}(h_i\eta_i+h_j\eta_j)$. At $\bv J=0$, where $y_{ij}=\pazocal{O}(\gamma)$, this difference is of order $\gamma^2$, so the two agree. For $\bv J\neq0$ it is of order $\gamma\alpha$, the same order as the correction terms of both estimators. Since~\eqref{eq:wc_fixed_full_eta_response} is complete at this order, the deformation correction in~\eqref{eq:J_TAP_gamma_inverse_chat} has error $\pazocal{O}(\gamma\alpha)$, compared with $\pazocal{O}(\gamma\alpha^2,\gamma^2)$ for~\eqref{eq:wc_fixed_full_eta_response}. In exact enumeration with $N=4$ at fixed escort moments, halving $\alpha$ from $0.05$ to $0.025$ reduces the error of the coefficient of $\gamma$ about fourfold for~\eqref{eq:wc_fixed_full_eta_response}, but only about twofold for~\eqref{eq:J_TAP_gamma_inverse_chat}.

\clearpage

\section{Exact-enumeration benchmark of deformed inverse recovery ($N=10$)}
\label{supp_sec:exact_inversion}

Deformation changes the fields and couplings needed to reproduce given escort moments. We test whether the closed-form inverse~\eqref{eq:Jij_inverse_main} recovers these changes using exact enumeration, which removes statistical sampling error from the comparison.
The undeformed reference is a heterogeneous network with $N=10$, $\beta=1$, couplings
\begin{equation}
    J_{ij}=\frac{J}{N} +\frac{1}{2 N}z_{ij}    
\end{equation}
at $J=1$, and fields $H_i=\bar H+0.5\,z_i$,  
where $z_i$ and $z_{ij}$ are independent standard-normal draws from NumPy's default generator with seed 21. 
We set $\bar H=-1.6435$ from the disorder-averaged calibration~\eqref{eq:supp_sk_calib_noisy}, which gives the exact mean activity $m=\mathbb E_{p_0}[K]/N=0.201$. 
The relation to the disorder-averaged calibration~\eqref{eq:supp_sk_calib_noisy} is discussed in Supplementary Note~\ref{supp_sec:meanfield_phase_calibration}.
We evaluate the reference escort moments $\eta_i^\ast$ and $\eta_{ij}^\ast$ by summing over all $2^N$ configurations, then hold them fixed as $\gamma'=N\beta\gamma$ varies. At each deformation, all fields and couplings are determined numerically so that the exact distribution reproduces these same moments. 
The plotted probabilities and observables are evaluated under the escort distribution. We compare these exact reference parameters with the estimates obtained from~\eqref{eq:Jij_inverse_main} at the prescribed $\gamma$, and with those obtained by setting $\gamma=0$ in the inverse. Both estimates use the same input moments. At $\gamma'=0$ they coincide, with a residual error from the weak-correlation expansion and local closure. \newline

The coupling correction explains why including deformation matters even when the input moments are exact. At fixed escort moments, the undeformed estimate is unchanged, while~\eqref{eq:Jij_inverse_main} adds the joint field and coupling response linear in $\gamma$. The leading coupling contribution is $\gamma h_i h_j$, where $h_i=\mathrm{logit}(\eta_i)$; the retained covariance-dependent terms modify this contribution. For the negative logits used here, this leading term raises the inferred couplings at positive deformation and lowers them at negative deformation. Setting $\gamma=0$ instead attributes the deformation-induced contribution to the correlations entirely to ordinary pairwise couplings. Including the deformation substantially reduces the coupling error at both displayed values $\gamma'=\pm0.4$ (Supplementary Fig.~\ref{fig:sim_invising_weakcorr_exact}c,d).
Field recovery depends more strongly on the sign of the deformation. Relative to the $\gamma=0$ inverse, the perturbative correction reduces the field error for $\gamma'>0$ but increases it for $\gamma'<0$.
Parameter errors affect the probability distribution through combined changes in fields and couplings, with different sensitivities for different parameter combinations. \newline 

Fixing the first two escort moments still allows the configuration probabilities, silence probability, entropy, and energy variance to vary with deformation. Each fitted escort distribution is evaluated at the deformation used in its inversion. The deformed inverse closely reproduces the configuration probabilities and captures the decline in silence probability, although it overestimates the latter (Supplementary Fig.~\ref{fig:sim_invising_weakcorr_exact}a,f). Entropy of the exact reference changes little across the sweep, and the $\gamma=0$ inverse is generally closer to the reference for this observable (Supplementary Fig.~\ref{fig:sim_invising_weakcorr_exact}g). The deformed inverse captures the decline in energy variance, although it slightly overestimates it, while the $\gamma=0$ inverse gives a constant value (Supplementary Fig.~\ref{fig:sim_invising_weakcorr_exact}h).

\clearpage

\section{Encoding invariance and regime dependence of negative deformation}
\label{supp_sec:deformation_sign_invariance}

The neural populations analyzed in the main text---cortical and subcortical---select a negative deformation, $\gamma'^\star_{\mathrm{mm}}<0$, associated with net suppression of coincident firing relative to the fitted pairwise models. Relating this result to the explosive transitions and enhanced memory capacity found at negative deformation in~\cite{aguilera2025explosive} requires accounting for the change from our $x_i \in \{0,1\}$ spike variables to $s_i \in \{-1,+1\}$ spins, since the transformation introduces an additive energy constant that affects the deformed distribution. \newline

The map $s_i=2x_i-1$ assigns spin values $+1$ and $-1$ to firing and silence within a time bin, respectively, rewriting the energy~\eqref{eq:energy} as $E(\bv{x})=E_s(\bv{s})+c$. The corresponding spin fields are $h_i=\tfrac12 H_i+\tfrac14\sum_{j\neq i}J_{ij}$, the couplings are $J^s_{ij}=J_{ij}/4$, and the energy shift is
\begin{equation}\label{eq:supp_encoding_constant}
    c=-\frac12\sum_i H_i-\frac14\sum_{i<j}J_{ij}.
\end{equation}
As in Supplementary Note~\ref{supp_sec:meanfield_phase_calibration}, these coupling coefficients include the $1/N$ factor written explicitly in~\cite{aguilera2025explosive}. Since $c$ is the uniform average of $E$ over all configurations, the full-support condition $1-\gamma\beta E(\bv{x})>0$ implies $1-\gamma\beta c>0$, allowing the constant to be absorbed through
\begin{equation}
    \bigl[1-\gamma\beta(E_s+c)\bigr]^{1/\gamma}=(1-\gamma\beta c)^{1/\gamma}\bigl(1-\gamma\beta_{\rm eff}E_s\bigr)^{1/\gamma},\qquad \beta_{\rm eff}=\frac{\beta}{1-\gamma\beta c}>0.
    \label{eq:supp_shift_beta_eff}
\end{equation}
The constant prefactor cancels upon normalization, leaving $\gamma$ unchanged and preserving the sign of $\gamma'=N\beta\gamma$, whose magnitude rescales with the inverse temperature. This transformation also preserves the state-dependent inverse temperature entering the mean-field activation function, 
\begin{equation}
    \beta'(\bv{x})=\beta/[1-\gamma\beta E(\bv{x})]=\beta_{\rm eff}/[1-\gamma\beta_{\rm eff}E_s(\bv{s})].
\end{equation}

For a single stored pattern $\bv{\xi}$, with $\xi_i\in\{-1,+1\}$, the zero-field associative-memory model has couplings $J^s_{ij}=J_s\xi_i\xi_j/N$, where $J_s>0$ sets their strength. Its exact energy is $E_s=-\tfrac12 J_s(Nm_s^2-1)$, where $m_s=N^{-1}\sum_i\xi_i s_i$ measures the overlap between the network state and the stored pattern. For $\gamma<0$, increasing $|m_s|$ lowers the energy and raises $\beta'$, reinforcing retrieval and producing an explosive transition with hysteresis when the feedback is sufficiently strong~\cite{aguilera2025explosive}. \newline

For inhibitory fields $H_i<0$, activating a set of neurons $A$ costs 
\begin{equation}
    E(A)=\sum_{i\in A}|H_i|-\sum_{i<j,\;i,j\in A}J_{ij}
\end{equation}
relative to the silent state. When this cost is positive for every nonempty $A$, silence is the ground state and negative deformation gives $\beta'(A)<\beta$ on all active configurations. For homogeneous fields $H_i=H$ and couplings $J_{ij}=J/N$, the energy per neuron approaches $e(m)=-(Hm+Jm^2/2)$ as $N\to\infty$, where $m=N^{-1}\sum_i x_i$ is the fraction of active neurons. Increasing activity therefore raises the energy and, for $\gamma<0$, lowers $\beta'$ wherever $H+Jm<0$. For $J>0$, the mean-field fixed-point condition excludes a majority-active solution with $m>1/2$ when $H\leq-J$, although several fixed points with $m<1/2$ may remain (Supplementary Note~\ref{supp_sec:meanfield_phase_calibration}).

\begin{figure}[p]
    \centering
    \includegraphics[width=\linewidth]{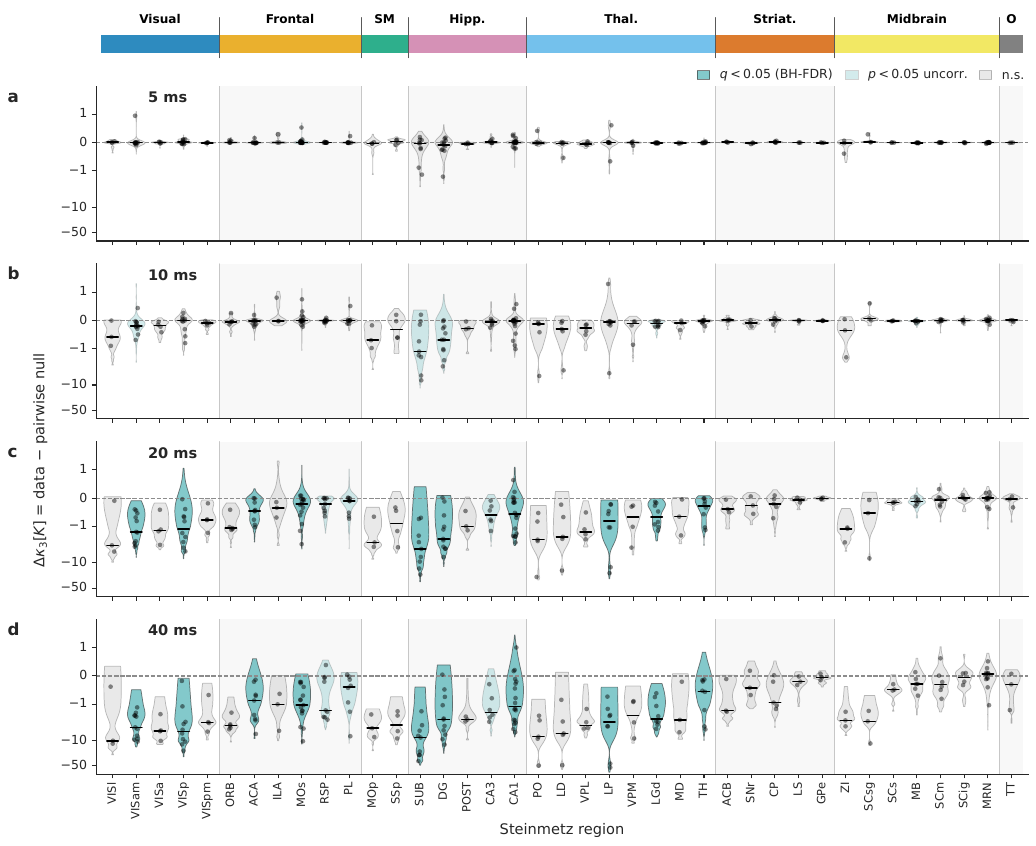}
    \caption{\textbf{Brain-wide third-cumulant deviations from the variance-matched pairwise null across bin sizes.}
    The unlettered top strip shows the anatomical-group guide for the $39$ recorded Steinmetz regions, using the grouping and display order of the eight groups listed in the \hyperref[sec:regional_selection]{Methods} (SM, sensorimotor; O, other); regions are arranged by anatomical group and ordered within each group by the $20$~ms median $\Delta\kappa_3[K]$.
    \textbf{(a)}~Spontaneous-block analysis with $5$~ms bins. \textbf{(b)}~The same analysis with $10$~ms bins. \textbf{(c)}~The same analysis with $20$~ms bins. \textbf{(d)}~The same analysis with $40$~ms bins. In each panel, $\Delta\kappa_3[K]$ is the third-cumulant residual defined in the \hyperref[sec:regional_selection]{Methods}; its null is the full-Boltzmann pairwise model matched to all firing rates and pairwise moments. Violin shapes show all random $N=30$ subsamples in the region; black points are per-session medians, and black horizontal bars show the across-session median. Significance is the equal-session-weighted cluster sign-flip test (two-sided), applied per region. Strongly colored violins mark regions that survive Benjamini--Hochberg false-discovery-rate control across all $39$ regions ($q<0.05$); lightly tinted violins are nominally significant before correction ($p<0.05$, uncorrected) but do not survive it; gray violins are not significant. No region survives correction at $5$ or $10$~ms; the same $10$ regions survive at both $20$ and $40$~ms, all suppressive, so the corrected higher-order signal emerges only at the coarser bins. Several large-effect regions cannot reach significance regardless of effect size because the sign-flip $p$-value is floored at $2/2^{n_{\mathrm{sess}}}$ by the number of sessions. The vertical axis uses a symmetric-log scale to display the increasingly long suppressive tails at larger bin sizes while retaining resolution around zero.}
    \label{fig:supp_kappa3_bins_violin}
\end{figure}

In the weak-correlation example of Supplementary Fig.~\ref{fig:supp_broaden_atlas}, negative deformation broadens the escort count distribution when only its mean is held fixed, whereas fixing the pairwise escort moment as well preserves the count variance and allows changes in higher cumulants. For triplets with negative fields and positive couplings,~\eqref{eq:three_body_coupling} gives a negative leading-order coefficient $\theta_{ijk}^{\mathrm{eff}}$ at $\gamma<0$ whenever
\begin{equation}
    |H_i|J_{jk}+|H_j|J_{ik}+|H_k|J_{ij}>J_{ij}J_{ik}+J_{ij}J_{jk}+J_{ik}J_{jk}.
\end{equation}
The negative field--coupling contributions then outweigh the positive coupling--coupling contributions, so the term $\theta_{ijk}^{\mathrm{eff}}x_i x_j x_k$ in the log-probability contributes a penalty for coincident firing of the three neurons. These interaction coefficients depend on the binary encoding, since the transformation redistributes contributions among interaction orders. The negative third-cumulant residual observed relative to the variance-matched pairwise model is consistent with this leading-order prediction of suppressive triplet interactions (Supplementary Fig.~\ref{fig:supp_kappa3_bins_violin}).

\clearpage

\onecolumngrid

\begin{figure}[p]
    \centering
    \includegraphics[width=\linewidth]{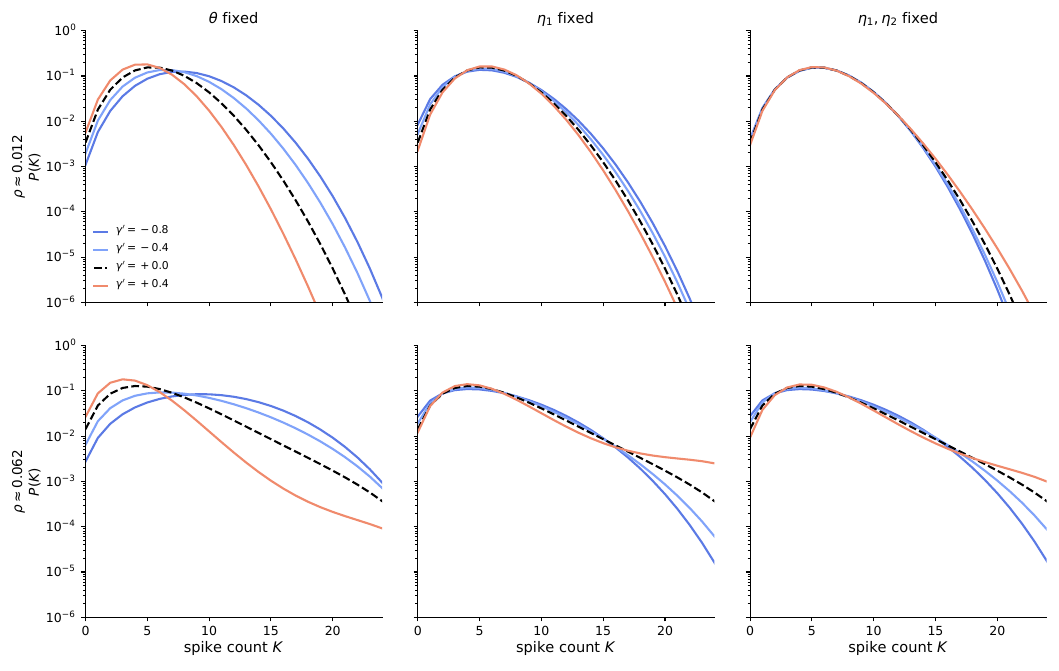}
    \caption{\textbf{How the deformation reshapes the homogeneous spike-count distribution under three calibrations.} Escort count distribution $P(K)$ of the homogeneous curved network ($N=30$, $\eta_1=0.2$), sweeping an illustrative range $\gamma'\in\{-0.8,-0.4,0,+0.4\}$. \textbf{Columns:} the quantity held fixed as $\gamma'$ varies---\emph{(left)} the natural parameters $\theta=(H,J)$ (no recalibration; both escort moments drift), \emph{(middle)} the rate $\eta_1$ only (field recalibrated, coupling held; variance free), \emph{(right)} both the rate $\eta_1$ and the pairwise moment $\eta_2$ (field and coupling recalibrated; equivalently, correlation and variance pinned at fixed $\eta_1$). \textbf{Rows:} two reference correlations, $\rho\approx0.012$ (top) and $\rho\approx0.062$ (bottom), on either side of the correlation at which the rate-fixed variance effect reverses sign. In the middle column the variance broadens for $\gamma'<0$ at weak correlation but for $\gamma'>0$ at strong correlation; in the right column the variance is fixed and the third cumulant increases monotonically with $\gamma'$.}
    \label{fig:supp_broaden_atlas}
\end{figure}

\clearpage

\begin{figure}[p]
    \centering
    \includegraphics[width=\linewidth]{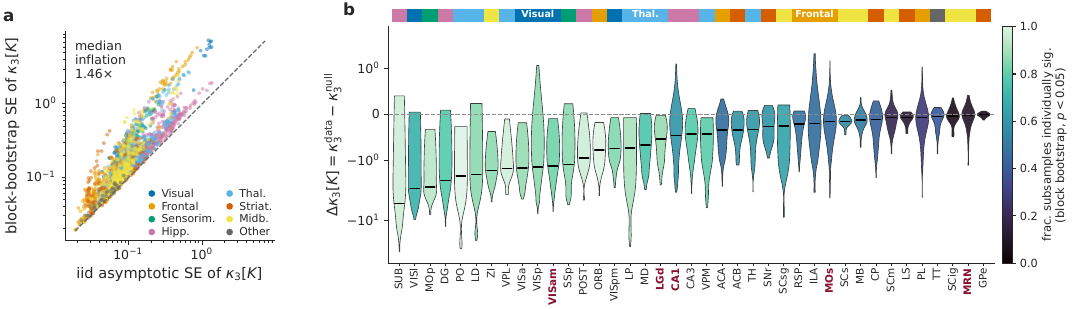}
    \caption{\textbf{Every $N=30$ subsample tested individually for suppressive higher-order structure, with resampling.} For each random $N=30$ subsample ($2271$ subsamples across the $39$ analyzed regions, $20$~ms bins) the population spike-count third cumulant is tested against its own variance-matched pairwise ($\gamma'=0$) full-Boltzmann null, one-sided suppressive, by a moving-block bootstrap ($B=800$) of the population-count series $K(t)=\sum_i x_i(t)$ that preserves its temporal autocorrelation. The block length was five integrated autocorrelation times of $K(t)$ (minimum $20$ bins; median $3.8$~s), and the fitted null value was held fixed.
    \textbf{(a)}~Moving-block-bootstrap standard error of $\kappa_3[K]$ (ordinate) against the naive i.i.d.\ asymptotic standard error (abscissa) per subsample; log--log, dashed identity; color, anatomical group.
    \textbf{(b)}~Third-cumulant residual $\Delta\kappa_3[K]=\kappa_3^{\mathrm{data}}-\kappa_3^{\mathrm{null}}$ per subsample, regions ordered by median (most suppressive left; symmetric-log $y$); each violin shaded by the fraction of its subsamples individually block-significant ($p<0.05$; mako). Top strip, anatomical-group guide (colors as in a); the five representative regions of Fig.~\ref{fig:hoi_survey_effscore} in bold.}
    \label{fig:supp_perreg_kappa3}
\end{figure}

\clearpage

\clearpage

\begin{figure}[p]
    \centering
    \includegraphics[width=\linewidth]{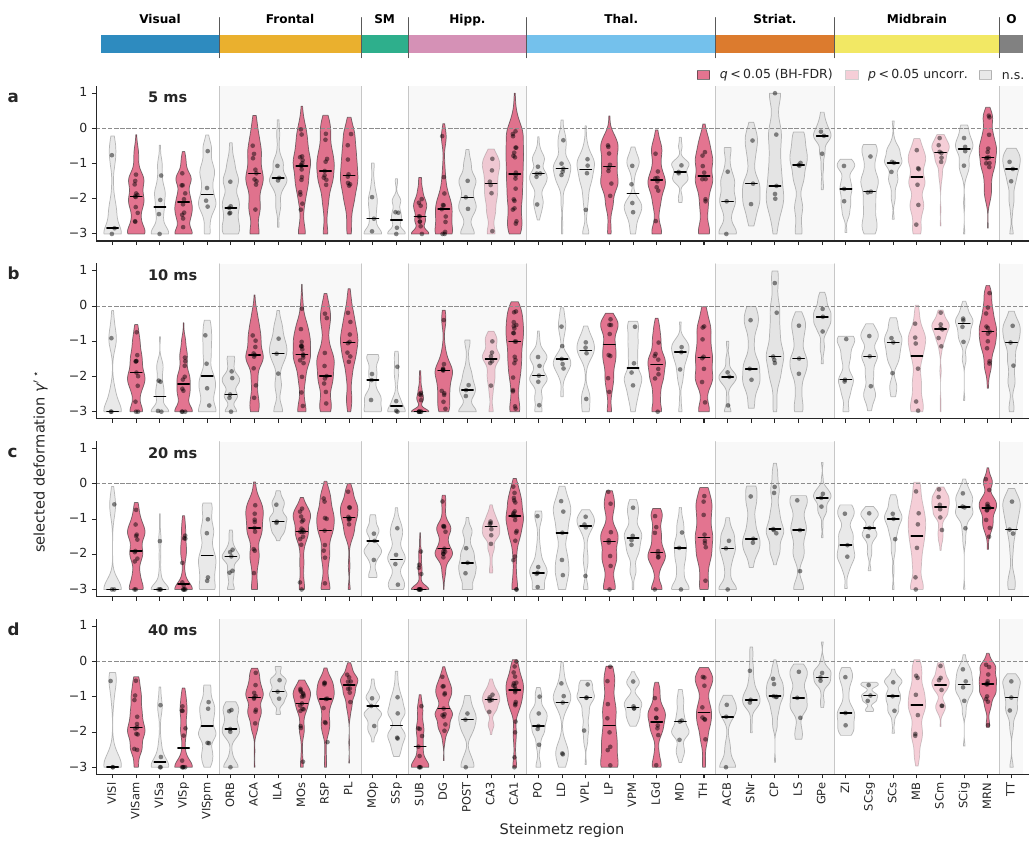}
    \caption{\textbf{Brain-wide deformation selected by the moment-matching estimator across bin sizes.} Companion to Fig.~\ref{fig:supp_kappa3_bins_violin}, with the same $39$ regions, anatomical grouping, top-strip labels (SM, sensorimotor; O, other) and display order (regions ordered within each group by the $20$~ms median $\Delta\kappa_3[K]$). For every random $N=30$ subsample the deformation is estimated by the moment-matching method (\hyperref[sec:efficient_score]{Methods}): a pairwise ($\gamma'=0$) full-Boltzmann null is fitted, the deformation score is projected onto the complement of the pairwise sufficient statistics, and $\gamma'^\star_{\mathrm{mm}}$ is read off by least-squares minimization of the profiled-statistic discrepancy over Monte-Carlo deformed fits, the grid minimum refined by a local quadratic (sub-grid) interpolation (linear within an end interval)---using sampling-based inverses for \emph{both} the pairwise and the deformed models, with no perturbative closed form.
    Panels \textbf{(a)}, \textbf{(b)}, \textbf{(c)}, and \textbf{(d)} show $\gamma'^\star_{\mathrm{mm}}$ at $5$, $10$, $20$, and $40$~ms bins, respectively. Violin shapes show all subsamples; black points are per-session medians and black bars the across-session median. Significance is assessed by the same equal-session-weighted cluster sign-flip test (two-sided), applied directly to the displayed moment-matching estimates. Strongly colored violins mark regions that survive Benjamini--Hochberg false-discovery-rate control across all $39$ regions ($q<0.05$); lightly tinted violins are nominally significant before correction ($p<0.05$) only; gray violins are not significant. The same $13$ regions survive correction at \emph{every} bin size, which is every region with enough sessions for the test to survive correction. The search is bounded to $\gamma'\in[-3,1]$; estimates reaching the lower search boundary accumulate at $\gamma'=-3$, so the magnitude there is set by the edge of the search range while the suppressive \emph{direction} is robust. Consistent with Fig.~\ref{fig:supp_kappa3_bins_violin}, the moment-matching estimate is suppressive across essentially all regions. Its depth varies nonmonotonically across bin sizes, unlike the third-cumulant significance trend.}
    \label{fig:supp_gammamap_bins}
\end{figure}

\clearpage

\clearpage

\begin{figure}[p]
    \centering
    \includegraphics[width=\linewidth]{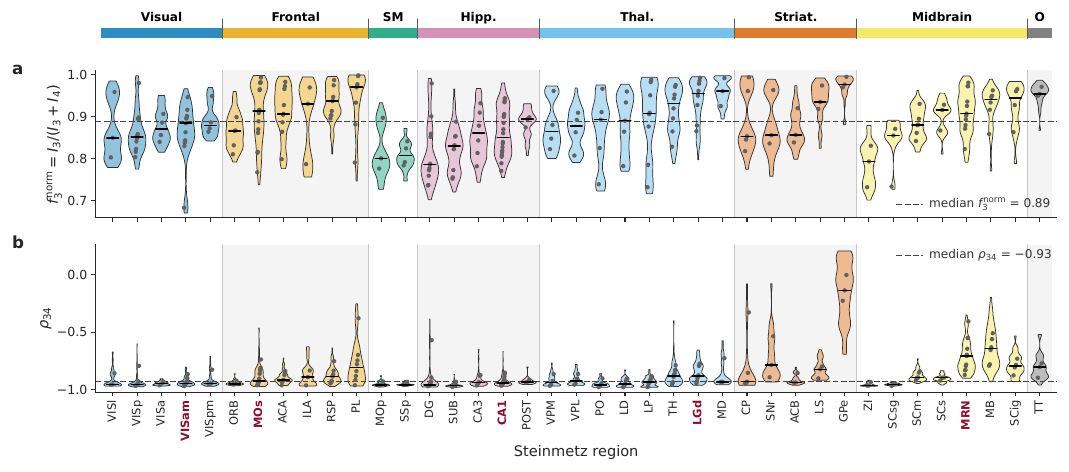}
    \caption{\textbf{Efficient-score third--fourth fractions.}
    For every random $N=30$ subsample in the $20$~ms all-region analysis ($2271$ subsamples across $39$ Steinmetz regions), the local deformation score at the pairwise null was decomposed as $q=-u^2/2=q_{\le2}+q_3+q_4$, where $q_3$ and $q_4$ contain the distinct three- and four-body monomials after binary reduction. Each component was projected off the pairwise nuisance span $F=(x_i,x_ix_j)$ under the fitted pairwise null, giving $s_0^\perp=q_3^\perp+q_4^\perp$.
    Regions are arranged by anatomical group (SM, sensorimotor; O, other) and ordered within each group by the median third-order norm fraction; bold acronyms mark the representative regions in Fig.~\ref{fig:hoi_survey_effscore}. In panels \textbf{(a)} and \textbf{(b)}, violins show all subsamples, gray points are per-session medians, and black bars are medians across sessions.
    \textbf{(a)}~Bounded third-order norm fraction $I_3/(I_3+I_4)$ with $I_j=\mathrm{Var}_0(q_j^\perp)$. The median across all subsamples is $0.89$, so the local efficient-score direction is mostly third-order in norm.
    \textbf{(b)}~Correlation $\rho_{34}=\mathrm{Cov}_0(q_3^\perp,q_4^\perp)/\sqrt{I_3I_4}$. The median is $-0.93$, showing that the fourth-order component usually cancels rather than independently reinforces the third-order direction.
    Because $q_3^\perp$ and $q_4^\perp$ are not orthogonalized against each other, panel (a) is a bounded norm diagnostic rather than an additive information fraction. The decomposition is local at $\gamma'=0$; a finite deformation can generate constrained interactions of every order up to $N$.}
    \label{fig:supp_effscore_order_fraction}
\end{figure}

\clearpage

\clearpage

\clearpage

\begin{figure}[p]
    \centering
    \includegraphics[width=0.72\linewidth]{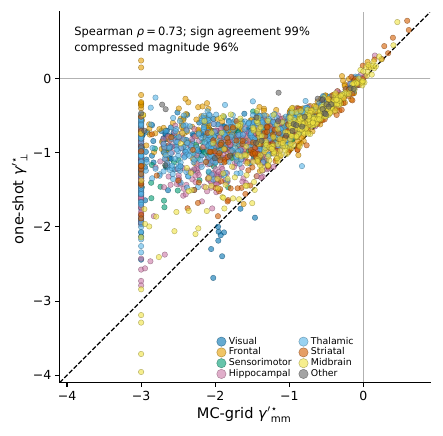}
    \caption{\textbf{The one-shot estimate recovers the direction but compresses the magnitude of the inferred curvature.} Comparison of the one-shot efficient-score estimate $\gamma'^\star_\perp$~\eqref{eq:efficient_score_estimate} (a single projection, no grid search) with the moment-matching estimate $\gamma'^\star_{\mathrm{mm}}$ (labeled MC-grid) across $2271$ random $N=30$ subpopulations from all $39$ analyzed regions ($20$~ms bins). Points are colored by anatomical group; the vertical accumulation at $\gamma'^\star_{\mathrm{mm}}=-3$ marks estimates reaching the lower boundary of the search grid. Dashed line, identity. The Spearman rank correlation, sign agreement, and fraction for which $|\gamma'^\star_\perp|<|\gamma'^\star_{\mathrm{mm}}|$ are annotated.}
    \label{fig:tap_mcmc_validation}
\end{figure}

\clearpage

\clearpage

\clearpage

\clearpage

\end{document}